\documentclass[useAMS,usenatbib]{mn2e}
\pdfoutput=1
\usepackage{graphicx}
\usepackage{color}
\usepackage{times}
\usepackage{amsmath,amssymb}
\title[The SAMI Pilot Survey: The Fundamental and Mass Planes]
{The SAMI Pilot Survey: The Fundamental and Mass Planes in Three Low-Redshift Clusters}
\author[N. Scott et al.]{Nicholas Scott,$^{1,2}$\thanks{E-mail: nscott@physics.usyd.edu.au}, L. M. R. Fogarty$^{1,2}$ , Matt S. Owers$^{3,4}$, Scott M. Croom$^{1,2}$,
\newauthor  Matthew Colless$^{5}$, Roger L. Davies$^{6}$, S. Brough$^{3}$, Michael B. Pracy$^{1}$,
\newauthor   Joss Bland-Hawthorn$^{1}$, D. Heath Jones$^{4,7}$, J. T. Allen$^{1,2}$, Julia J. Bryant$^{1,2,3}$,
\newauthor Luca Cortese$^{8}$, Michael Goodwin$^{3}$, Andrew W. Green$^{3}$, Iraklis S. Konstantopoulos$^{3}$, 
\newauthor J.S. Lawrence$^{3,4}$, Samuel Richards$^{1,2,3}$, Rob Sharp$^{5}$.
\\
$^{1}$ Sydney Institute for Astronomy, School of Physics, University of Sydney, NSW 2006, Australia.\\ 
$^{2}$ ARC Centre of Excellence for All-Sky Astrophysics (CAASTRO).\\
$^{3}$ Australian Astronomical Observatory, PO Box 915, North Ryde, NSW 1670, Australia.\\
$^{4}$ Department of Physics and Astronomy, Macquarie University, Sydney, NSW, 2109, Australia. \\
$^{5}$ Research School of Astronomy and Astrophysics, Australian National University, Canberra ACT 2611, Australia. \\
$^{6}$ Sub-Dept. of Astrophysics, Department of Physics, University of Oxford, Denys Wilkinson Building, Keble Rd., Oxford, OX1 3RH, UK. \\
$^{7}$ School of Physics, Monash University, Clayton, VIC 3800, Australia. \\
$^{8}$ Centre for Astrophysics and Supercomputing, Swinburne University of Technology, Hawthorn, VIC 3122, Australia.
}
\date{}
\pagerange{\pageref{firstpage}--\pageref{lastpage}} \pubyear{2014}

\def\LaTeX{L\kern-.36em\raise.3ex\hbox{a}\kern-.15em
    T\kern-.1667em\lower.7ex\hbox{E}\kern-.125emX}

\begin{document}

\label{firstpage}

\maketitle

\begin{abstract}
Using new integral field observations of 106 galaxies in three nearby clusters we investigate how the intrinsic scatter of the Fundamental Plane depends on the way in which the velocity dispersion and effective radius are measured. Our spatially resolved spectroscopy, combined with a cluster sample with negligible relative distance errors allows us to derive a Fundamental Plane with minimal systematic uncertainties. From the apertures we tested, we find that velocity dispersions measured within a circular aperture with radius equal to one effective radius minimises the intrinsic scatter of the Fundamental Plane. Using simple yet powerful Jeans dynamical models we determine dynamical masses for our galaxies. Replacing luminosity in the Fundamental Plane with dynamical mass, we demonstrate that the resulting Mass Plane has further reduced scatter, consistent with zero intrinsic scatter. Using these dynamical models we also find evidence for a possibly non-linear relationship between dynamical mass-to-light ratio and velocity dispersion. 
\end{abstract}


\section{Introduction}
\label{sec:intro}

Early-type galaxies occupy a thin two-dimensional surface in the three-dimensional parameter space of velocity dispersion, $\sigma$, effective radius, R$_e$ and mean effective surface brightness, $\langle \mu_e \rangle$. This surface is known as the Fundamental Plane (FP), and was first identified by \citet{Djorgovski:1987} and \citet{Dressler:1987}. The FP is usually expressed in the form: $\log \mathrm{R}_e = a \log \sigma + b\ \langle \mu_e \rangle + c$, however in this work we use the form more suited to studies of galaxy evolution, $\log L = \alpha \log \sigma + \beta \log R_e + \gamma$. This form of the the FP has the advantage that measurements of the three parameters are essentially uncorrelated, and is therefore easier to interpret in terms of galaxy evolution.

Since its discovery, numerous studies have examined the FP. Initial studies utilised samples of a few hundred objects \citep[e.g.][]{Jorgensen:1996,Hudson:1997,Scodeggio:1997,Pahre:1998,Gibbons:2001,Colless:2001} to great effect, pinning down the coefficients of the plane, as well as constraining its thickness. \citet{Jorgensen:1996} used a sample of $\sim 200$ galaxies from 10 clusters to find a plane of the form: $\log \mathrm{R}_e = a \log \sigma + b\ \langle \mu_e \rangle + c$. They found substantial variation between individual clusters, with a in the range 0.59 to 1.59 and b in the range -0.87 to -0.57.  \citet{DOnofrio:2008} used a greatly expanded sample to study the FP in 57 clusters, finding a similar variation in the FP coefficients between clusters. They noted that this variation was dependent on the distribution of galaxy properties within a given sample, and, in particular, on the luminosity distribution of galaxies within each cluster.

\citet{Mobasher:1999} and \citet{Pahre:1998} extended the study of the FP into the near-infrared, again finding a tight plane with coefficients similar, but not identical, to those in the optical. \citet{LaBarbera:2010} compared the optical and NIR FPs in a large, homogeneous sample of galaxies, finding only small variations in the FP coefficients with wavelength. Another source of variation in the FP coefficients comes from the choice of fitting method, most commonly through least-squares minimizations of the direct residuals, or the residuals orthogonal to the plane \citep[direct Maximum Likelihood fitting of 3D Gaussian models has been used to find similar forms of the FP, e.g.][]{Colless:2001,Magoulas:2012}. \citet{Bernardi:2003b} find that the direct method gives an a coefficient $\sim 1.2$ (consistent with \citet{Jorgensen:1996}, whereas the orthogonal approach yields a coefficient $\sim 1.4$. Recent studies have used samples of thousands \citep{NigocheNetro:2009} or tens of thousands of galaxies \citep{Hyde:2009,LaBarbera:2010,Magoulas:2012} to confirm these trends. Despite these variations, many studies have repeatedly found a FP with small scatter of order 0.1 dex, and it is this small scatter that is the primary appeal of the FP as a tool for understanding galaxy evolution.

\begin{figure*}
\includegraphics[width=3.25in,clip,trim=30 10 15 15]{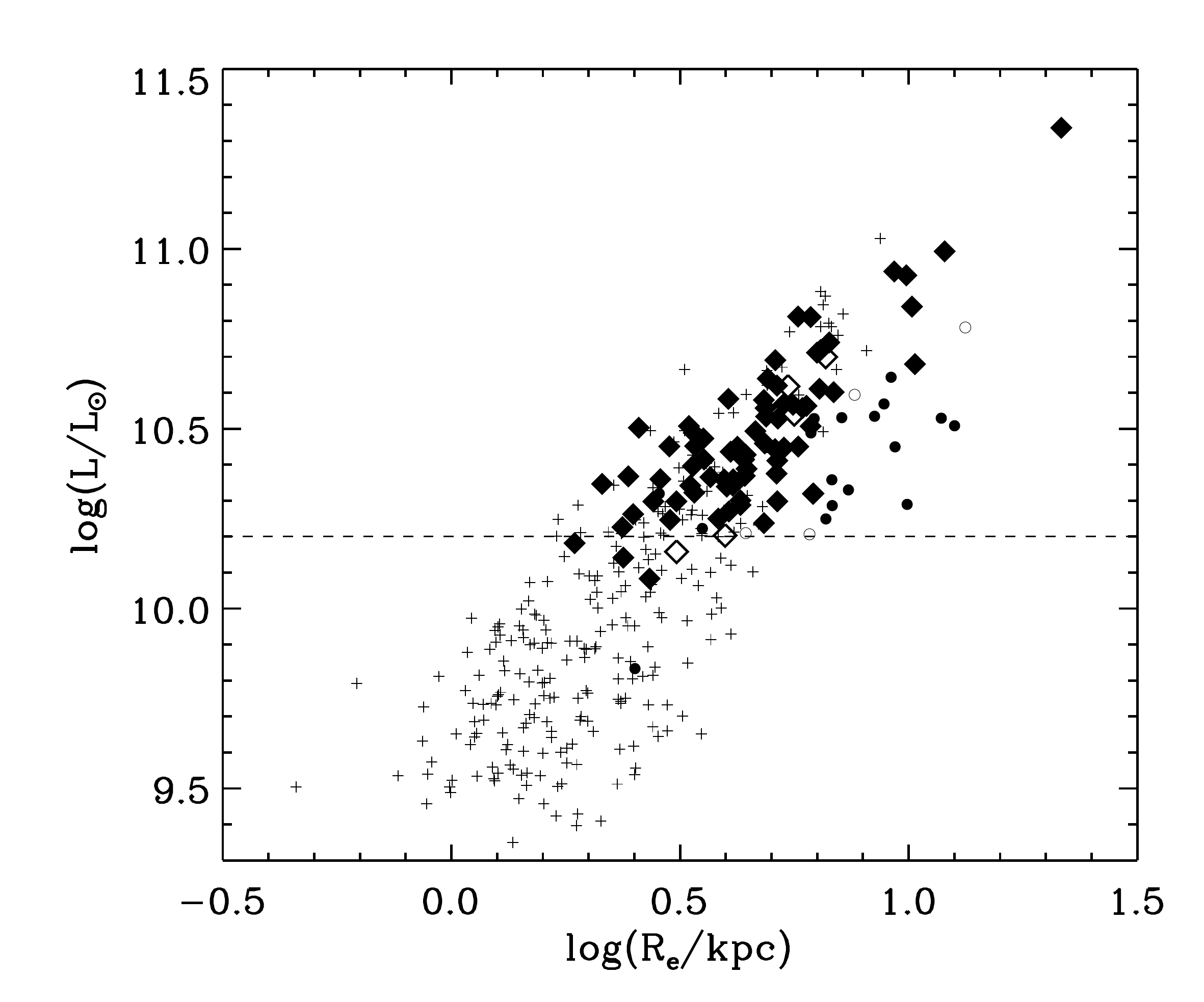}
\includegraphics[width=3.25in,clip,trim=20 10 15 15]{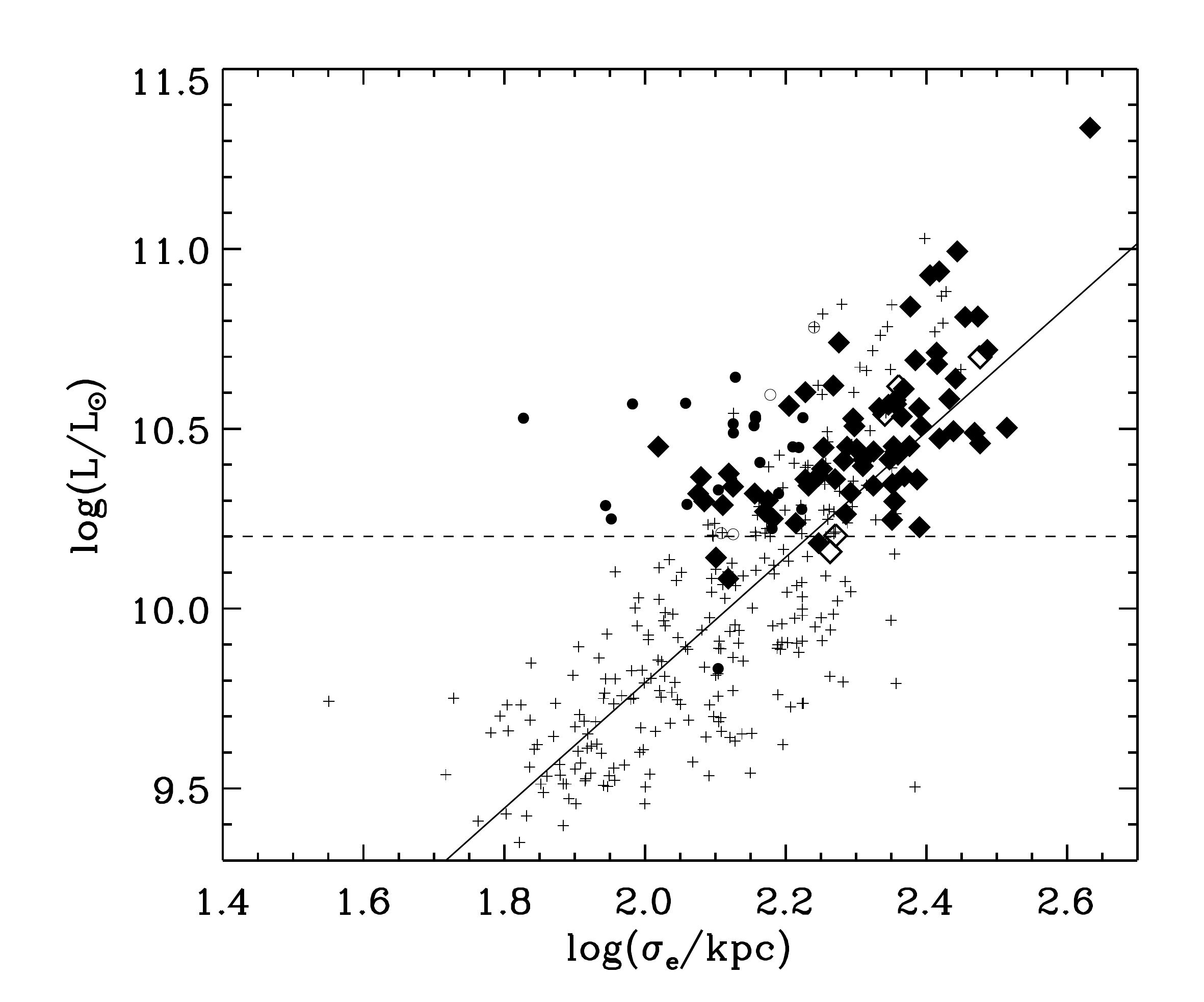}
\caption{Left panel: The distribution of our sample in the luminosity--radius plane. Filled diamonds represent early-type cluster members, open diamonds are early-type non-members. Filled circles are late-type members, open circles are late-type non-members.  For comparison we also show data points from the ATLAS$^\mathrm{3D}$ survey (crosses), and a fit to the ATLAS$^\mathrm{3D}$ data (solid line). The dashed line indicates our absolute magnitude selection limit of M$_r < -20.9$. Right panel: As left panel but for the luminosity -- velocity dispersion plane.  Relative to the volume-limited sample of the ATLAS$^\mathrm{3D}$ survey, the SAMI galaxies are significantly more luminous, but, in the region of overlap, have similar distributions.} 
\label{fig:FJ_K}
\end{figure*}

The FP provides insight into the structural properties and formation of early-type galaxies in two complementary ways. First, the Virial Theorem predicts that, if they are a family of homologous, dynamically relaxed systems, then early-type galaxies should lie on the Virial Plane, the FP with coefficients $\alpha = 2$ and $\beta = 1$. The observed FP is tilted with respect to the Virial Plane, with different slopes depending on the sample and the chosen photometric band. Two principal explanations have been suggested for this tilt: either early-type galaxies are not a homologous family and their structural properties vary systematically along the plane \citep[e.g.][]{Graham:1997}, or the stellar mass-to-light ratio (M/L$_\star$) varies in a similar way \citep[e.g.][]{Renzini:1993}. The second way in which FP analysis has informed theories of galaxy evolution has been to search for additional galaxy properties that play a controlling role in early-type galaxy evolution by searching for parameters that correlate with a galaxy's deviation from or position within the plane \citep{Graves:2009,Magoulas:2012}.

Beyond improved number statistics, two recent innovations have been applied to the study of the FP, both motivated by the Virial Theorem. The predictions of the Virial Theorem apply strictly to global properties of a galaxy, whereas the observed quantities in the FP are not truly global. In particular, the majority of FP studies have measured velocity dispersions in a small central aperture. Some studies have attempted to account for this by `correcting' $\sigma$ to either a fixed physical aperture or to a common fraction of a galaxy's radius, however such corrections introduce an additional source of scatter into the plane. The advent of the widespread use of Integral Field Spectrographs (IFSs), which provide two-dimensional spectroscopy over significant fields-of-view has removed the need for this correction, allowing either comparable apertures to be used, or, in the best cases, an aperture large enough to provide a reasonable measure of the global $\sigma$. Examples of FP studies that have used two-dimensional spectroscopy are \citet{Cappellari:2006,Jeong:2009,Falcon-Barroso:2011,Scott:2012} and \citet{Cappellari:2013a}.

The second innovation came from noting that the Virial Theorem applies to the total dynamical mass, not to the total luminosity, L. In the majority of studies the stellar light has been used as a proxy for the total mass. Several studies have used either dynamical modelling of spatially resolved spectroscopy \citep{Thomas:2007,Cappellari:2013a} or strong lensing \citep{Bolton:2007,Auger:2010} to directly determine the dynamical mass and substitute this into the FP. This version of the FP is often known as the Mass Plane (MP), and has been found to have both reduced intrinsic scatter and coefficients closer to the Virial coefficients than the standard FP.

In this study we use IFS observations from the Sydney--AAO Multi-Object IFS (SAMI) Pilot Survey \citep{Fogarty:2014a} to study the FP and MP in three clusters at redshift $\sim 0.05$. The principal advance in this study is in combining spatially resolved spectroscopy, which is largely free from bias or uncertainty associated with aperture corrections, with a sample selected from cluster environments, where the relative distance errors between galaxies in the same cluster are negligible. This is the largest study of the FP and MP in clusters using IFS data to date, and also the first to fully constrain the massive ($M > 10^{11} M_\odot$) end of the  galaxy distribution.

In Section \ref{sec:sample} we describe the selection of our sample, the IFS observations and the complimentary imaging. In the same section we describe the measurement of the FP parameters, R$_e$, $\sigma$ and $\langle \mu_e \rangle$. In Section \ref{sec:fp} we present our best-fitting planes. In Section \ref{sec:modelling} we describe the use of Jeans dynamical models to determine M$_{JAM}$, the dynamical mass as measured from Jeans models, for each of the galaxies and in Section \ref{sec:mass_results} we present the resulting MP for our sample. Finally, in  Section \ref{sec:conclusion} we summarise our conclusions.

\begin{table}
\caption{Properties of the three clusters from which our sample is drawn.}
\label{tab:cluster_props}
\centering
\begin{tabular}{lccc}
\hline
Cluster & Number of & Redshift & Distance \\
& ETG Targets & & (Mpc)\\
\hline
Abell 85 & 30 & 0.055 & 232\\
Abell 168 & 23 & 0.045 & 179\\
Abell 2399 & 44 & 0.058 & 243\\
\hline
\end{tabular}
\end{table}

\section{Sample}
\label{sec:sample}
The SAMI Pilot Survey is a study of 106 galaxies in three $z \sim 0.05$ clusters using the SAMI instrument \citep{Croom:2012} on the 3.9m Anglo-Australian Telescope at Siding Spring Observatory. The pilot survey is a precursor to the SAMI Galaxy Survey \citep[SGS][]{Bryant:2015} of $\sim 3400$ galaxies with the same instrument. While there is some overlap between the galaxies of the Pilot Survey and the SGS, the selection criteria are very different and the two samples cannot be combined in a trivial fashion. The pilot survey sample (hereafter simply the sample) was selected from three Abell clusters; Abell 85, Abell 168 and Abell 2399. Some general properties of the three clusters are summarised in Table \ref{tab:cluster_props}, but are described in detail in \citet{Fogarty:2014a}. The initial selection was drawn from the X-ray selected catalogue of \citet{Wang:2011} and included all galaxies within 1$^\circ$ of the cluster centres, in the redshift range $ 0.025 < z < 0.085$ and having an absolute r-band magnitude M$_r < -20.25$ mag in the New York University Value Added Galaxy Catalogue \citep[NYU--VAGC][]{Blanton:2005}. For an $h = 0.72$ cosmology, this corresponds to M$_r < -20.9$ mag, which is approximately 2 mag brighter than the magnitude limit for the ATLAS$^\mathrm{3D}$ survey \citep{Cappellari:2013a}. From this selection, useful observations of 106 targets were obtained (a further 6 targets were severely affected by astrometric issues and were unusable). These galaxies span a range of morphological type, stellar mass and local environmental density within the cluster \citep[for details again see][]{Fogarty:2014a}. After observation, it was determined from a caustics analysis that 9 galaxies were not cluster members, reducing the sample to 97. 

\begin{figure}
\includegraphics[width=3.25in]{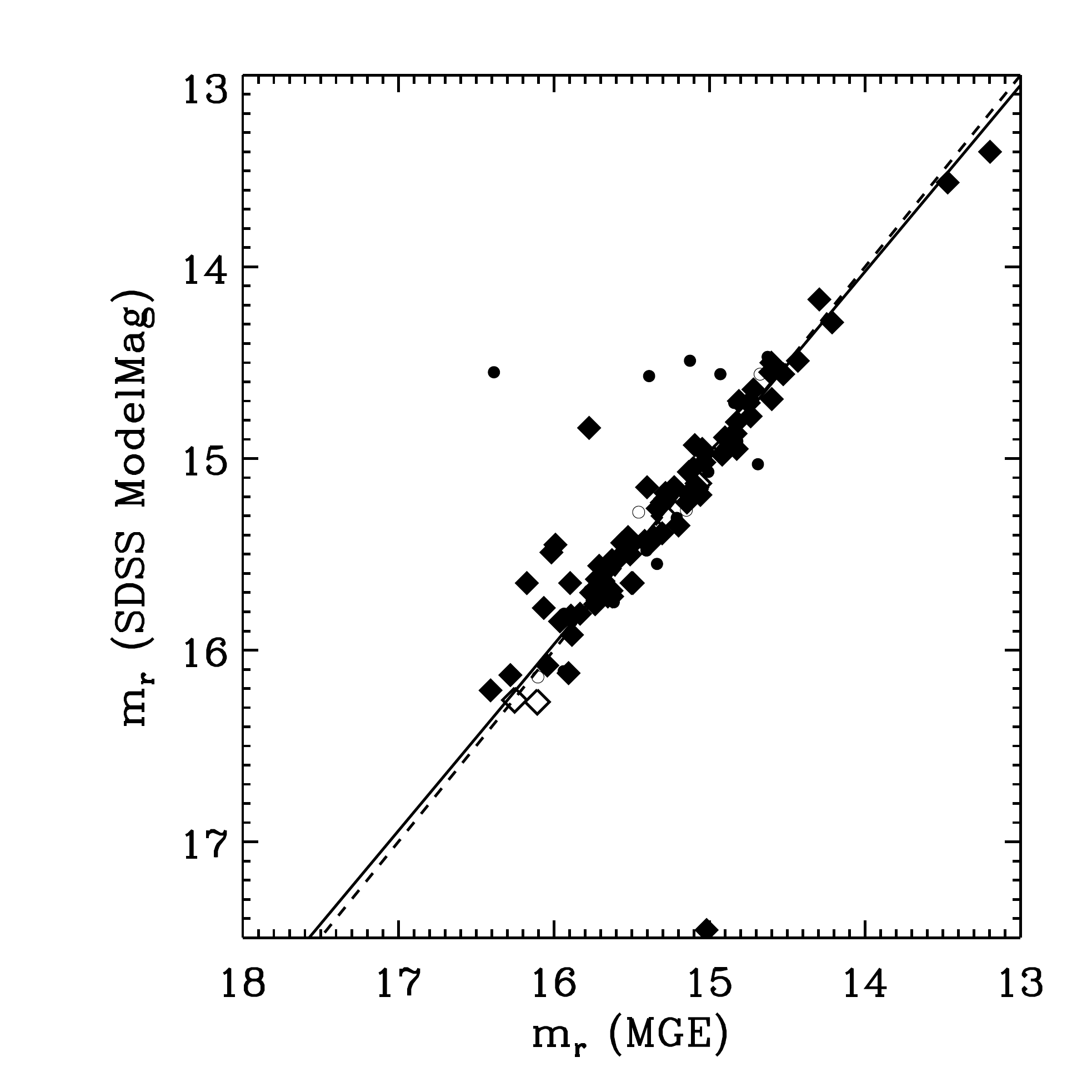}
\caption{Comparison of {\it r-}band magnitudes derived from our MGE model fits to SDSS ModelMag magnitudes (symbols as per Fig. \ref{fig:FJ_K}). The solid line shows the best-fitting relation between the two sets of magnitudes, while the dashed line shows the 1:1 correlation. The scatter about the correlation, after rejecting the most prominent outliers, is 0.11 mag.}
\label{fig:mag_comp}
\end{figure}

Based upon visual inspection of {\it gri} Sloan Digital Sky Survey Data Release 8 \citep[SDSS DR8][]{Aihara:2011} colour images we morphologically classified the sample into early and late-type galaxies based on the presence of spiral arms, or, in edge-on galaxies, the presence of a prominent, galaxy-scale dust lane. We determined that 74 out of the 97 cluster members (76 per cent) are early-type galaxies, with the remaining 23 (24 per cent) being late-types. Of the 9 non-cluster members, 5 are early-type galaxies and 4 are late-type galaxies. When determining the galaxy scaling relations we use only data for the 74 early-type cluster members. However, we do indicate the position of the remaining galaxies with measurable parameters.

In Figure \ref{fig:FJ_K} we show the distribution of our galaxies in the radius -- luminosity plane (left panel) and the velocity dispersion -- luminosity plane (right panel). The derivation of these quantities is described in detail in the following section. We also indicate the position of galaxies from the ATLAS$^\mathrm{3D}$ survey (small crosses), a volume-limited sample with an {\it r-}band magnitude limit of M$_r \sim -18.6$ mag. Above our magnitude limit (indicated by the dashed line) the SAMI pilot sample and ATLAS$^\mathrm{3D}$ sample have similar distributions, with the SAMI pilot galaxies being, on average, larger (by 22 per cent) but with similar dispersion at fixed luminosity, though the mean offset in size is within the scatter of the distributions.

Because of these differences between our sample and the robustly-selected ATLAS$^\mathrm{3D}$ sample and the relatively small range in luminosity and dispersion probed by our sample, we largely restrict our analysis to differential determinations of the FP and MP within our sample. The absolute determination of the FP and MP coefficients, and the comparison to those quantities in other samples is significantly affected by our sample selection, making those quantities challenging to interpret physically.

The upcoming SAMI Galaxy Survey will be largely unaffected by these issues. The SAMI Galaxy Survey sample is a factor of 35 larger, and spans a luminosity range that is an order of magnitude greater than the Pilot sample, resulting in significantly more accurate measurements of the coefficients and observed scatter. In addition, because of its better characterised sample \citep[see][]{Bryant:2015} accounting for the influence of the selection on the FP and MP coefficients will be much simpler than for the Pilot sample, when the SAMI Galaxy Survey is complete.

\section{Data and derived quantities}
\subsection{Photometry}

Total {\it r-}band luminosities, L, effective radii, R$_e$ and mean effective surface brightnesses, $\langle \mu_e \rangle$ were measured from $r-$band SDSS Data Release 8 (DR8) images. For each galaxy a Multi-Gaussian Expansion (MGE) model \citep{Emsellem:1994} was constructed from the SDSS $r-$band image using the procedure of \citet{Cappellari:2002}. \citet{Scott:2013} demonstrated that these models accurately capture total luminosities and surface brightness distributions for galaxies with a broad range of photometric and morphological properties. These models have the advantage that no assumption needs to be made about the functional form of a galaxy's surface brightness profile at large radii. The extrapolation of the models to large radii does, however, depend on the depth of the input photometry, but this is a common drawback of most techniques. In the implementation of \citet{Scott:2013}, the MGE models of barred galaxies are constrained in such a way as to have the flattening of the outer disk, except for extreme cases of the most pronounced bars where the disk flattening cannot be well recovered.

The MGE model represents the surface brightness of a galaxy as a sum of $n$ two dimensional Gaussians, $j$, with varying peak surface brightness, $\Sigma_j$, dispersion, $\sigma_j$ and axial ratio, $q_j$. If the galaxy major axis is aligned with the $x$ axis and the surface brightness distribution is centred at $x,y = (0,0)$, the surface brightness at a given spatial position is given by:
\begin{equation}
	\Sigma (x,y) = \sum^n_{j=1} \Sigma_j exp \left[ - \frac{1}{2\sigma_j^2} \left( x^2 + \frac{y^2}{q^2_j} \right) \right].
\end{equation}
We express the $\sigma_j$ in terms of physical radii (in kpc) using the Hubble flow-corrected distances to each of our three clusters from the NASA/IPAC Extragalactic Database. The total galaxy luminosity is then given by the sum of the total luminosity in each of the Gaussian components:

\begin{equation}
	L_{Tot} = \sum_{j=1}^n L_j = \sum_{j=1}^n 2 \pi \Sigma_j \sigma_j^2 q_j.
\end{equation}

\begin{figure}
\includegraphics[width=3.5in]{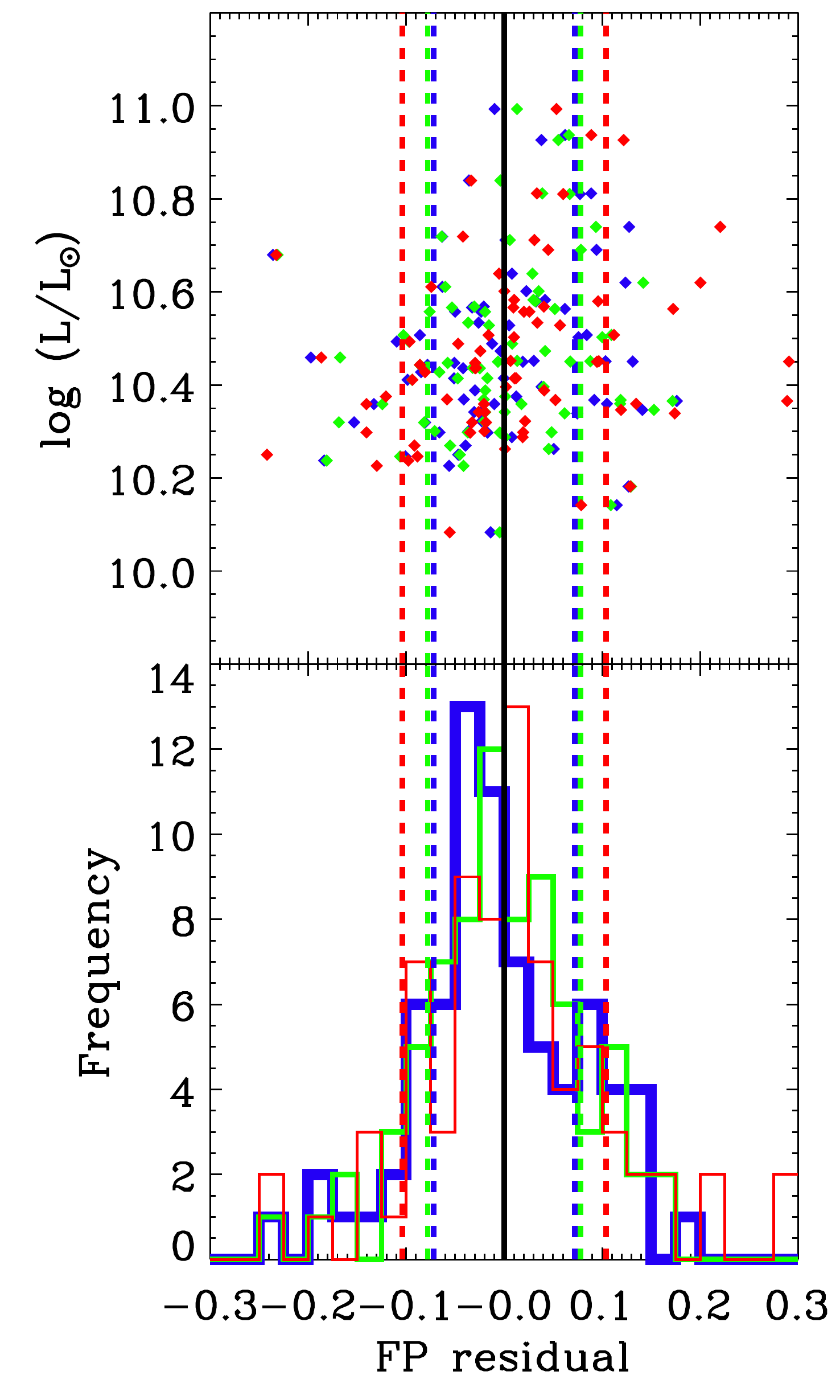}
\caption{Residuals ($\log L - \alpha \log \sigma - \beta \log R_e - \gamma$) from the three different variations on the FP described in the text. Green points and histogram: `IFS' FP. Blue points and histogram: `central' FP. Red points and histogram: `ellipse' FP. The dashed lines indicate the corresponding rms scatter. Upper panel: luminosity vs residual from the FP. Lower panel: histogram of residuals for each FP. The residuals for the central and IFS FPs are similar, but the ellipse FP has significantly larger scatter.}
\label{fig:fp_resids}
\end{figure}

We define the effective radius, R$_e$, as the radius which encloses half of the total luminosity, L$_{Tot}$, of the {\it circularised} MGE model (the MGE model where each Gaussian component has q$_{obs} = 1$), and the dispersions, $\sigma$, are scaled such that the peak and total luminosities of each component are the same as the original model. Following \citet{Cappellari:2013a} we uniformly multiply these MGE-derived effective radii by a factor 1.35. This ensures consistency with previous studies, in particular the r$^{1/4}$ growth-curve measurements of \citet{RC3}. We use a {\sc Python} implementation of the routine {\it find\_galaxy.pro}\footnote{available as part of the MGE package from http://purl.org/cappellari/software} to determine the effective ellipticity, $\epsilon_e$ and position angle, PA$_e$ at 1 R$_e$. This is done by determining the second moments of the luminosity distribution of the isophote whose area, A, is equal to $\pi$R$_e$. 

\begin{figure}
\includegraphics[width=3.25in]{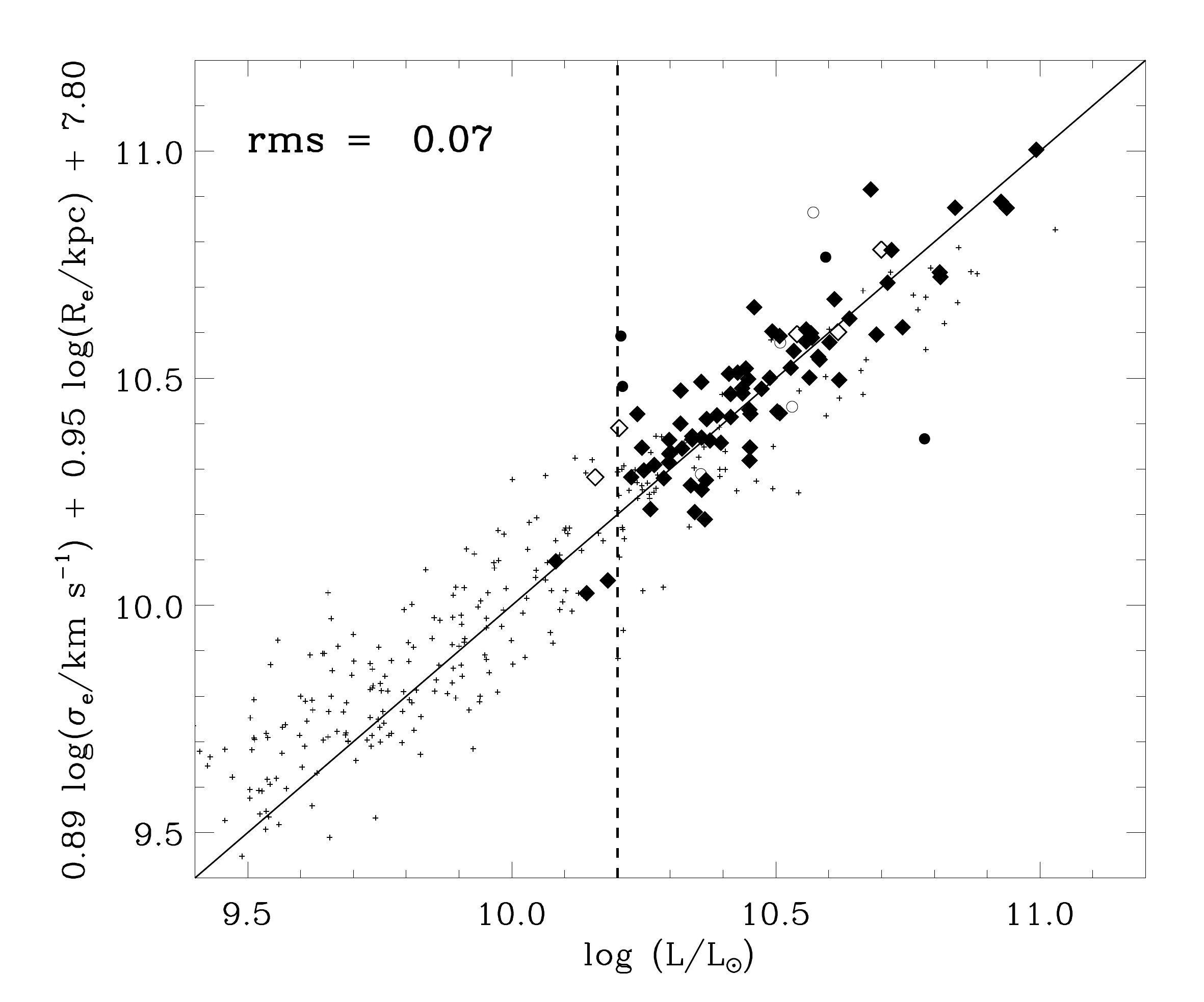}
\caption{Our preferred version of the FP, the IFS FP, determined using L, R$_e$ and $\sigma_e$. Symbols as per Figure \ref{fig:FJ_K}. The dashed line indicates the lower magnitude limit of our sample. Below the lower magnitude limit of the SAMI pilot sample the ATLAS$^\mathrm{3D}$ data deviate significantly from our best-fitting plane.}
\label{fig:fp_comp}
\end{figure}

Uncertainties in our L$_{Tot}$ measurements were determined through comparison to SDSS ModelMag magnitudes. This comparison is shown in Figure \ref{fig:mag_comp}. The scatter about the 1:1 relation, after rejecting the most extreme outliers, is 0.11 mag, or 10 per cent in individual measurements of L. The most extreme outlier, where the SDSS ModelMag is $\sim 2.5$ mags fainter than our MGE magnitude, has several close companions, and the SDSS ModelMag pipeline has failed to properly identify the target galaxy. The majority of the other outliers, where the SDSS ModelMags are brighter than our MGE magnitudes, are for galaxies with prominent non-axisymmetric structures (principally bars), where the MGE model fails to properly reproduce the surface brightness. The measured uncertainty represents an upper bound to the true uncertainty on our measurement --- some of the scatter is likely due to the SDSS ModelMag magnitudes being derived from {\it either} exponential or de Vaucouleurs profile fits, whereas our MGE models allow for any profile shape. This is consistent with the uncertainty given in \citet{Scott:2013}, who compared MGE-derived magnitudes from SDSS {\it r-}band photometry to other literature magnitudes, finding an uncertainty of 10 per cent in individual measurements. No comparable set of R$_e$ measurements exist for us to compare our own measurements to, so we cannot directly estimate the uncertainty as for the luminosities. However, \citet{Cappellari:2013a} performed a comparison of a similar sample of MGE-derived R$_e$ measurements and a set of literature R$_e$ values derived from traditional curve-of-growth estimates. They found an uncertainty of 10 per cent in their individual measurements, and, given the identical measurement technique and imaging used in this study, we adopt this value as representative of the uncertainty on our own R$_e$ measurements.

\subsection{Spectroscopy}
The SAMI instrument is a multi-object IFS that uses an innovative fused-fibre `hexabundle' design to obtain two dimensional spectroscopy of up to 13 targets simultaneously across a 1$^\circ$ diameter field of view. Each hexabundle consists of 61 1.6 arcsec diameter fibres closely packed into an approximately circular grid, with the entire hexabundle having a diameter of 15 arcsec. All 13 hexabundles feed the double-armed AAOmega spectrograph \citep{Sharp:2006}. For these observations a spectral resolution of R $\sim 1700$ was selected for the blue arm, giving a wavelength coverage of 3700--5700 \AA. The red arm observations were not used in this analysis. 

\begin{figure}
\includegraphics[width=3.25in]{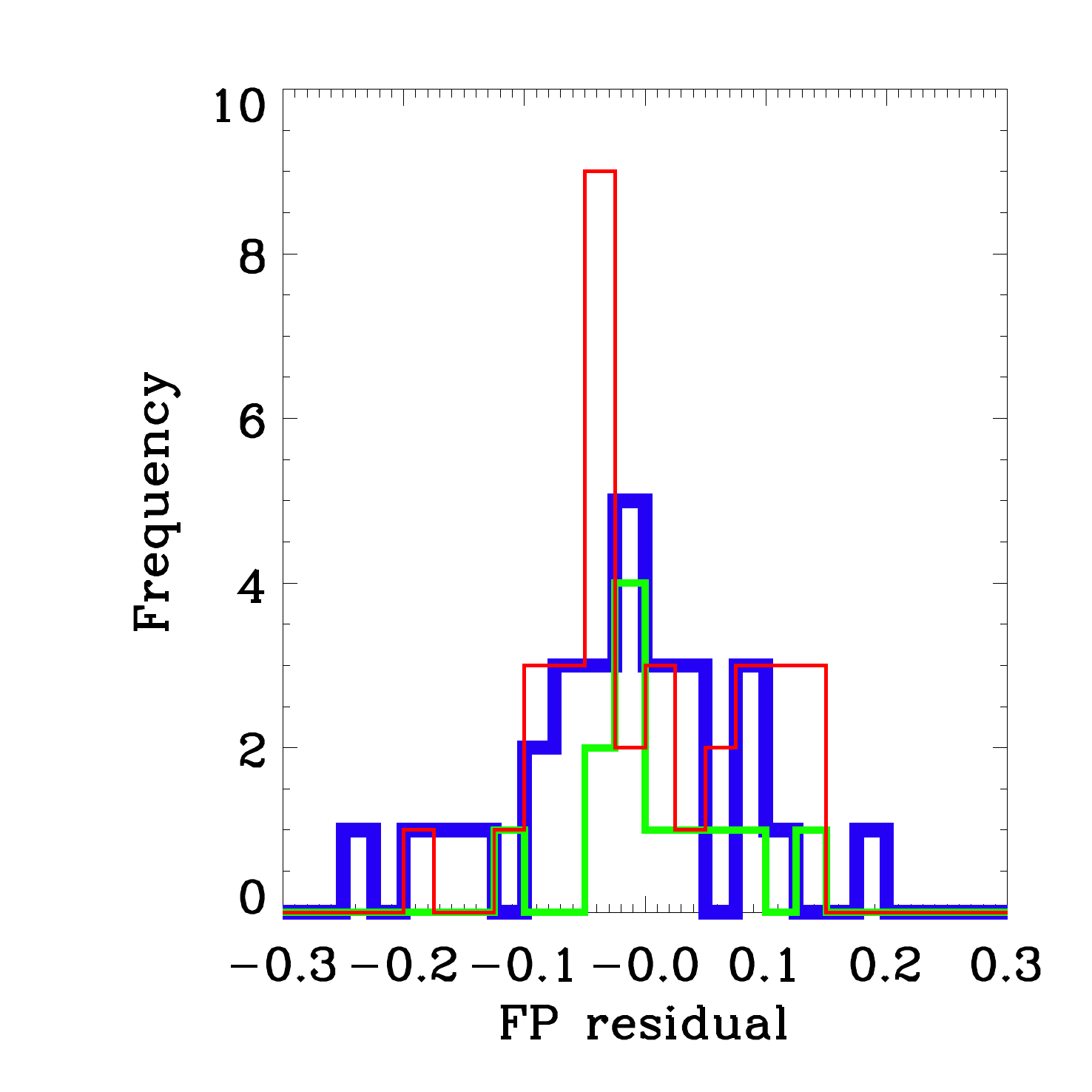}
\caption{Residuals from the IFS FP for each of the three clusters in our sample. The blue, green and red lines indicate Abell 85, Abell 168 and Abell 2399 respectively. The scatter in each cluster is comparable to that for the full sample.}
\label{fig:cluster_fps}
\end{figure}

The data were reduced using the {\sc 2dFdr} data reduction software, combined with a dedicated SAMI data reduction script written in {\sc Python}. This included the standard steps of bias subtraction, flat-field correction, fibre extraction, wavelength calibration, sky subtraction, telluric correction and cosmic ray removal, as well as reconstruction of the three-dimensional data cube from the row-stacked fibre spectra. General SAMI data reduction is described in \citet{Allen:2015} and \citet{Sharp:2015}, with the specific reduction of the Pilot sample data described in \citet{Fogarty:2014a}.

\subsubsection{Stellar Kinematics}

The SAMI data were used to derive two sets of kinematic information for each galaxy: (i) maps of the luminosity-weighted mean line-of-sight stellar velocity and velocity dispersion and (ii) the luminosity-weighted mean line-of-sight stellar velocity dispersions (measured within several different apertures). $\sigma$ was measured within three apertures: (i) a circular aperture of radii R$_e$ ($\sigma_e$), (ii) a circular aperture of radii R$_e$/8 ($\sigma_{e/8}$) and (iii) an elliptical aperture with ellipticity $\epsilon$ (as defined above) and a major axis radius R$_{e,maj} = \mathrm{R}_e/\sqrt{1 - \epsilon}$ ($\sigma_{e,ell}$). The central R$_e$/8 $\sigma$ measurements are broadly representative of classical studies of the FP, which typically used a small aperture of fixed physical size and then apply an empirical correction based on the relative size of the galaxy and the spectroscopic aperture. For three galaxies with the largest physical sizes our spectroscopy does not sample out to 1 R$_e$. We therefore apply an aperture correction following \citet{Cappellari:2006}, though the magnitude of this correction is less than 10 per cent for all three objects. 

 A single spectrum was constructed for each galaxy by summing the spectra from all spaxels within the given aperture. A single variance spectrum for each galaxy was constructed in the same way. The {\sc Python} penalised Pixel Fitting (pPXF) algorithm of \citet{Cappellari:2004} was used to determine all stellar kinematic quantities. pPXF uses a penalised maximum likelihood method to first construct an optimal template spectrum from a subset of a library of stellar spectra, then convolves this optimal template with a line-of-sight velocity distribution (LOSVD) to match the observed galaxy spectrum. The LOSVD is parameterised by a Gaussian, corresponding to the velocity, V, and velocity dispersion, $\sigma$. The library of stellar templates was composed of the 985 MILES stellar spectra \citep{Sanchez-Blazquez:2006,Falcon-Barroso:2011b}. In addition to the template spectra, a fourth order additive polynomial was included in each fit to account for the effects of dust extinction and residual flux calibration errors. The three sets of velocity dispersions and the associated uncertainties are given in Table \ref{tab:sample}. The derivation of the kinematic maps will be described in detail in Fogarty et al., in prep., but largely follows the same procedure as that described here, with the exception that the LOSVD was parameterised by the first four moments of a Gauss-Hermite expansion. .

\section{The Fundamental Plane}
\label{sec:fp}

\begin{table}
\caption{Coefficients and uncertainties of the best-fitting FPs for i) each of the different FP determinations described in the text and ii) each of the individual clusters. For the clusters we give only our preferred IFS FP of the form: $\log L = \alpha \log \sigma_e + \beta \log R_e + \gamma$. We give these coefficients primarily for use with determining the residuals from the various FPs. Because of our sample selection, these coefficients are not applicable to FPs representative of the global galaxy population.}
\label{tab:cluster_fps}
\centering
\begin{tabular}{l|ccccccc}
\hline
Plane & $\alpha$ & err($\alpha$) & $\beta$ & err($\beta$) & $\gamma$ & err($\gamma$) & rms \\
\hline
Central & 0.80 & 0.07 & 0.95 & 0.05 & 4.64 & 0.25 & 0.08 \\
IFS & 0.89 & 0.08 & 0.96 & 0.05 & 4.44 & 0.26 & 0.07\\
Ellipse & 1.15 & 0.10 & 0.83 & 0.06 & 4.30 & 0.33 & 0.10 \\
\hline
Abell 85 & 0.80 & 0.13 & 0.85 & 0.09 & 5.07 & 0.41 & 0.08 \\
Abell 168 & 0.97 & 0.37 & 0.78 & 0.12 & 4.97 & 0.81 & 0.05 \\
Abell 2399 & 0.92 & 0.13 & 1.12 & 0.10 & 3.72 & 0.49 & 0.08 \\
\hline
\end{tabular}
\end{table}

We determined several different variations of the Fundamental Plane, using the radii, luminosity and dispersion measurements described in the previous section. These variations are: i) a central FP using $\sigma_{e/8}$, L and R$_e$ (hereafter the `central plane'), ii) an integral field FP using $\sigma_e$, L and R$_e$ (the `IFS plane') and iii) an integral field FP that accounts for galaxy shape using $\sigma_{e,ell}$, L and R$_{e,maj}$ (the `ellipse plane'). 

We determine the best-fitting plane in each case using the Python routine {\sc lts\_planefit}\footnote{Available from http://purl.org/cappellari/software}, which is fully described in \citet{Cappellari:2013a}. This routine minimises the squared residuals while iteratively clipping outliers, finding a robust global solution for the coefficients of the best-fitting plane. The coefficients of the three planes are given in Table \ref{tab:cluster_fps}. The upper panel of figure \ref{fig:fp_resids} shows the residuals for each version of the FP plotted against L, with the lower panel showing a histogram of the residuals for each plane. The IFS FP has the lowest rms scatter in the $\log L$ direction, 0.07, though it is only 9 per cent smaller than that for the central FP. This is unsurprising given the typical seeing of our observations ($\sim 2$ arcsec) means the central, R$_e/8$ aperture is significantly contaminated by light from larger radii.

In contrast, the ellipse FP has 40 per cent larger scatter than either the central or IFS FPs. Both $\sigma_{e,ell}$, and R$_{e,maj}$ contribute to this increase in scatter. The use of $\sigma_{e,ell}$ increases the scatter by 0.1 dex, while the use of R$_{e,maj}$ increases the scatter by 0.25 dex. The increase in scatter due to the use of $\sigma_{e,ell}$ may be due to the inclusion of more spaxels from the edge of the field-of-view, which typically have higher variance due to the dither strategy of the observations. The increase in scatter due to R$_{e,maj}$ may have a physical cause, as R$_{e,maj}$ is independent of galaxy inclination, whereas the variation of R$_e$ and $\sigma$ (in any shape aperture) with inclination are typically anti-correlated, reducing the dependency of the observed scaling relations on inclination. 

These rms uncertainties are determined after excluding outliers --- of which there are 2--3 for each plane. For comparison, the {\it r-}band FP of \citet{Cappellari:2013a} has an observed rms scatter of 0.1 dex in $\log L$ (also determined after removing outliers). When seeking to minimise the scatter in the FP, adopting a large circular aperture that is scaled to reflect the size of the target galaxy is the optimal approach.

\subsection{Sample selection and the Fundamental Plane}

Our preferred plane, the IFS plane, which has the lowest scatter, is shown in Figure \ref{fig:fp_comp}. The early-type cluster members are indicated with large black diamonds. As previously, we indicate the position of galaxies from the ATLAS$^\mathrm{3D}$ sample with small crosses and our magnitude limit with a dashed line. Below our magnitude limit the ATLAS$^\mathrm{3D}$ data deviate significantly from our plane, indicating that our sample selection significantly affects the coefficients we derive for our best-fitting plane.

This dependence of the FP coefficients on the magnitude limit of a sample was clearly identified by \citet{DOnofrio:2008}, \citet{Hyde:2009} and \citet{NigocheNetro:2009}. These authors note that this variation is due to the distribution of galaxies within the plane. \citet[their figure 11]{DOnofrio:2008} and \citet[their figure 7]{Hyde:2009} found that as the lower magnitude limit of a sample increases the FP coefficients decrease systematically. Extrapolating the result of \citet{Hyde:2009}, we expect the a coefficient for our sample (with its {\it r-}band magnitude limit of -20.9) to be $\sim 0.3$ lower than for the ATLAS$^\mathrm{3D}$ sample (with its {\it r-}band magnitude limit of $\sim -19$). This is consistent with the difference between the ATLAS$^\mathrm{3D}$ coefficient $\alpha = 1.25$ and our best-fitting FP coefficient of $\alpha = 0.89$.

\begin{figure}
\includegraphics[width=3.25in]{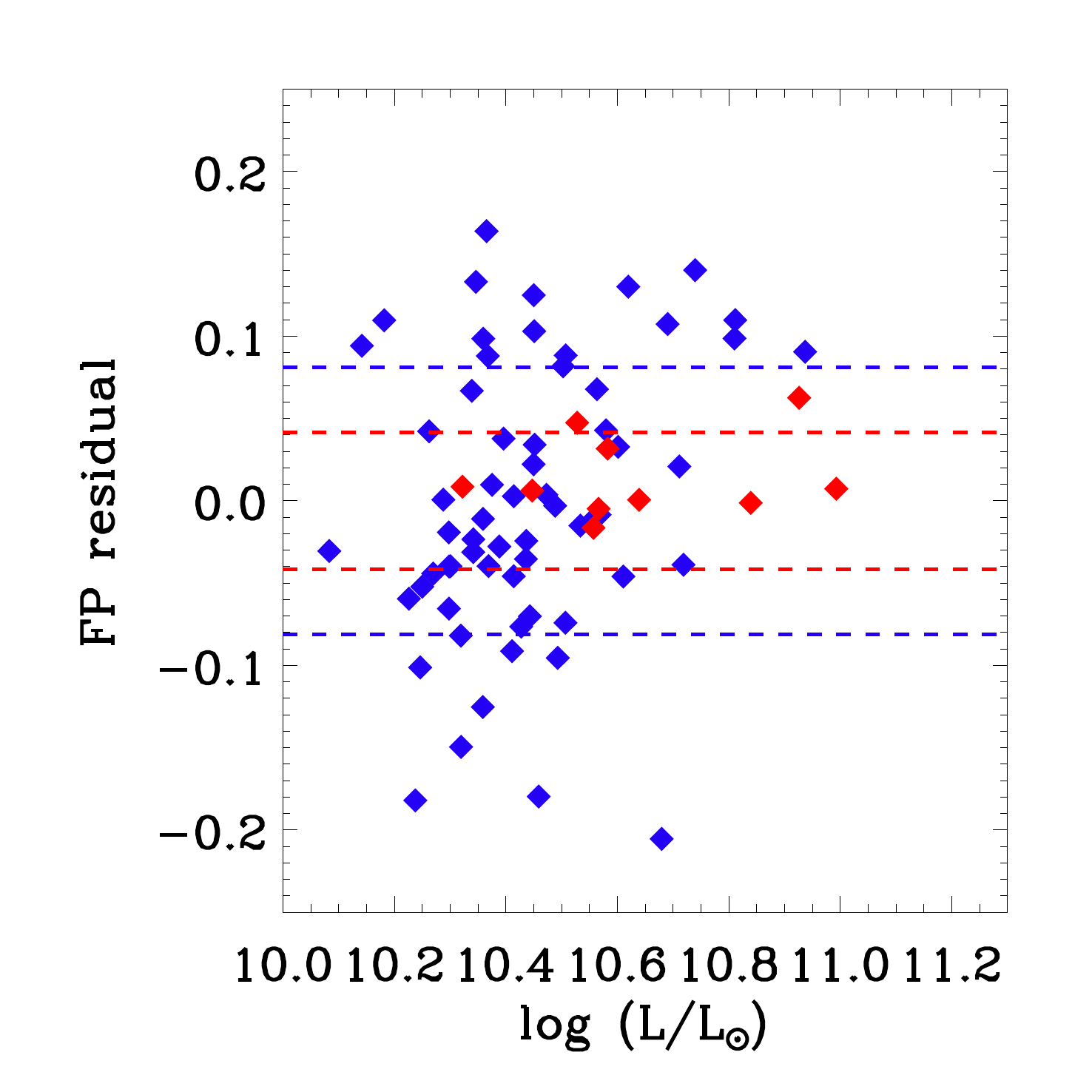}
\caption{Residuals from the best-fitting FPs for the fast (blue) and slow (red) rotator samples vs. luminosity. The dashed lines indicate the rms scatter for the two samples. The SR FP has significantly reduced scatter compared to full sample, consistent with no intrinsic scatter. The FR FP is essentially the same as that for the full sample.}
\label{fig:sami_fp_frsr}
\end{figure}

\subsection{Velocity dispersion from higher order Gauss-Hermite LOSVDs}

In addition to the standard approach of parameterising the LOSVD as a Gaussian, we also measured velocity dispersions from a LOSVD parameterisation which included the higher-order moments $h_3$ and $h_4$, $\sigma_{e,h}$. The relationship between $\sigma_e$ and $\sigma_{e,h}$ is slightly non-linear, with $\sigma_e \propto \sigma_{e,h}^{0.9}$, and a small scatter of 0.02 dex. As might be expected from the relationship between $\sigma_e$ and $\sigma_{e,h}$, the FP derived using $\sigma_{e,h}$ has a decreased $\alpha$ coefficient with respect to our preferred plane. The higher-order moments FP has the form: $\log L = 0.79 \log \sigma_e + 0.96 \log R_e + 4.66$.

\subsection{Late-type galaxies and cluster non-members}

 While we do not include the late-type galaxies and non-cluster members in the determination of the FP, we do indicate their position relative to the best-fitting plane. The late-type cluster members (small, filled black circles) have a relatively small mean offset from the FP of 0.03 dex, consistent within the scatter of the relation. They do however show a much larger scatter, having rms residuals in the L direction of 0.14 dex ---  double that for the early-type cluster members. Increased observational errors will account for some of this increased scatter (lower S/N in the IFS spectra, increased uncertainty in R$_e$ and L as the MGE models are less able to reproduce the details of the surface brightness profiles of the late-types), though there likely remains increased intrinsic scatter compared to the early-type galaxies. Both the early- and late-type non-members show more significant mean offsets from the FP, as expected if their true distances are different to the cluster galaxies. However, given the observed scatter in the plane, and with some established cluster members having large rms residuals from the plane, the non-member galaxies cannot be classified as such based on the FP alone. The uncertainty on the FP-derived distance at the distances of our clusters is $\sim 35$ Mpc, significantly larger than the typical depth of the clusters, therefore it is unsurprising that we cannot identify cluster members from the FP alone.

\begin{figure*}
\includegraphics[width=6.5in]{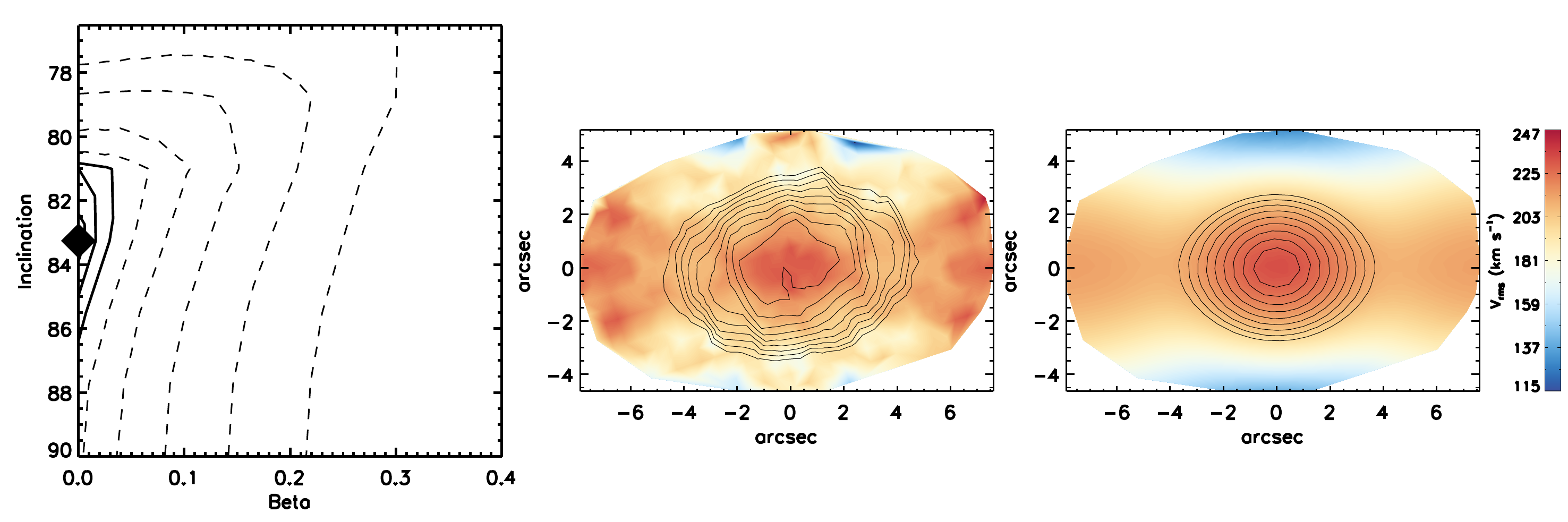}
\caption{Example of a JAM model fit to our SAMI stellar kinematics. The centre panel shows the observed $v_{rms} = \sqrt (v^2 + \sigma^2)$ for J011446.94+003128.8, while the right panel shows the best-fitting $v_{rms}$ predicted by our JAM model. The contours in the two panels show the observed and MGE model surface brightness distributions respectively. The left panel shows the reduced $\chi^2$ contours for the full range of inclinations, $i$ and anisotropies, $\beta$ explored by our models, with the black diamond indicating the location of the best-fitting model. The solid contours indicate the 1, 2 and 3 $\sigma$ confidence levels, with the dashed contours showing subsequent factors of two increase in $\chi^2$. The lower limit on the inclination is imposed by the input photometric model.}
\label{fig:jam_modelling}
\end{figure*}

Having determined that the integral-field version of the FP shows the smallest scatter of the three variations we examined, we now explore how this plane varies between the three different clusters within our sample. Separating the FP into separate clusters removes any scatter due to relative distance errors between the three clusters. We determine the individual cluster FPs exactly as described in the previous section. The residuals from the best-fitting planes are shown in Figure \ref{fig:cluster_fps} and the coefficients of the three planes are shown in Table \ref{tab:cluster_fps}. The scatter about each of the three planes is comparable to the full sample FP. If we substitute angular sizes as opposed to physical sizes in kilo parsecs into the best-fitting FP we can derive relative distances between the clusters. Using the distance to Abell 85 of 232 Mpc, we infer a distance of 167$^{+30}_{-26}$ Mpc to Abell 168 and a distance of 240$^{+44}_{-37}$ Mpc to Abell 2399, consistent with the distances given in Table \ref{tab:cluster_props}.

\begin{figure}
\includegraphics[width=3.25in]{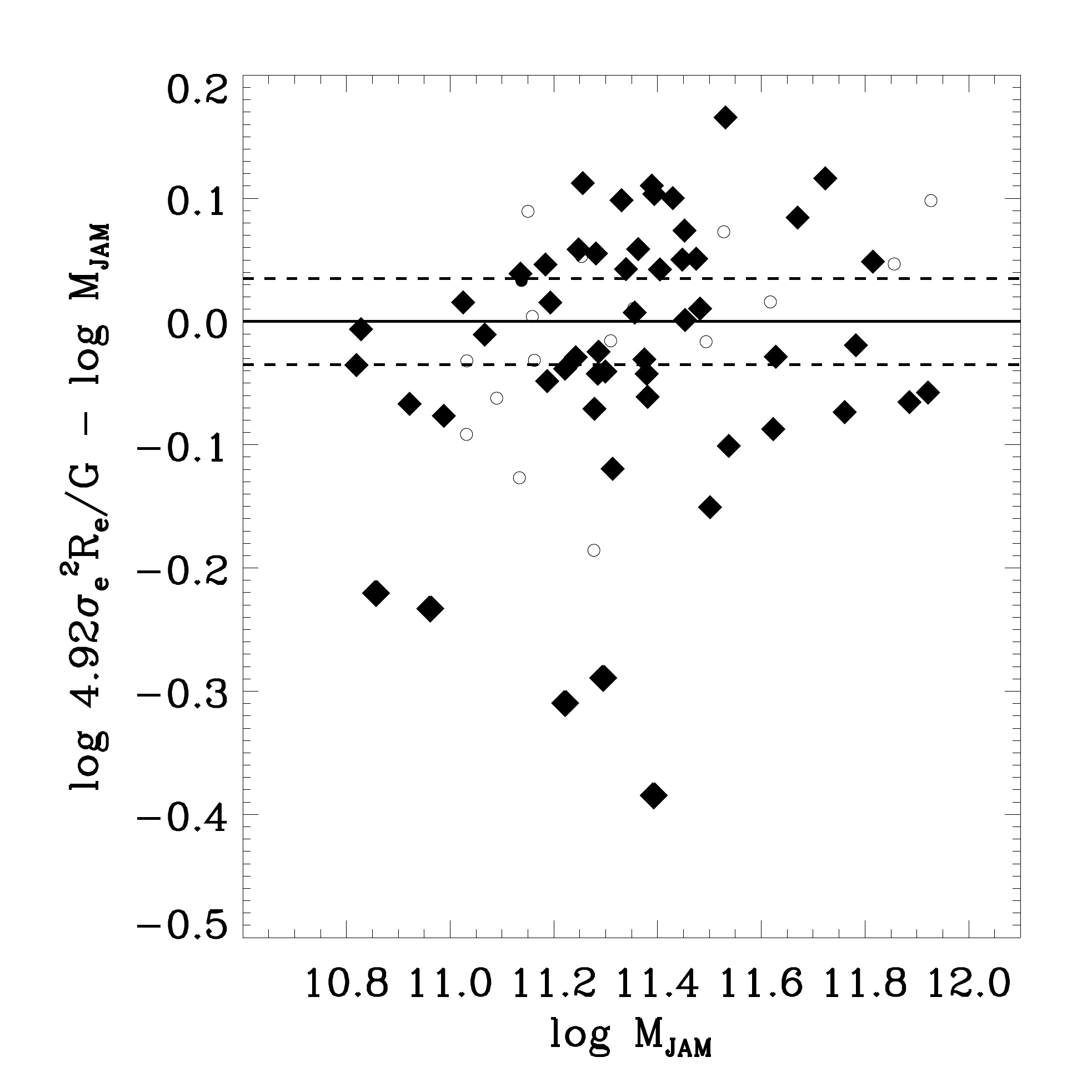}
\caption{Comparison of JAM model masses, M$_{JAM}$ to those derived from a simple Virial estimator, M$_{Vir} = \alpha \sigma_e^2 R_e / G$. Symbols as per Figure \ref{fig:mag_comp}. We find a tight correlation between the two mass measurements, with an rms scatter of 0.06 dex. We find a best-fitting scaling factor, $\alpha = 4.95 \pm 0.50$, consistent with the values of \citet{Cappellari:2006} and \citet{Scott:2009}. The five galaxies for which M$_\mathrm{JAM}$ is significantly larger than M$_\mathrm{Vir}$ are excluded from the MP analysis as described in the text.}
\label{fig:mjam_mvir}
\end{figure}
\begin{figure}
\includegraphics[width=3.25in]{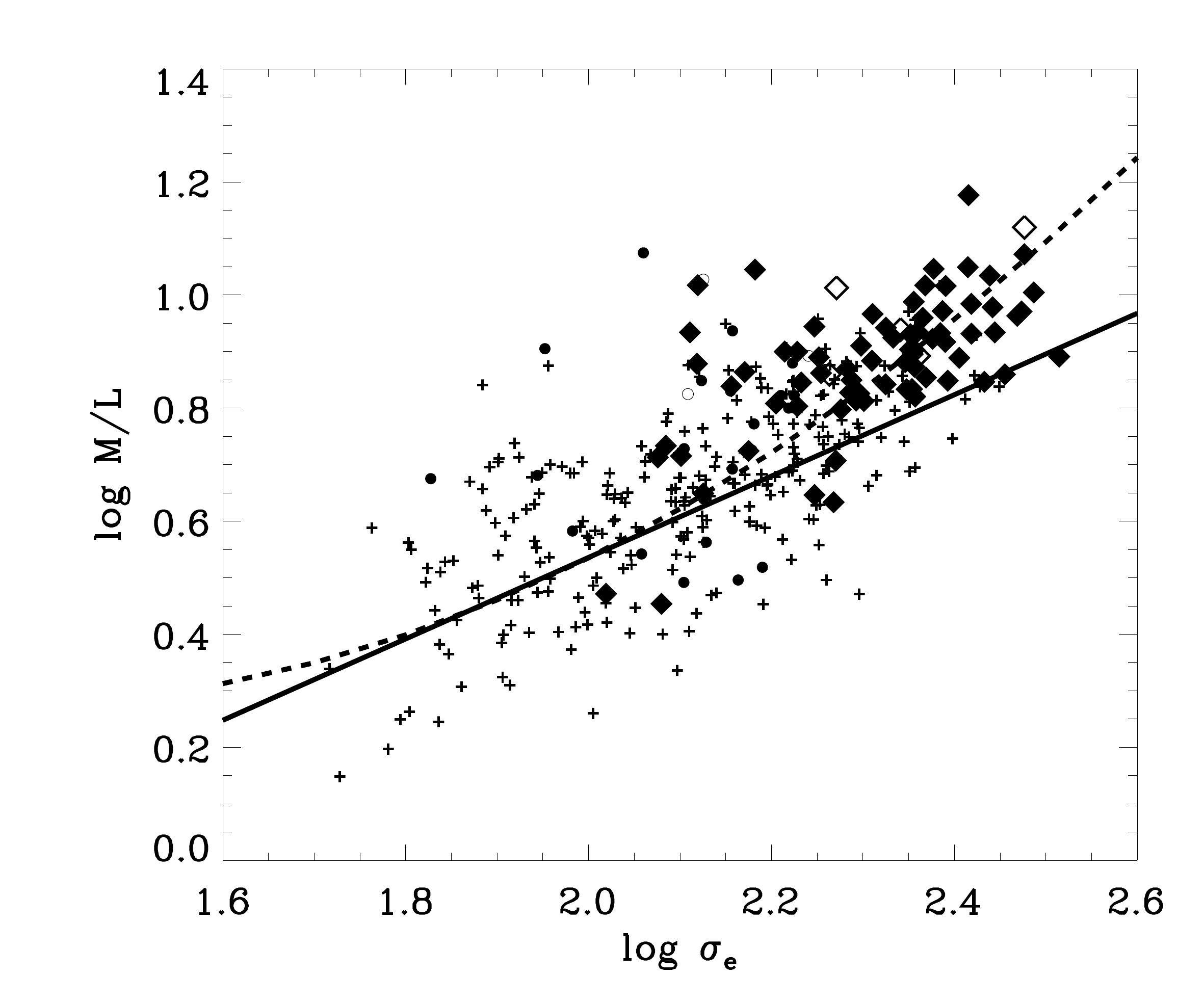}
\caption{The relationship between $\sigma_e$ and dynamical mass-to-light ratio, M/L$_{JAM}$ for our data (symbols as per Figure \ref{fig:mag_comp}) and the ATLAS$^\mathrm{3D}$ survey \citep[crosses][]{Cappellari:2011a}. The solid line shows a fit to the ATLAS$^\mathrm{3D}$ sample. The dashed line shows the curved relation of \citet{Zaritsky:2006}, derived from a sample with a much broader range in $\sigma$, but with significantly less accurate M/L.}
\label{fig:ml_sigma}
\end{figure}
\begin{figure}
\includegraphics[width=3.25in]{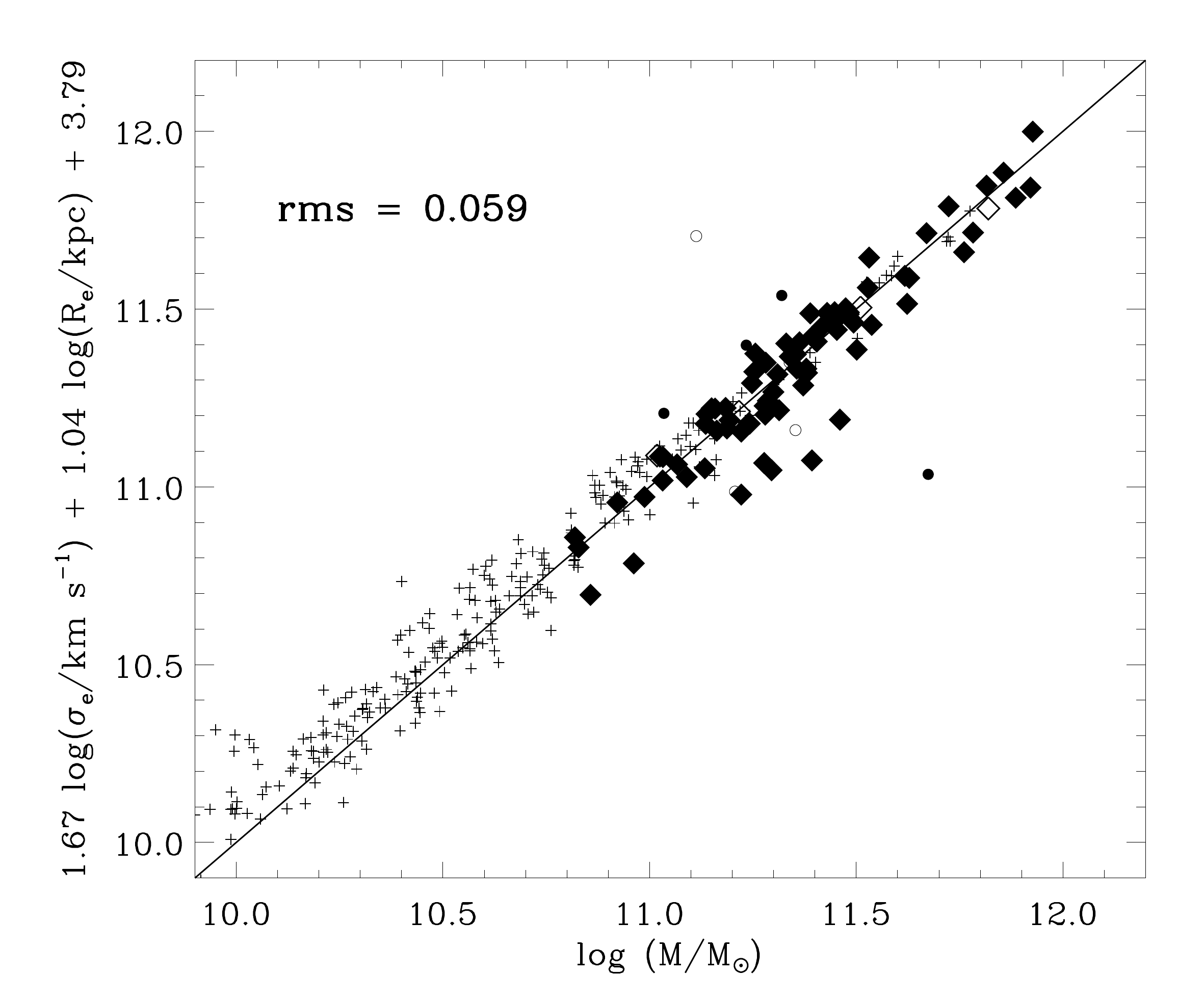}
\caption{The best-fitting Mass Plane (M$_{JAM}$ -- $\sigma_e$ -- $R_e$) for our full sample. Symbols as per Figure \ref{fig:mag_comp}. The MP is consistent with having no intrinsic scatter. The small crosses indicate the position of galaxies from the ATLAS$^\mathrm{3D}$ sample. Below 10$^{11.4}$ M$_\odot$ the ATLAS$^\mathrm{3D}$ data deviate significantly from our best-fitting plane, though at the high-mass end this disagreement is less significant.}
\label{fig:sami_mp}
\end{figure}

We also examine how the FP depends on kinematic type. In \citet{Fogarty:2014a} we classified our galaxies into fast rotators (FRs) and slow rotators (SRs) based on the morphology of their velocity and velocity dispersion fields. SRs and FRs are thought to have had significantly different evolutionary histories \citep[see e.g.][]{Emsellem:2011}, and this may manifest in their respective FPs. We determine the FP for each kinematic sub-sample as above, with the residuals from the resulting planes shown in Fig. \ref{fig:sami_fp_frsr}. The FR FP does not differ significantly from the full sample FP --- this is unsurprising given that the majority of our galaxies are FRs. The SR FP has significantly reduced observed scatter, a factor of two smaller than for the full sample. This is consistent with the SR FP having no intrinsic scatter. The coefficients of the SR FP differ from those of the full sample, though this is due to the higher average luminosity of the SRs compared to the full sample, rather than any intrinsic difference in the scaling relation of SRs. When we restrict the FR sample to the luminosity range of the SRs and redetermine the plane we find coefficients consistent with those of the SR plane, but with larger rms scatter about the plane. The observed scatter in the SR plane corresponds to a distance uncertainty of 8 per cent. However, given the small number of SRs in our sample, only 11 objects, the observed uncertainty is not necessarily a good measure of the true scatter in the population. To robustly estimate the uncertainty of distance measurements from the SR-only FP a larger population of SRs is required.

\section{Dynamical Modelling}
\label{sec:modelling}

Dynamical masses were determined using the Jeans Anisotropic MGE (JAM) modelling method of \citet{Cappellari:2008}. This modelling method makes empirically motivated assumptions about the internal structure and dynamics of galaxies, which restrict the range of model solutions. This allows dynamical masses to be determined from the first two moments of the line-of-sight velocity distribution, V and $\sigma$. This method does not fit the small-scale details of the kinematics, but instead makes a prediction based on the observed photometry (parameterised by the MGE models) and two further parameters, the inclination, $i$ and the anisotropy, $\beta$. The simplicity of the models is an advantage when applied to somewhat noisy data such as our own, as spurious features of the kinematics do not strongly affect the predicted mass.

In practice, the JAM models make a prediction for the second moments of the velocity distribution, v$_{rms} = \sqrt{v^2 + \sigma^2}$, based on the observed surface brightness distribution of a galaxy, $i$ and $\beta$. We sampled values of $\beta$ from 0.0 to 0.4, in steps of 0.025. This range is empirically motivated by the more detailed dynamical models of \citet{Cappellari:2006}. We sampled a range in $i$ from 90 degrees (edge on) to a minimum $i$ dictated by the roundest Gaussian component of the MGE model for each galaxy in steps of 2 degrees. $\beta$ and $i$ essentially determine the shape of the v$_{rms}$ field. For each value of $\beta$ and $i$ a best-fitting M/L is found by scaling the model v$_{rms}$ field such that the median value of v$_{rms}$ in the model and in the observations are the same. The best-fitting model is determined by computing $\chi^2$ for each value of $\beta$ and $i$ and finding the minimum value. This process is illustrated in Fig. \ref{fig:jam_modelling}, where the left panel shows contours of $\chi^2$ for the explored parameter space, and the centre and right panels show the observed and best-fitting model v$_{rms}$ maps respectively. 

\begin{figure}
\includegraphics[width=3.5in]{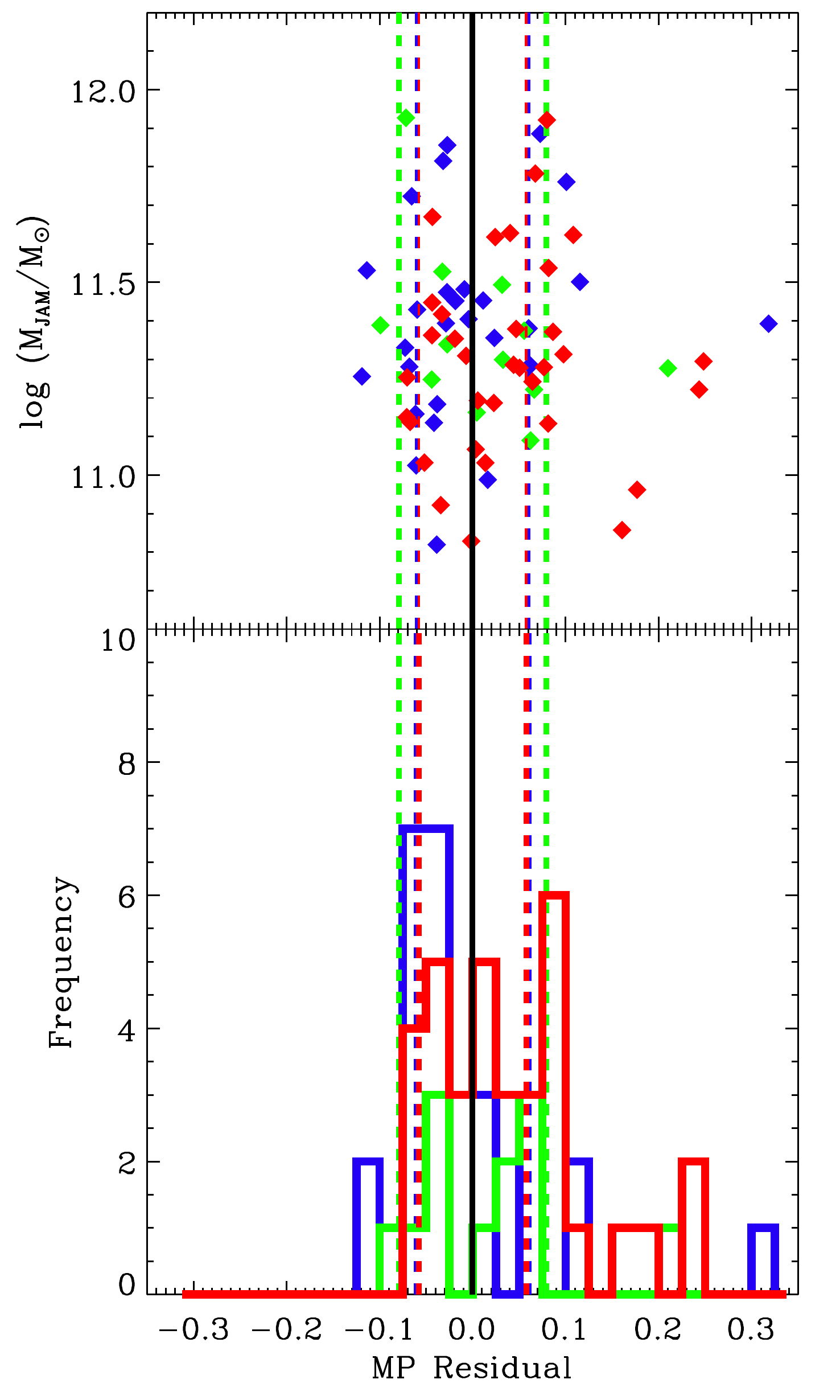}
\caption{Residuals from the best-fitting MP for each of the three clusters in our sample. Colours as in Figure \ref{fig:cluster_fps}. The dashed lines indicate the rms scatter for each of the three clusters. All three MPs are consistent with having zero intrinsic scatter, given the measurement uncertainties and the observed rms scatter.}
\label{fig:cluster_mps}
\end{figure}

\section{Dynamical Mass Scaling Relations}
\label{sec:mass_results}

\subsection{Virial vs. JAM masses}

We begin by comparing our dynamical masses derived from JAM modelling, M$_{JAM}$ to those derived from a simple Virial mass estimator, M$_{Vir} = \alpha \sigma^2 R_e/G$, where $\alpha$ is an empirically derived constant. This comparison is shown in Figure \ref{fig:mjam_mvir}. The best-fitting normalisation, $\alpha$, for our sample is $\alpha = 4.95 \pm 0.50$. This is consistent with the value of 5.0 found by \citet{Cappellari:2006}. If we allow for a non-linear relationship between M$_{JAM}$ and M$_{Vir}$ we find a best-fitting relation of: $\log M_{JAM} = (0.93 \pm 0.06) \log \sigma^2 R_e/G + (0.79 \pm 0.69)$. This favours a slightly non-linear relation with $M_{JAM}$. The observed rms scatter is 0.06 dex, or 22 per cent, after excluding the most extreme outliers, as described below. 

There are five galaxies whose JAM masses are significantly larger ($ > 0.2$ dex) than the Virial estimate. These galaxies are all close to edge-on and have significant spheroid components. The same issue occurs in the ATLAS$^\mathrm{3D}$ sample -- the six galaxies where $M_{JAM}$ is significantly larger than $M_{Vir}$ are also all close to edge-on with significant spheroidal components. It is likely that the deprojection of the MGE surface brightness model fails to capture the true three-dimensional structure of the galaxy, resulting in significant errors in M$_{JAM}$. We reject these galaxies as outliers (as was done in \citet{Cappellari:2013a}), and they are not included in any of the determinations of the best-fitting MP that follow. 

\subsection{The M/L -- $\sigma$ relation}

We also examine the correlation between our dynamical mass-to-light ratios, M/L$_{JAM}$ and $\sigma_e$, which is shown in Figure \ref{fig:ml_sigma}. We find a significant linear correlation within our data (solid black points), with:

\begin{equation}
\log M/L = (0.64 \pm 0.06) \log \sigma_e - (0.60 \pm 0.13).
\end{equation}

When we compare to data from the ATLAS$^\mathrm{3D}$ survey \citep[][crosses]{Cappellari:2013a} we find that our sample is offset above the ATLAS$^\mathrm{3D}$ relation. Combining both datasets, which expands the range in $\sigma_e$ sampled, gives a steeper linear relation than either of the individual samples:

\begin{equation}
\log M/L = (0.85 \pm 0.04) \log \sigma_e - (1.11 \pm 0.08),
\end{equation}
which is consistent with the finding of \citet{Cappellari:2006}, that the relation steepens when they restricted their fit to galaxies with $\sigma > 100 \mathrm{km\ s}^{-1}$. An alternative description of the relation comes from \citet{Zaritsky:2006}, who, using data with a much broader range in $\sigma$ ($\sim 10 - 1000$ km s$^{-1}$) but indirectly determined M/Ls, identified a curved relation between M/L and $\sigma$. The relation of \citet[rescaled to the {\it r-}band]{Zaritsky:2006} is shown in Figure \ref{fig:ml_sigma} as the dashed line. This curved relation provides a good description of the data, however given the relatively narrow range in $\sigma_e$ spanned by our combined SAMI and ATLAS$^\mathrm{3D}$ sample we cannot distinguish which of the curved and linear relations provides the better fit.
 
\subsection{The Mass Plane}

Following Section \ref{sec:fp} we derive the MP for our data using the {\sc lts\_planefit} routine. Here we focus only on the IFS version of the plane. This plane is shown in Figure \ref{fig:sami_mp} and has rms residuals in the $\log M$ direction of 0.059 dex. The observed scatter in the MP is entirely accounted for by the measurement errors on the three observational quantities --- the MP is consistent with having no intrinsic scatter. The coefficients of the best-fitting mass plane, $\alpha = 1.67$ and $\beta = 1.04$ are closer to the Virial expectation ($\alpha = 2, \beta =1$) than for any of the FPs investigated here, however $\alpha$ still differs significantly (3 standard deviations) from the theoretical value.

The early-type non-member galaxies are fully consistent with the MP, having negligible mean offset from the plane, and rms scatter consistent with the cluster member population. In contrast, the late-type galaxies have significantly increased scatter of 0.17 dex in $M$, a factor three larger than for the early-type galaxies. This is consistent with the increase in scatter of the late-types compared to the early-types in the FP, and is again likely a combination of increased intrinsic scatter and increased measurement errors. 

We also indicate the position of galaxies from the ATLAS$^\mathrm{3D}$ survey with small crosses. Below 10$^{11.4}$ M$_\odot$ the ATLAS$^\mathrm{3D}$ data deviate significantly from our best-fitting plane, though at higher masses this disagreement is reduced. This implies that the coefficients our our best-fitting MP are significantly affected by our sample selection.

\begin{table}
\caption{Coefficients and uncertainties of the best-fitting MPs for i) our full sample and ii) each of the individual clusters. The MP is of the form: $\log M_{JAM} = \alpha \log \sigma_e + \beta \log R_e + \gamma$. We give these coefficients primarily to derive residuals from the MPs. Because of our sample selection these coefficients are not applicable to MPs for volume-limited samples representative of the global galaxy population.}
\label{tab:cluster_mps}
\centering
\begin{tabular}{l|ccccccc}
\hline
Sample & $\alpha$ & err($\alpha$) & $\beta$ & err($\beta$) & $\gamma$ & err($\gamma$) & rms\\
\hline
All & 1.67 & 0.11 & 1.04 & 0.07 & 3.17 & 0.35 & 0.059 \\
\hline
Abell 85 & 1.63 & 0.17 & 1.08 & 0.10 & 3.05 & 0.49 & 0.060 \\
Abell 168 & 1.95 & 0.52 & 0.72 & 0.16 & 3.80 & 1.10 & 0.055 \\
Abell 2399 & 1.81 & 0.18 & 1.10 & 0.13 & 2.60 & 0.69 & 0.055 \\
\hline
\end{tabular}
\end{table}

We also determine the best-fitting MP separately in each of the three clusters in our sample. The residuals from the three MPs are shown in Figure \ref{fig:cluster_mps} and the corresponding coefficients are given in Table \ref{tab:cluster_mps}. As with the FP, we can use the MP to infer the relative distance between the clusters. Using the distance to Abell 85 of 232 Mpc, we infer a distance of 157$^{+23}_{-20}$ Mpc to Abell 168 and a distance of 260$^{+28}_{-25}$ Mpc to Abell 2399, again consistent with the distances given in Table \ref{tab:cluster_props}. Our MP derived distances have uncertainties $\sim 50$ per cent smaller than those derived from the FP.

\section{Conclusions}
\label{sec:conclusion}

In this paper we have presented a study of the two-- and three--parameter scaling relations of a sample of 74 early-type galaxies observed with the SAMI integral field spectrograph. Utilising integral field spectroscopy to measure a spatially resolved velocity dispersion, combined with selecting galaxies from three massive clusters (to eliminate additional scatter due to relative distance uncertainties between the target galaxies) allows us to measure the FP with minimal systematic uncertainties. 

We find an extremely tight FP, with observed rms scatter 0.072. The scatter about the FP is minimised by measuring $\sigma$ in a large circular aperture of radius R$_e$, demonstrating that integral field spectroscopy has an important role to play in measuring galaxy scaling relations. Separating our galaxies by kinematic type, we find that the slow rotator FP is consistent with having zero intrinsic scatter.

We utilise spatially resolved maps of velocity and velocity dispersion, combined with wide-field imaging to construct Jeans dynamical models for all galaxies in our sample, deriving dynamical masses, M$_{JAM}$ and dynamical mass-to-light ratios, M/L. Replacing the total luminosity L with the dynamical mass, M$_{JAM}$, we find a Mass Plane for our full sample that again has no intrinsic scatter. The coefficients of this best-fitting Mass Plane differ significantly from the Virial theorem expectation. In addition we find evidence for a curved relationship between M/L and $\sigma_e$.

While the sample used in this study is relatively small compared to some recent measurements of the FP, the small observed scatter we find for our best-fitting FP demonstrated the power of integral field spectroscopy. The SAMI Galaxy Survey \citep{Bryant:2015} will observe $\sim 3400$ galaxies, with $\sim 800$ of these selected from a small number of clusters. With this statistically significant sample of galaxies with both integral field spectroscopy observations and well-determined relative distances we will be able to measure the FP with minimal systematic and sample-driven uncertainties. In addition, this sample will allow us to construct dynamical models for thousands of galaxies, an order of magnitude more than existing studies.

\section{Acknowledgements}
The authors would like to thank the anonymous referee for their helpful comments that significantly improved the paper from its first submission. We would also like to thank Michele Cappellari for his advice on the correct usage of the JAM modelling code and for discussion on the results of this work.

The SAMI Galaxy Survey is based on observations made at the Anglo-Australian Telescope. The Sydney-AAO Multi-object Integral field spectrograph (SAMI) was developed jointly by the University of Sydney and the Australian Astronomical Observatory. The SAMI input catalogue is based on data taken from the Sloan Digital Sky Survey, the GAMA Survey and the VST ATLAS Survey. The SAMI Galaxy Survey is funded by the Australian Research Council Centre of Excellence for All-sky Astrophysics (CAASTRO), through project number CE110001020, and other participating institutions. The SAMI Galaxy Survey website is http://sami-survey.org/

This work was supported by the Astrophysics at Oxford grants (ST/H002456/1 and ST/K00106X/1) as well as visitors grant (ST/H504862/1) from the UK Science and Technology Facilities Council. RLD acknowledges travel and computer grants from Christ Church, Oxford and support from the Australian Astronomical Observatory Distinguished Visitors programme, the ARC Centre of Excellence for All Sky Astrophysics, and the University of Sydney during a sabbatical visit.

 This research has made use of the NASA/IPAC Extragalactic Database (NED) which is operated by the Jet Propulsion Laboratory, California Institute of Technology, under contract with the National Aeronautics and Space Administration.
 
 This research made use of Montage, funded by the National Aeronautics and Space Administration's Earth Science Technology Office, Computational Technologies Project, under Cooperative Agreement Number NCC5-626 between NASA and the California Institute of Technology. The code is maintained by the NASA/IPAC Infrared Science Archive. 
 
 This research made use of Astropy, a community-developed core Python package for Astronomy (Astropy Collaboration, 2013, http://www.astropy.org).

\bibliographystyle{mn2e}
\bibliography{sami_pilot_mass_modelling}

\appendix

\section{Sample Table}

\begin{table*}
\caption{Fundamental Plane parameters and other properties of the 106 galaxies in our sample.}
\label{tab:sample}
\centering
\begin{tabular}{lcccccccccc}
\hline
Galaxy name & Luminosity & R$_e$ & $\epsilon$ & $\sigma_c$ & $\sigma_e$ & $\sigma_{e,ell}$ & M/L & JAM fit & Type & Cluster\\
 & (log$_{10}$ L$_\odot$) & (arcsec) & & km s$^{-1}$ & km s$^{-1}$ & km s$^{-1}$ & & quality & & \\
  (1) & (2) & (3) & (4) & (5) & (6) & (7) & (8) & (9) & (10) & (11)\\
\hline
J003906.77-084758.3 & 10.4 & 2.4 & 0.09 & 119$\pm$3 & 120$\pm$2 & 119$\pm$2 & 2.8$\pm$0.2 & 0 & F & 85 \\
J004001.68-095252.4 & 10.5 & 2.3 & 0.28 & 246$\pm$4 & 262$\pm$3 & 261$\pm$3 & 8.5$\pm$0.2 & 1 & F & 85 \\
J004004.88-090302.6 & 10.4 & 3.4 & 0.36 & 195$\pm$4 & 191$\pm$3 & 195$\pm$3 & 7.4$\pm$0.3 & 1 & F & 85 \\
J004018.68-085257.1 & 10.3 & 2.8 & 0.45 & 166$\pm$4 & 149$\pm$2 & 145$\pm$2 & 5.3$\pm$0.2 & 1 & F & 85 \\
J004046.47-085005.0 & 10.6 & 3.9 & 0.06 & 243$\pm$5 & 215$\pm$3 & 215$\pm$3 & 8.4$\pm$0.3 & 1 & S & 85 \\
J004101.87-091233.1 & 10.5 & 3.2 & 0.15 & 299$\pm$8 & 299$\pm$5 & 298$\pm$5 & 11.8$\pm$0.5 & 2 & F & 85 \\
J004112.21-091010.2 & 10.5 & 2.2 & 0.06 & 244$\pm$3 & 247$\pm$3 & 247$\pm$3 & 7.1$\pm$0.2 & 1 & F & 85 \\
J004112.79-093203.7 & 10.4 & 4.5 & 0.39 & 126$\pm$7 & 132$\pm$6 & 126$\pm$6 & 7.1$\pm$0.7 & 2 & L & 85 \\
J004122.06-095240.8 & 10.7 & 6.8 & 0.41 & 275$\pm$5 & 260$\pm$5 & 259$\pm$4 & 15.0$\pm$0.6 & 1 & F & 85 \\
J004128.56-093426.7 & 10.4 & 3.4 & 0.42 & 212$\pm$4 & 199$\pm$3 & 198$\pm$3 & 6.5$\pm$0.2 & 1 & F & 85 \\
J004130.29-091545.8 & 10.6 & 3.9 & 0.62 & 125$\pm$7 & 114$\pm$6 & 110$\pm$6 & 3.5$\pm$0.4 & 2 & L & 85 \\
J004130.42-091406.7 & 10.5 & 1.7 & 0.30 & 335$\pm$6 & 327$\pm$4 & 323$\pm$4 & 7.8$\pm$0.3 & 1 & F & 85 \\
J004131.25-094151.0 & 10.3 & 4.1 & 0.09 & 150$\pm$3 & 143$\pm$3 & 143$\pm$3 & 6.9$\pm$0.4 & 0 & F & 85 \\
J004133.41-090923.4 & 10.3 & 2.2 & 0.11 & 202$\pm$3 & 196$\pm$2 & 196$\pm$2 & 6.5$\pm$0.2 & 2 & S & 85 \\
J004134.89-092150.5 & 10.5 & 2.0 & 0.26 & 248$\pm$5 & 225$\pm$3 & 224$\pm$3 & 6.8$\pm$0.2 & 1 & F & 85 \\
J004143.00-092621.9 & 10.8 & 6.7 & 0.45 & 231$\pm$4 & 238$\pm$5 & 236$\pm$4 & 11.1$\pm$0.6 & 0 & S & 85 \\
J004148.22-091703.1 & 10.5 & 2.2 & 0.26 & 301$\pm$6 & 294$\pm$4 & 295$\pm$4 & 9.2$\pm$0.3 & 1 & F & 85 \\
J004150.17-092547.4 & 10.7 & 4.2 & 0.18 & 335$\pm$7 & 306$\pm$4 & 304$\pm$4 & 10.1$\pm$0.5 & 2 & F & 85 \\
J004150.46-091811.2 & 11.3 & 14.2 & 0.24 & 419$\pm$12 & 429$\pm$14 & 400$\pm$12 & 15.1$\pm$1.3 & 2 & S & 85 \\
J004152.16-093014.8 & 10.7 & 4.2 & 0.21 & 275$\pm$5 & 259$\pm$3 & 258$\pm$4 & 11.2$\pm$0.3 & 2 & F & 85 \\
J004153.50-092943.9 & 10.6 & 4.5 & 0.66 & 166$\pm$4 & 169$\pm$3 & 153$\pm$3 & 7.9$\pm$0.3 & 2 & F & 85 \\
J004200.64-095004.0 & 10.4 & 3.4 & 0.77 & 134$\pm$7 & 131$\pm$4 & 133$\pm$4 & 10.4$\pm$0.7 & 2 & F & 85 \\
J004205.86-090240.7 & 10.4 & 2.9 & 0.09 & 210$\pm$4 & 204$\pm$3 & 203$\pm$3 & 9.3$\pm$0.8 & 2 & F & 85 \\
J004215.91-093252.0 & 10.3 & 2.2 & 0.30 & 206$\pm$4 & 211$\pm$3 & 209$\pm$3 & 7.0$\pm$0.2 & 1 & F & 85 \\
J004218.75-091528.4 & 10.4 & 2.6 & 0.26 & 252$\pm$5 & 243$\pm$3 & 245$\pm$3 & 9.4$\pm$0.3 & 2 & F & 85 \\
J004233.86-091040.5 & 10.6 & 3.2 & 0.09 & 258$\pm$4 & 245$\pm$3 & 245$\pm$3 & 8.3$\pm$0.2 & 1 & F & 85 \\
J004233.99-095442.2 & 10.6 & 3.2 & 0.04 & 240$\pm$6 & 227$\pm$3 & 226$\pm$3 & 7.5$\pm$0.3 & 1 & F & 85 \\
J004242.26-085528.1 & 10.3 & 2.6 & 0.15 & 137$\pm$2 & 133$\pm$2 & 131$\pm$2 & 4.5$\pm$0.2 & 2 & F & 85 \\
J004244.68-093316.2 & 10.6 & 2.7 & 0.33 & 301$\pm$5 & 270$\pm$3 & 271$\pm$3 & 7.0$\pm$0.2 & 2 & S & 85 \\
J004310.12-095141.2 & 10.9 & 6.5 & 0.08 & 259$\pm$7 & 254$\pm$5 & 253$\pm$4 & 7.7$\pm$0.3 & 2 & S & 85 \\
J011327.21+000908.9 & 10.5 & 10.8 & 0.12 & 153$\pm$2 & 142$\pm$3 & 152$\pm$3 & 6.8$\pm$0.5 & 0 & L & 168 \\
J011346.32+001820.6 & 10.5 & 6.1 & 0.61 & 171$\pm$4 & 167$\pm$4 & 159$\pm$3 & 6.6$\pm$0.3 & 2 & L & 168 \\
J011415.78+004555.2 & 9.8 & 2.2 & 0.28 & 114$\pm$5 & 127$\pm$4 & 126$\pm$4 & 3.1$\pm$0.2 & 1 & L & 168 \\
J011421.54+001046.9 & 10.6 & 4.5 & 0.08 & 246$\pm$5 & 227$\pm$3 & 227$\pm$3 & 6.6$\pm$0.2 & 2 & F & 168 \\
J011425.68+003209.8 & 10.4 & 8.0 & 0.32 & 126$\pm$7 & 162$\pm$9 & 142$\pm$9 & 6.6$\pm$0.8 & 0 & L & 168 \\
J011430.80+001928.3 & 10.5 & 7.2 & 0.36 & 140$\pm$3 & 143$\pm$2 & 146$\pm$2 & 4.9$\pm$0.2 & 0 & L & 168 \\
J011443.86+001709.6 & 10.2 & 5.6 & 0.42 & 74$\pm$6 & 89$\pm$5 & 89$\pm$5 & 8.0$\pm$1.0 & 2 & L & 168 \\
J011446.94+003128.8 & 10.5 & 2.9 & 0.32 & 264$\pm$5 & 237$\pm$3 & 237$\pm$2 & 8.4$\pm$0.2 & 2 & F & 168 \\
J011454.21+003026.5 & 10.3 & 2.1 & 0.40 & 202$\pm$4 & 193$\pm$2 & 197$\pm$2 & 6.7$\pm$0.3 & 1 & F & 168 \\
J011454.25+001811.8 & 10.5 & 3.9 & 0.18 & 285$\pm$5 & 274$\pm$3 & 273$\pm$3 & 10.8$\pm$0.3 & 2 & F & 168 \\
J011456.26+000750.4 & 10.4 & 4.2 & 0.37 & 169$\pm$5 & 165$\pm$3 & 167$\pm$3 & 6.3$\pm$0.3 & 2 & L & 168 \\
J011457.59+002550.8 & 11.0 & 10.2 & 0.10 & 281$\pm$4 & 278$\pm$4 & 286$\pm$4 & 8.6$\pm$0.3 & 1 & S & 168 \\
J011459.61+001533.1 & 10.4 & 2.1 & 0.37 & 229$\pm$5 & 233$\pm$3 & 227$\pm$3 & 7.1$\pm$0.2 & 0 & F & 168 \\
J011503.63+002418.7 & 10.4 & 2.9 & 0.36 & 133$\pm$5 & 145$\pm$1 & 144$\pm$1 & 3.1$\pm$0.1 & 2 & L & 168 \\
J011507.33+002756.8 & 10.4 & 3.5 & 0.45 & 159$\pm$3 & 169$\pm$2 & 168$\pm$2 & 6.4$\pm$0.2 & 1 & F & 168 \\
J011508.73+003433.5 & 10.3 & 1.8 & 0.25 & 228$\pm$5 & 224$\pm$3 & 222$\pm$3 & 7.4$\pm$0.3 & 2 & F & 168 \\
J011515.78+001248.4 & 10.5 & 4.2 & 0.05 & 253$\pm$5 & 231$\pm$3 & 231$\pm$3 & 9.1$\pm$0.3 & 1 & F & 168 \\
J011516.77+001108.3 & 10.4 & 3.5 & 0.24 & 228$\pm$6 & 227$\pm$3 & 226$\pm$3 & 8.0$\pm$0.2 & 2 & F & 168 \\
J011531.18+001757.2 & 10.4 & 3.0 & 0.20 & 246$\pm$4 & 222$\pm$3 & 224$\pm$3 & 6.8$\pm$0.2 & 2 & F & 168 \\
J011603.31-000652.7 & 10.2 & 3.0 & 0.31 & 157$\pm$5 & 151$\pm$4 & 154$\pm$5 & 5.9$\pm$0.4 & 0 & L & 168 \\
J011605.60-000053.6 & 10.5 & 5.3 & 0.63 & 154$\pm$4 & 143$\pm$3 & 139$\pm$3 & 8.6$\pm$0.4 & 2 & L & 168 \\
J011612.79-000628.3 & 10.4 & 3.6 & 0.03 & 216$\pm$5 & 193$\pm$2 & 193$\pm$2 & 7.1$\pm$0.2 & 2 & F & 168 \\
J011623.61+002644.8 & 10.2 & 3.8 & 0.41 & 117$\pm$5 & 128$\pm$3 & 132$\pm$3 & 6.7$\pm$0.4 & 0 & L & 168$^\dagger$ \\
J011703.58+000027.4 & 10.3 & 2.4 & 0.39 & 163$\pm$7 & 154$\pm$5 & 155$\pm$5 & 3.3$\pm$0.2 & 0 & L & 168 \\
J215432.20-070924.1 & 10.3 & 3.2 & 0.33 & 93$\pm$5 & 121$\pm$4 & 125$\pm$4 & 5.4$\pm$0.4 & 2 & F & 2399 \\
\hline
\end{tabular}
\end{table*}
\begin{table*}
Table \ref{tab:sample} continued. \\
\centering
\begin{tabular}{lcccccccccc}
\hline
Galaxy name & Luminosity & R$_e$ & $\epsilon$ & $\sigma_c$ & $\sigma_e$ & $\sigma_{e,ell}$ & M/L & JAM fit & Type & Cluster\\
 & (log$_{10}$ L$_\odot$) & (arcsec) & & km s$^{-1}$ & km s$^{-1}$ & km s$^{-1}$ & & quality & &\\
 (1) & (2) & (3) & (4) & (5) & (6) & (7) & (8) & (9) & (10) & (11)\\
\hline
J215445.80-072029.1 & 10.4 & 2.7 & 0.32 & 222$\pm$4 & 211$\pm$3 & 211$\pm$3 & 8.8$\pm$0.3 & 1 & F & 2399 \\
J215447.94-074329.7 & 10.2 & 1.2 & 0.21 & 179$\pm$3 & 176$\pm$2 & 176$\pm$2 & 4.4$\pm$0.1 & 1 & F & 2399 \\
J215457.43-073551.3 & 10.5 & 3.5 & 0.12 & 232$\pm$4 & 219$\pm$3 & 219$\pm$3 & 8.7$\pm$0.3 & 0 & S & 2399$^\dagger$ \\
J215556.94-065337.9 & 10.6 & 5.8 & 0.32 & 172$\pm$3 & 134$\pm$3 & 136$\pm$3 & 3.7$\pm$0.2 & 1 & L & 2399 \\
J215604.08-071938.1 & 10.3 & 4.3 & 0.26 & 88$\pm$5 & 87$\pm$5 & 87$\pm$4 & 5.6$\pm$0.6 & 1 & L & 2399 \\
J215619.00-075515.6 & 10.2 & 1.9 & 0.18 & 240$\pm$5 & 224$\pm$3 & 222$\pm$3 & 8.0$\pm$0.3 & 1 & F & 2399 \\
J215624.56-081159.8 & 10.2 & 2.4 & 0.72 & 152$\pm$5 & 152$\pm$3 & 174$\pm$4 & 11.1$\pm$0.7 & 2 & F & 2399 \\
J215628.95-074516.1 & 10.5 & 3.6 & 0.01 & 121$\pm$4 & 104$\pm$3 & 105$\pm$3 & 3.0$\pm$0.2 & 2 & F & 2399 \\
J215634.44-075217.5 & 10.1 & 1.7 & 0.50 & 126$\pm$4 & 131$\pm$3 & 135$\pm$3 & 7.6$\pm$0.4 & 0 & F & 2399 \\
J215635.58-075616.9 & 10.4 & 2.8 & 0.41 & 170$\pm$4 & 176$\pm$3 & 182$\pm$3 & 8.8$\pm$0.4 & 2 & F & 2399 \\
J215636.04-065225.6 & 10.3 & 6.2 & 0.49 & 93$\pm$5 & 114$\pm$7 & 107$\pm$6 & 11.9$\pm$1.6 & 1 & L & 2399 \\
J215637.29-074043.0 & 10.7 & 4.2 & 0.07 & 215$\pm$7 & 188$\pm$3 & 189$\pm$3 & 6.3$\pm$0.2 & 2 & F & 2399 \\
J215643.13-073259.8 & 10.5 & 3.3 & 0.20 & 218$\pm$7 & 197$\pm$3 & 196$\pm$3 & 6.7$\pm$0.3 & 2 & S & 2399 \\
J215646.76-065650.3 & 10.7 & 4.1 & 0.31 & 317$\pm$7 & 299$\pm$4 & 299$\pm$4 & 13.2$\pm$0.4 & 1 & F & 2399$^\dagger$ \\
J215650.44-074111.3 & 10.3 & 2.5 & 0.56 & 155$\pm$4 & 148$\pm$3 & 152$\pm$3 & 7.3$\pm$0.3 & 2 & F & 2399 \\
J215653.48-075405.5 & 10.1 & 1.5 & 0.51 & 126$\pm$3 & 126$\pm$2 & 127$\pm$2 & 5.2$\pm$0.2 & 2 & F & 2399 \\
J215656.92-065751.3 & 10.2 & 2.0 & 0.17 & 185$\pm$5 & 183$\pm$4 & 180$\pm$4 & 7.2$\pm$0.4 & 2 & F & 2399$^\dagger$ \\
J215658.25-074910.7 & 10.2 & 3.0 & 0.11 & 165$\pm$4 & 163$\pm$4 & 164$\pm$4 & 7.9$\pm$0.5 & 1 & F & 2399 \\
J215658.51-074843.1 & 10.5 & 4.1 & 0.16 & 145$\pm$9 & 133$\pm$6 & 133$\pm$6 & 8.8$\pm$0.9 & 1 & L & 2399 \\
J215701.22-075415.2 & 10.4 & 3.4 & 0.33 & 188$\pm$6 & 179$\pm$3 & 182$\pm$3 & 7.3$\pm$0.3 & 2 & S & 2399 \\
J215701.35-074653.3 & 10.4 & 1.8 & 0.21 & 184$\pm$6 & 186$\pm$4 & 182$\pm$3 & 5.1$\pm$0.3 & 1 & F & 2399 \\
J215701.71-075022.5 & 10.9 & 5.8 & 0.29 & 278$\pm$4 & 262$\pm$4 & 263$\pm$3 & 9.6$\pm$0.3 & 2 & F & 2399 \\
J215716.83-075450.5 & 10.4 & 2.8 & 0.29 & 229$\pm$5 & 228$\pm$3 & 227$\pm$3 & 8.6$\pm$0.3 & 1 & F & 2399 \\
J215721.41-074846.8 & 10.5 & 3.8 & 0.16 & 216$\pm$7 & 198$\pm$3 & 197$\pm$3 & 8.1$\pm$0.4 & 1 & F & 2399 \\
J215723.40-075814.0 & 10.6 & 3.1 & 0.29 & 280$\pm$6 & 276$\pm$4 & 276$\pm$4 & 9.5$\pm$0.3 & 1 & S & 2399 \\
J215726.31-075137.7 & 10.3 & 3.9 & 0.28 & 117$\pm$5 & 119$\pm$4 & 128$\pm$4 & 5.2$\pm$0.4 & 2 & F & 2399 \\
J215727.30-073357.5 & 10.3 & 2.6 & 0.34 & 169$\pm$5 & 170$\pm$4 & 173$\pm$4 & 7.0$\pm$0.3 & 2 & F & 2399 \\
J215727.63-074812.8 & 10.3 & 2.0 & 0.48 & 219$\pm$4 & 226$\pm$3 & 228$\pm$3 & 9.7$\pm$0.4 & 1 & F & 2399 \\
J215728.65-073155.4 & 10.3 & 2.6 & 0.12 & 169$\pm$3 & 167$\pm$2 & 166$\pm$2 & 7.6$\pm$0.3 & 1 & L & 2399 \\
J215729.42-074744.5 & 10.8 & 3.6 & 0.45 & 373$\pm$19 & 297$\pm$4 & 296$\pm$4 & 9.4$\pm$0.3 & 2 & F & 2399 \\
J215733.30-074420.6 & 10.5 & 3.8 & 0.31 & 137$\pm$3 & 133$\pm$2 & 131$\pm$2 & 4.4$\pm$0.2 & 2 & L & 2399 \\
J215733.47-074739.2 & 10.8 & 3.8 & 0.33 & 318$\pm$8 & 285$\pm$4 & 287$\pm$4 & 7.2$\pm$0.2 & 1 & F & 2399 \\
J215733.72-072729.3 & 10.7 & 3.2 & 0.50 & 269$\pm$6 & 242$\pm$3 & 238$\pm$3 & 8.6$\pm$0.3 & 2 & F & 2399 \\
J215743.17-072347.5 & 10.6 & 3.2 & 0.11 & 181$\pm$3 & 185$\pm$2 & 185$\pm$2 & 4.3$\pm$0.2 & 0 & F & 2399 \\
J215743.23-074545.1 & 10.6 & 3.5 & 0.20 & 247$\pm$7 & 221$\pm$3 & 221$\pm$3 & 7.6$\pm$0.3 & 2 & S & 2399 \\
J215745.05-075701.8 & 10.2 & 1.5 & 0.38 & 250$\pm$4 & 245$\pm$3 & 249$\pm$3 & 10.4$\pm$0.4 & 1 & F & 2399 \\
J215753.00-074419.0 & 10.6 & 3.8 & 0.06 & 167$\pm$5 & 160$\pm$3 & 160$\pm$3 & 6.4$\pm$0.3 & 2 & F & 2399 \\
J215759.85-072749.5 & 10.6 & 5.6 & 0.13 & 118$\pm$5 & 96$\pm$3 & 98$\pm$3 & 3.8$\pm$0.3 & 1 & L & 2399 \\
J215806.62-080642.4 & 10.4 & 2.8 & 0.13 & 177$\pm$3 & 178$\pm$3 & 177$\pm$3 & 7.8$\pm$0.9 & 1 & F & 2399 \\
J215807.50-075545.4 & 10.6 & 4.0 & 0.39 & 245$\pm$10 & 233$\pm$4 & 236$\pm$4 & 10.4$\pm$0.4 & 1 & F & 2399 \\
J215810.04-074801.3 & 10.4 & 2.1 & 0.40 & 211$\pm$11 & 204$\pm$4 & 206$\pm$4 & 7.7$\pm$0.5 & 1 & F & 2399 \\
J215811.35-072654.0 & 10.3 & 1.7 & 0.24 & 229$\pm$4 & 226$\pm$3 & 229$\pm$3 & 7.9$\pm$0.3 & 1 & F & 2399 \\
J215826.28-072154.0 & 10.5 & 7.4 & 0.13 & 70$\pm$7 & 67$\pm$7 & 70$\pm$7 & 4.7$\pm$1.1 & 1 & L & 2399 \\
J215840.76-074939.8 & 10.2 & 2.5 & 0.32 & 199$\pm$6 & 186$\pm$4 & 187$\pm$4 & 10.3$\pm$0.5 & 1 & F & 2399$^\dagger$ \\
J215853.98-071531.8 & 10.8 & 8.4 & 0.64 & 170$\pm$3 & 173$\pm$4 & 166$\pm$3 & 7.8$\pm$0.4 & 0 & L & 2399$^\dagger$ \\
J215902.71-073930.0 & 10.3 & 2.7 & 0.52 & 130$\pm$5 & 129$\pm$3 & 125$\pm$3 & 8.6$\pm$0.6 & 2 & F & 2399 \\
J215910.35-080431.2 & 10.6 & 4.8 & 0.23 & 143$\pm$3 & 150$\pm$3 & 155$\pm$3 & 5.3$\pm$0.2 & 2 & L & 2399$^\dagger$ \\
J215924.41-073442.7 & 10.3 & 4.6 & 0.22 & 124$\pm$4 & 127$\pm$4 & 123$\pm$5 & 5.3$\pm$0.8 & 0 & L & 2399 \\
J215942.63-073028.6 & 10.2 & 3.8 & 0.45 & 145$\pm$9 & 133$\pm$6 & 133$\pm$6 & 10.7$\pm$1.1 & 2 & L & 2399$^\dagger$ \\
J215945.43-072312.2 & 10.6 & 3.4 & 0.11 & 231$\pm$4 & 229$\pm$3 & 229$\pm$3 & 7.8$\pm$0.3 & 2 & S & 2399$^\dagger$ \\
\hline
\end{tabular}
\\
Notes: Column (1): SDSS galaxy ID. Column (2): Total {\it r-}band luminosity. Column (3): Effective radius in arcseconds. Column (4): Ellipticity. Column (5): Velocity dispersion measured in a central R$_e$/8 circular aperture. Column (6): Velocity dispersion measured in a 1 R$_e$ circular aperture. Column (7): Velocity dispersion measured in an elliptical aperture with ellipticity $\epsilon$ and major axis radius R$_{e,maj}$. Column (8): Dynamical mass-to-light ratio derived from JAM modelling. Column (9): Kinematic morphological type -- L = late-type galaxy, S = slow rotator, F = fast rotator. Column (10): Quality of JAM model fit. Following \citet{Cappellari:2013a}, the JAM model fit qualities were classified as: 2 -- good fit, 1 -- adequate fit, 0 -- poor fit or bad data, therefore uncertainty on M/L may be underestimated. Column (11): Host cluster. Galaxies subsequently identified as non-members are indicates with $^\dagger$.
\end{table*}

\section{V$_{rms}$ maps and JAM model fits}
\begin{figure*}
\includegraphics[height=0.135\textheight]{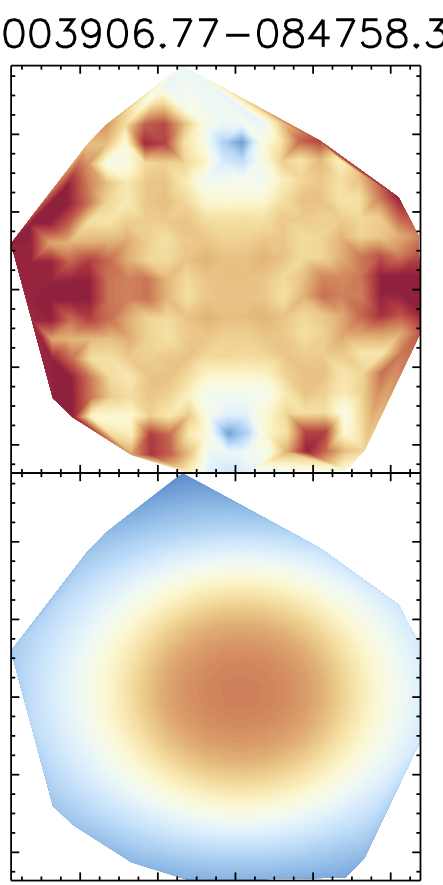}
\includegraphics[height=0.135\textheight]{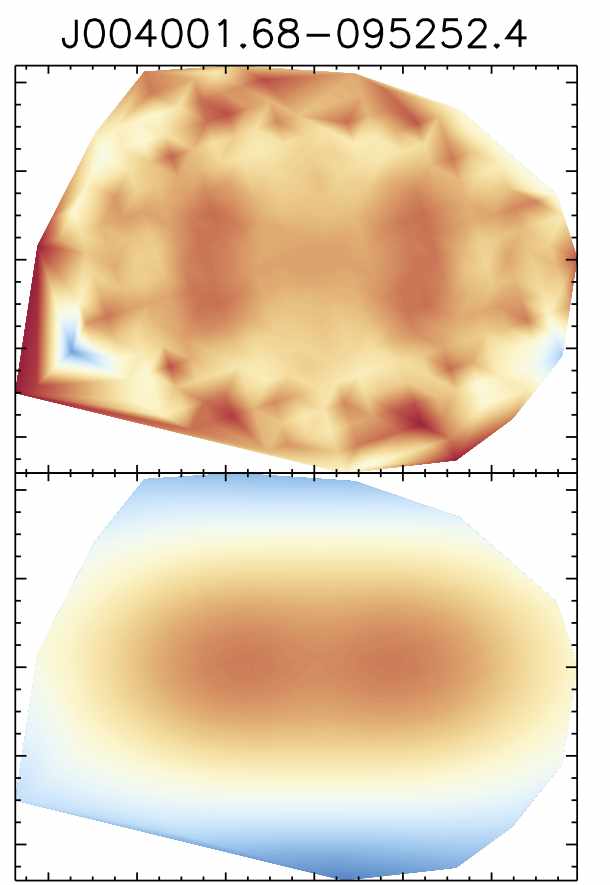}
\includegraphics[height=0.135\textheight]{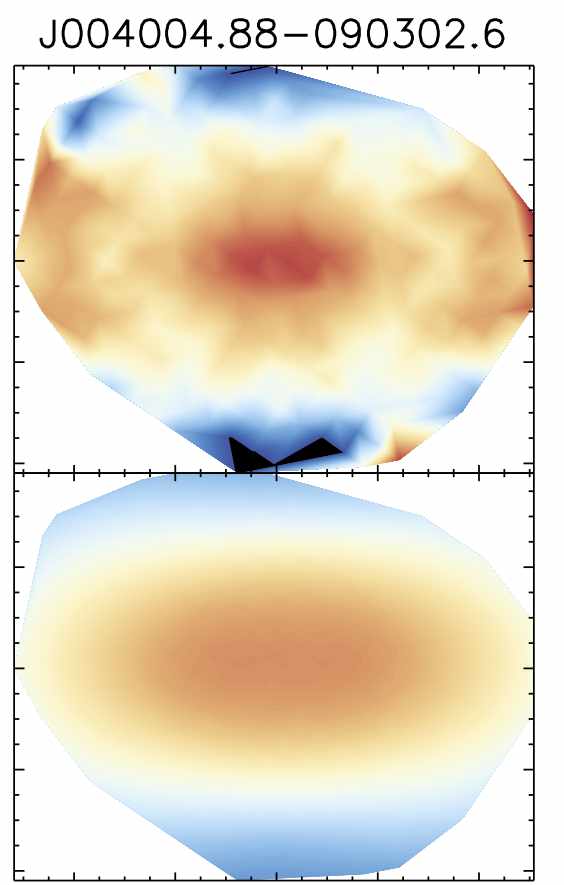}
\includegraphics[height=0.135\textheight]{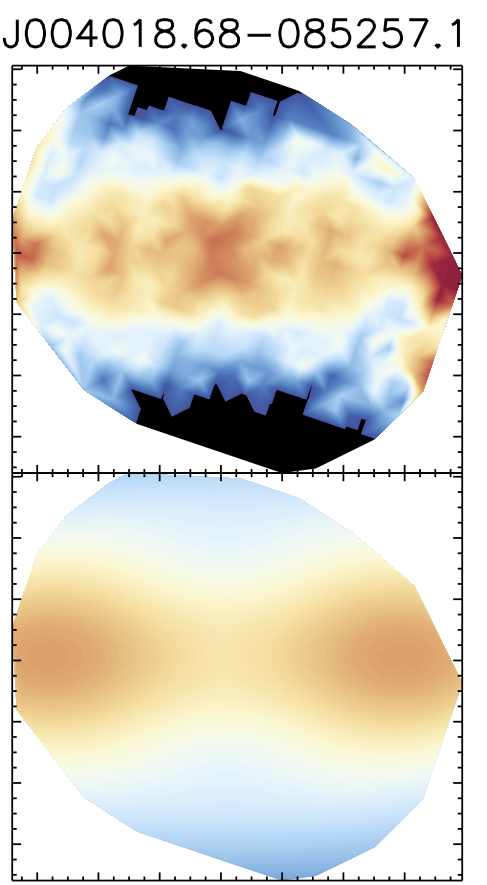}
\includegraphics[height=0.135\textheight]{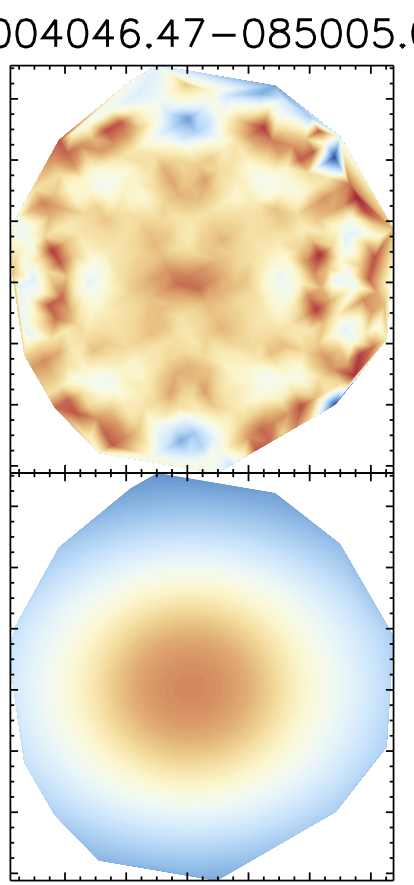}
\includegraphics[height=0.135\textheight]{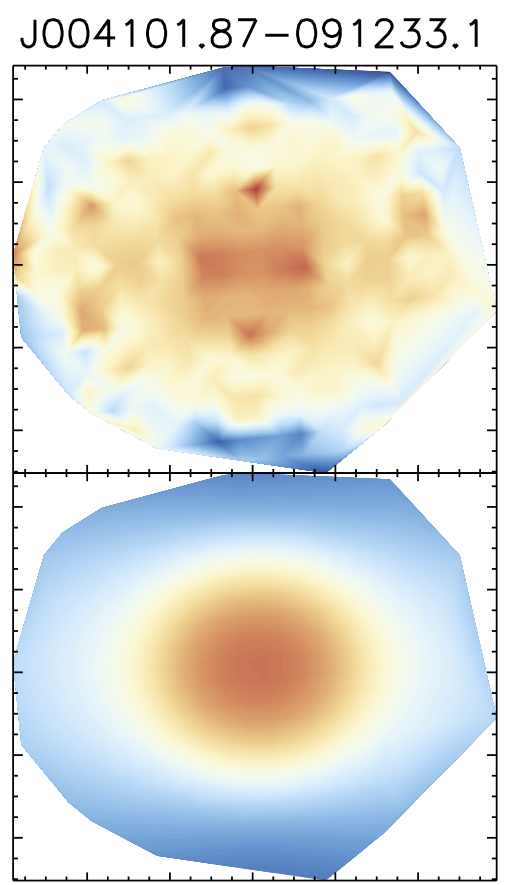}
\includegraphics[height=0.135\textheight]{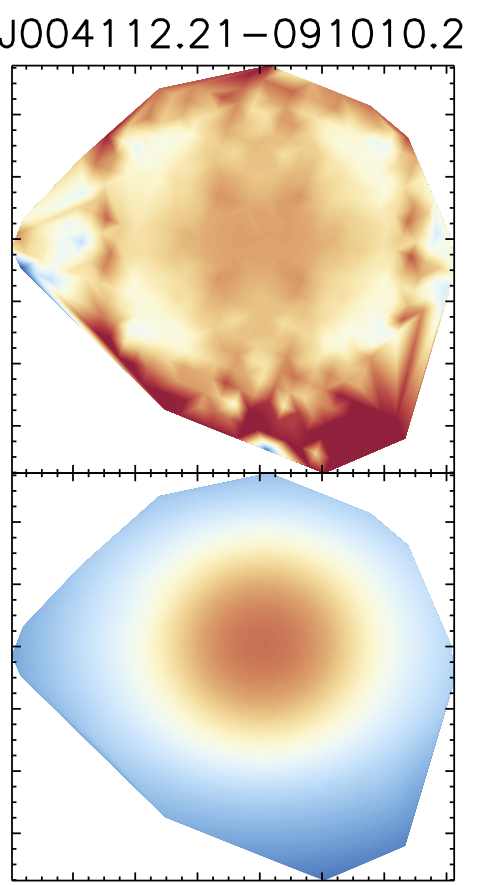}
\includegraphics[height=0.135\textheight]{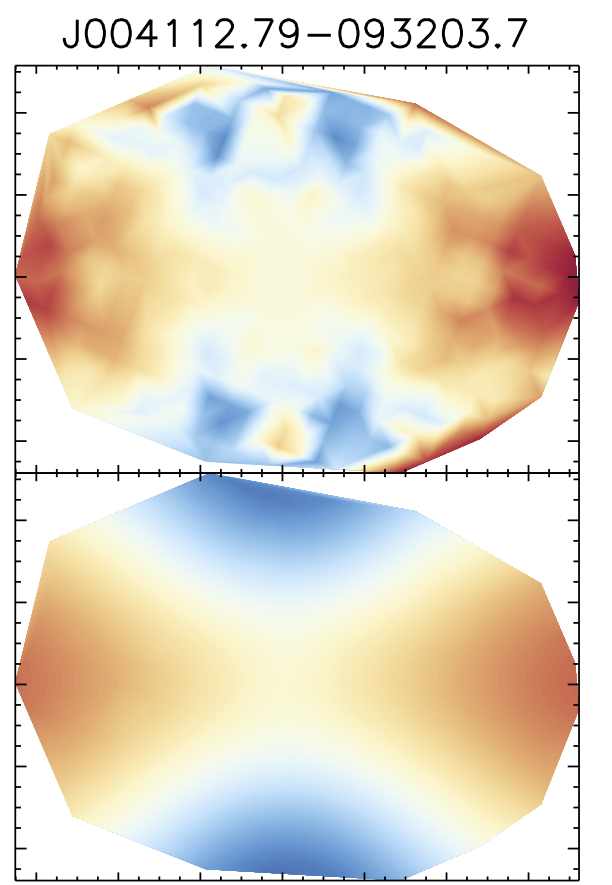}
\includegraphics[height=0.135\textheight]{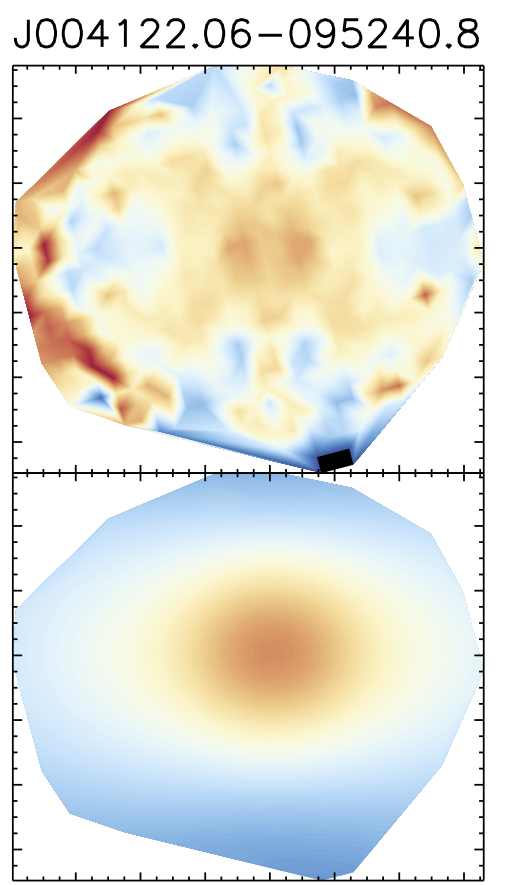}
\includegraphics[height=0.135\textheight]{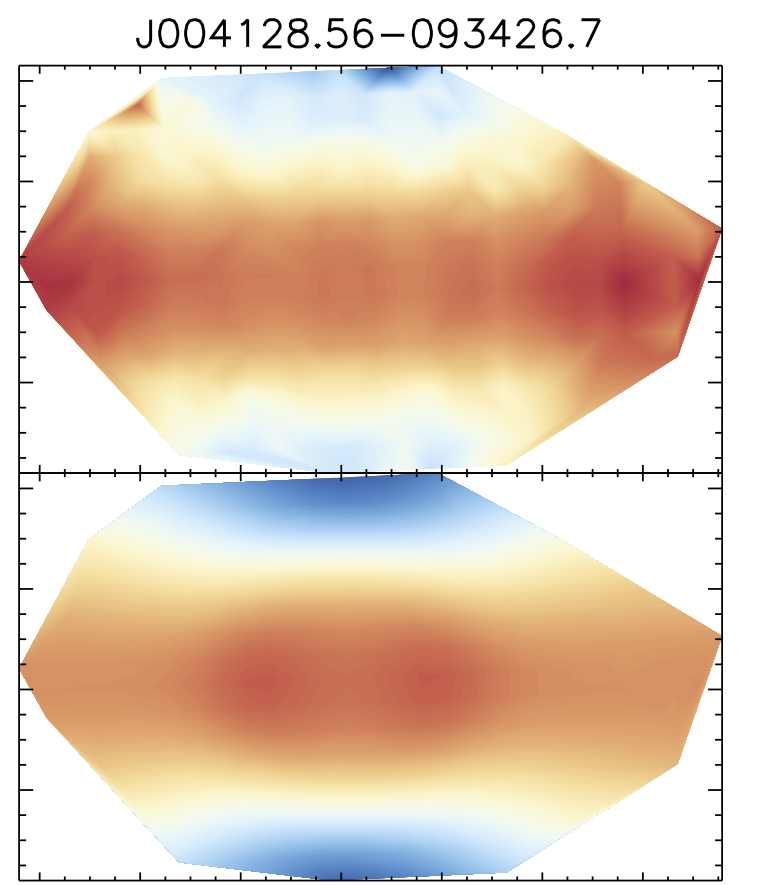}
\includegraphics[height=0.135\textheight]{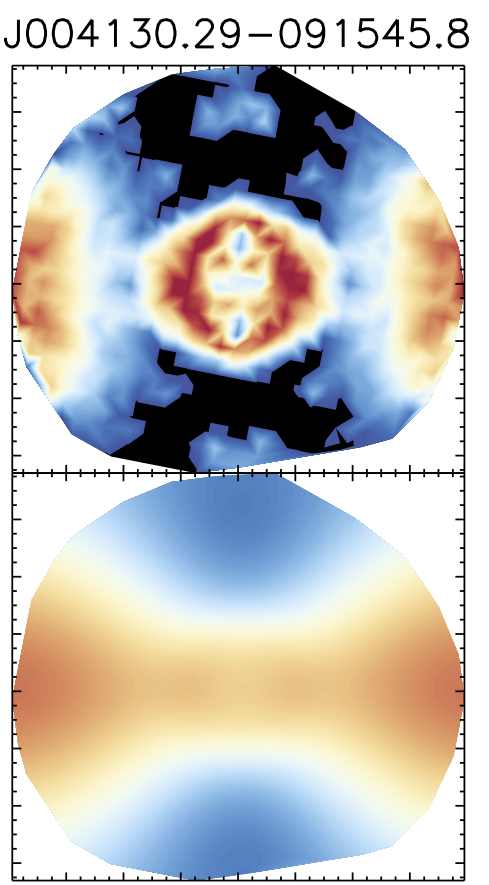}
\includegraphics[height=0.135\textheight]{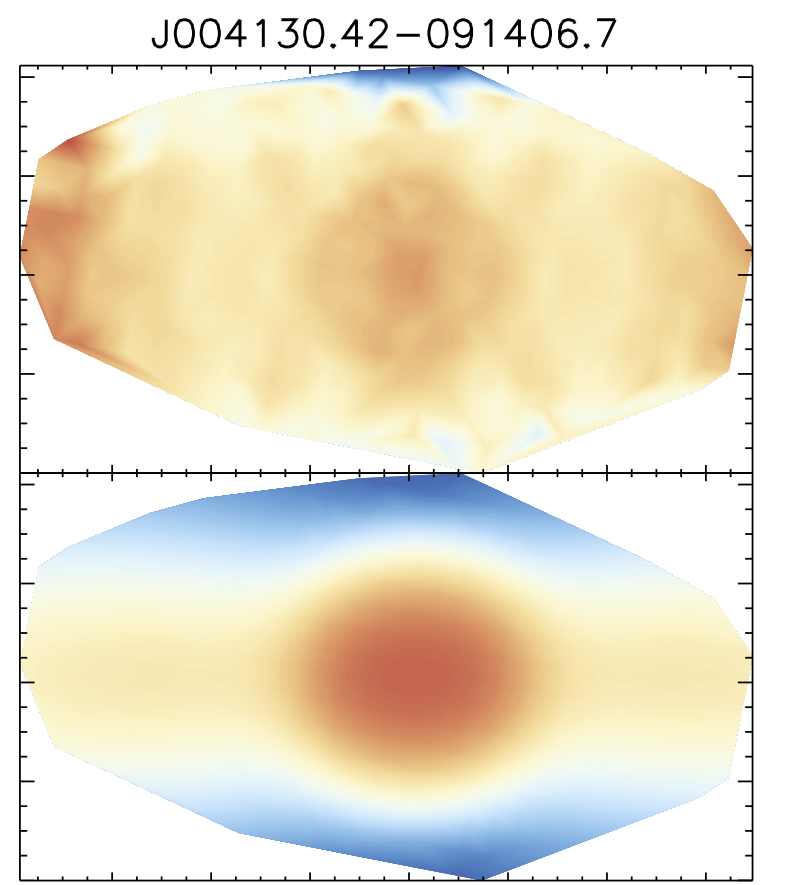}
\includegraphics[height=0.135\textheight]{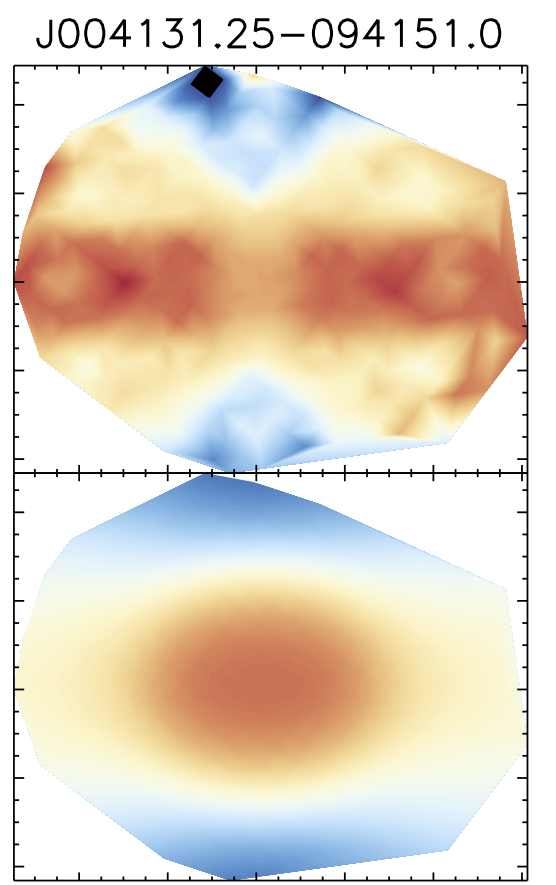}
\includegraphics[height=0.135\textheight]{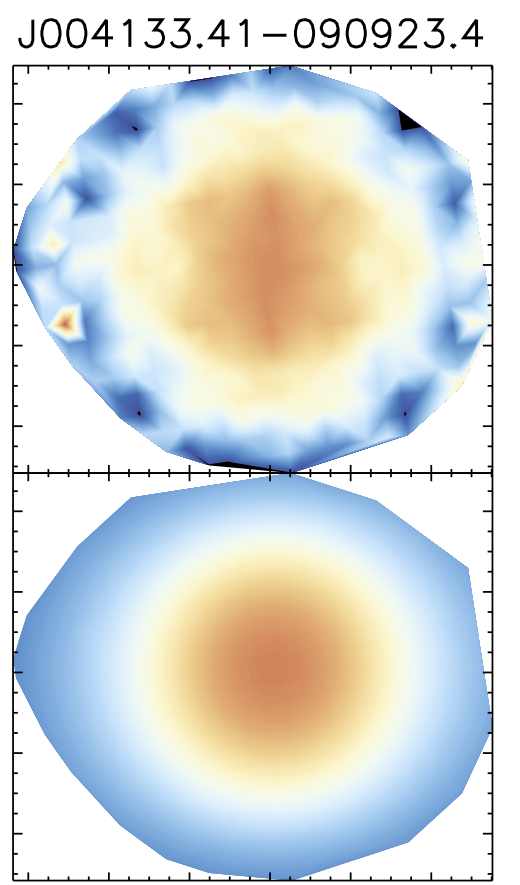}
\includegraphics[height=0.135\textheight]{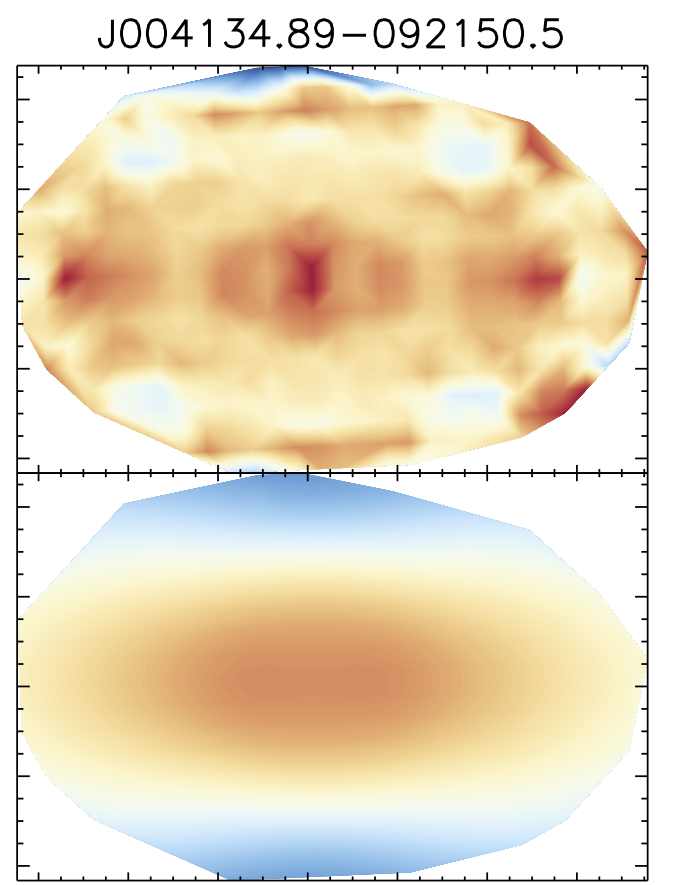}
\includegraphics[height=0.135\textheight]{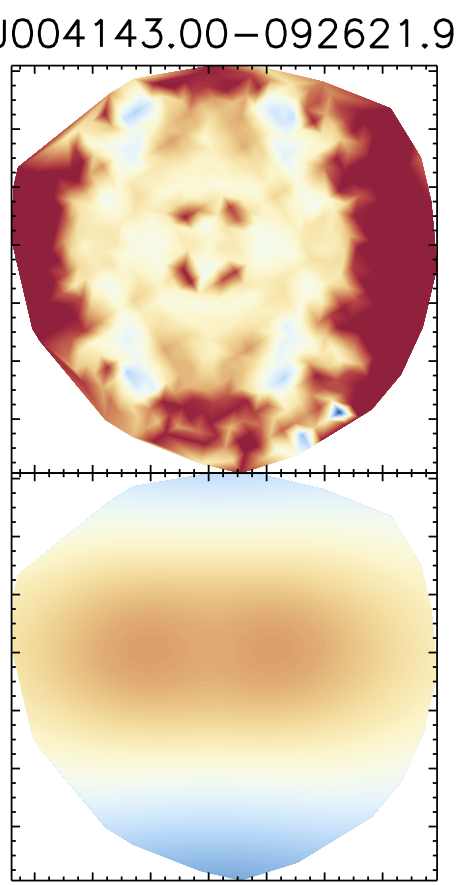}
\includegraphics[height=0.135\textheight]{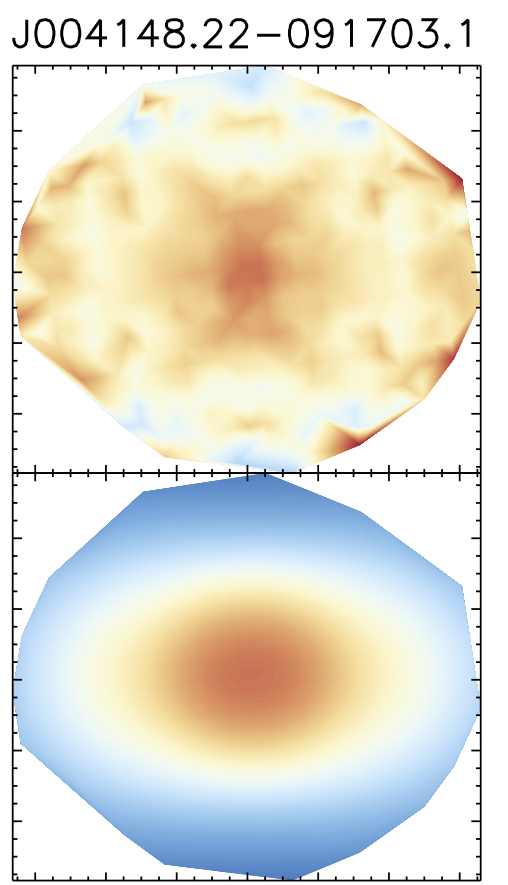}
\includegraphics[height=0.135\textheight]{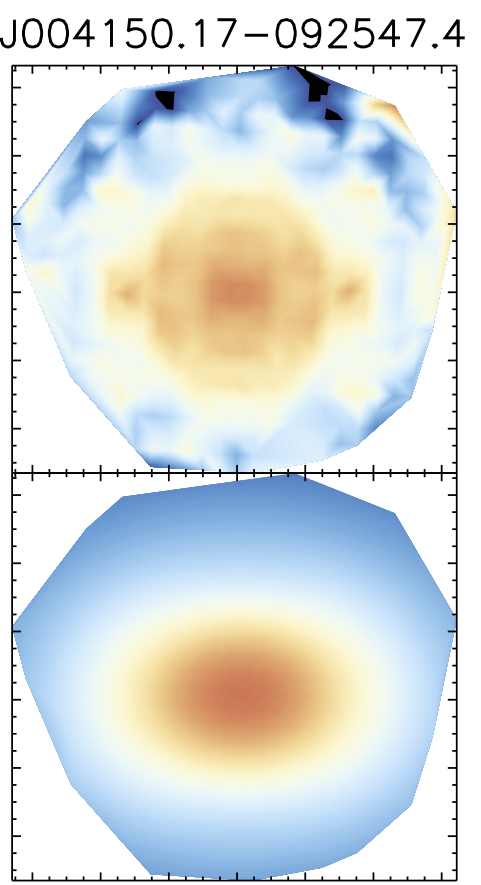}
\includegraphics[height=0.135\textheight]{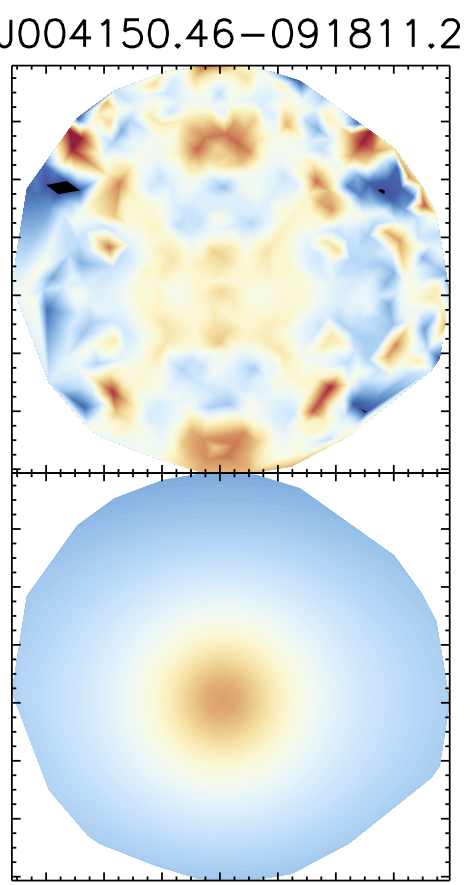}
\includegraphics[height=0.135\textheight]{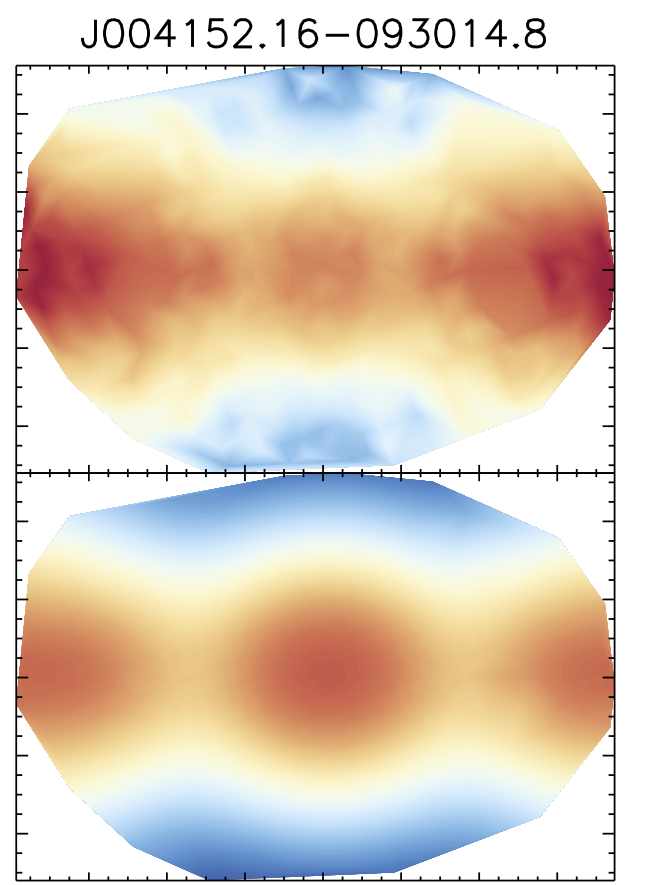}
\includegraphics[height=0.135\textheight]{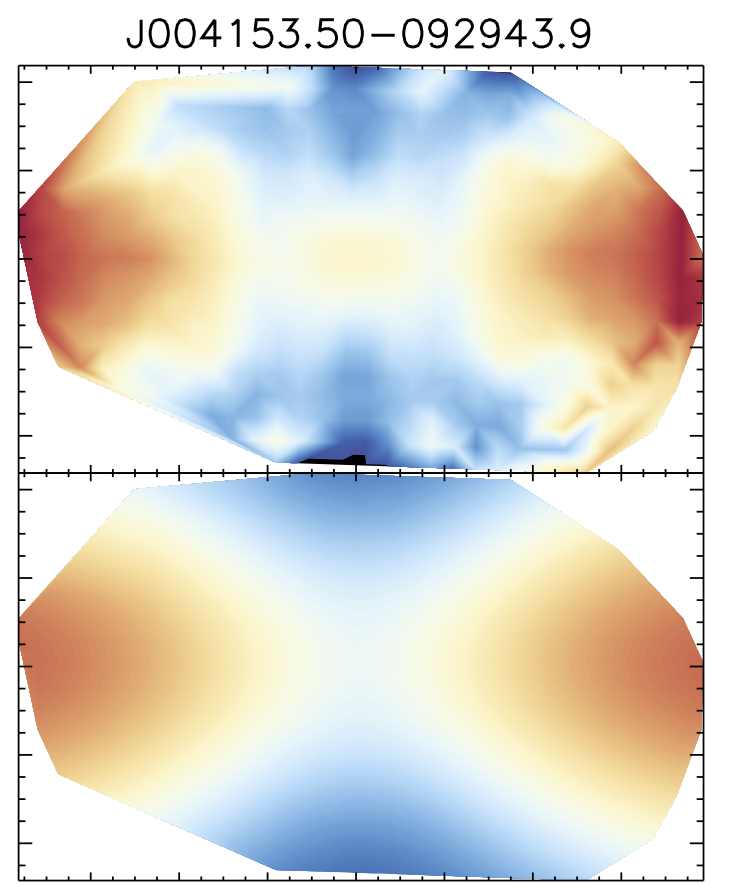}
\includegraphics[height=0.135\textheight]{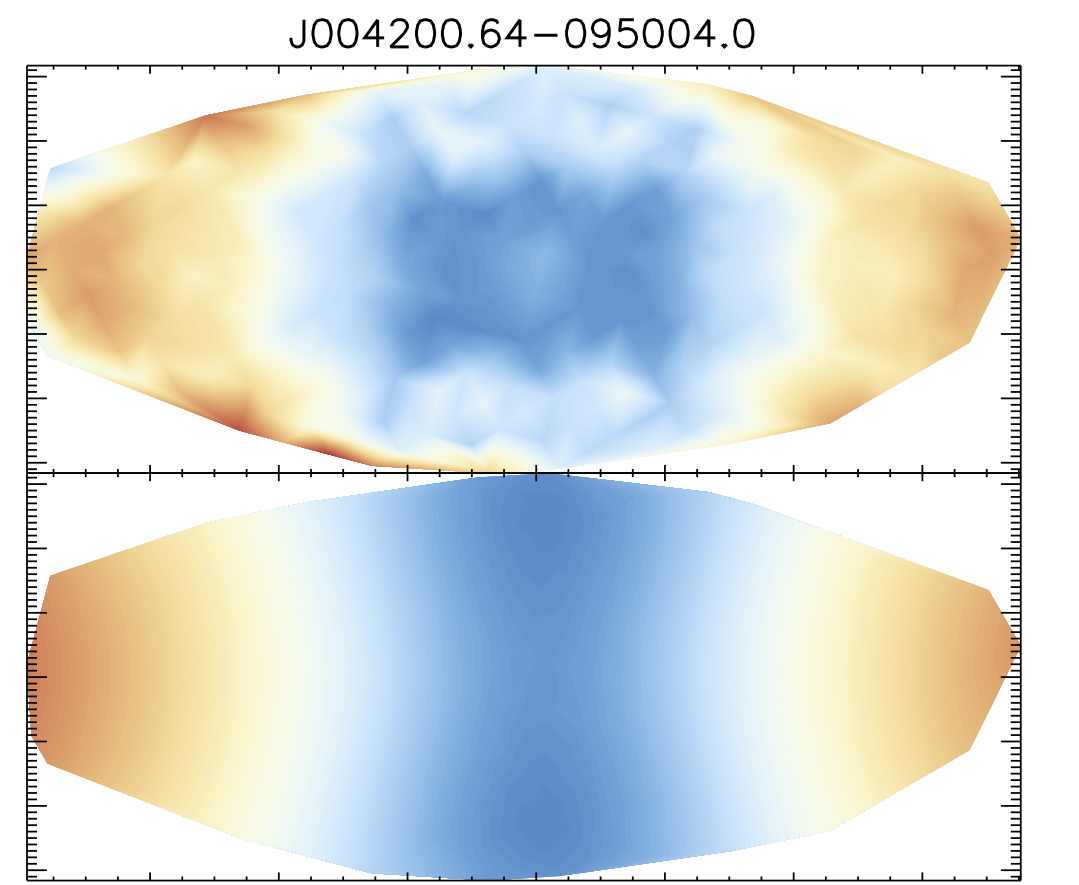}
\includegraphics[height=0.135\textheight]{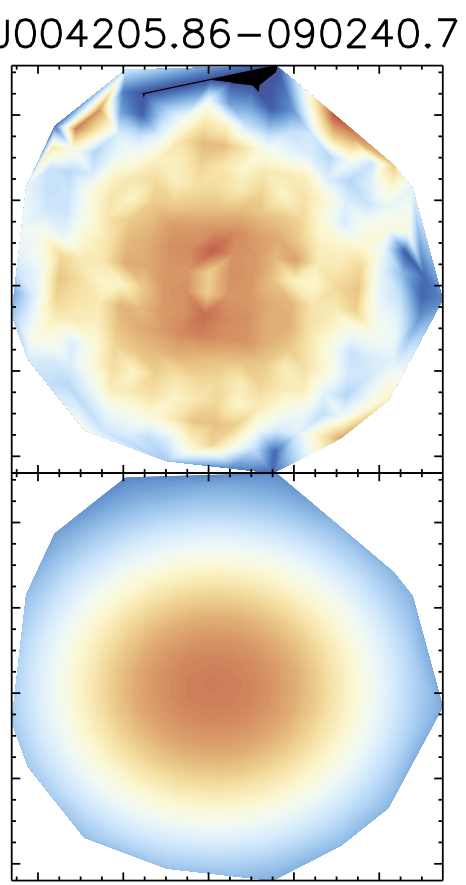}
\includegraphics[height=0.135\textheight]{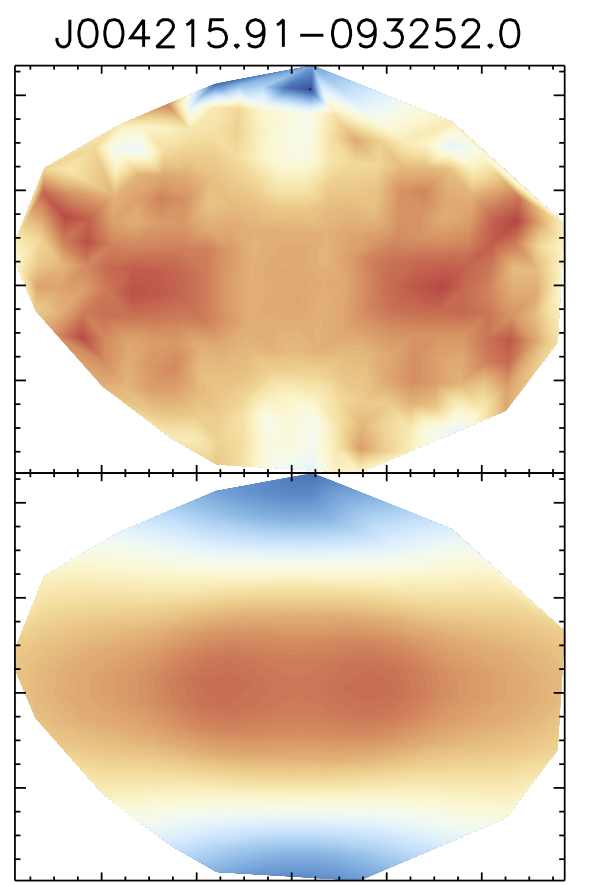}
\includegraphics[height=0.135\textheight]{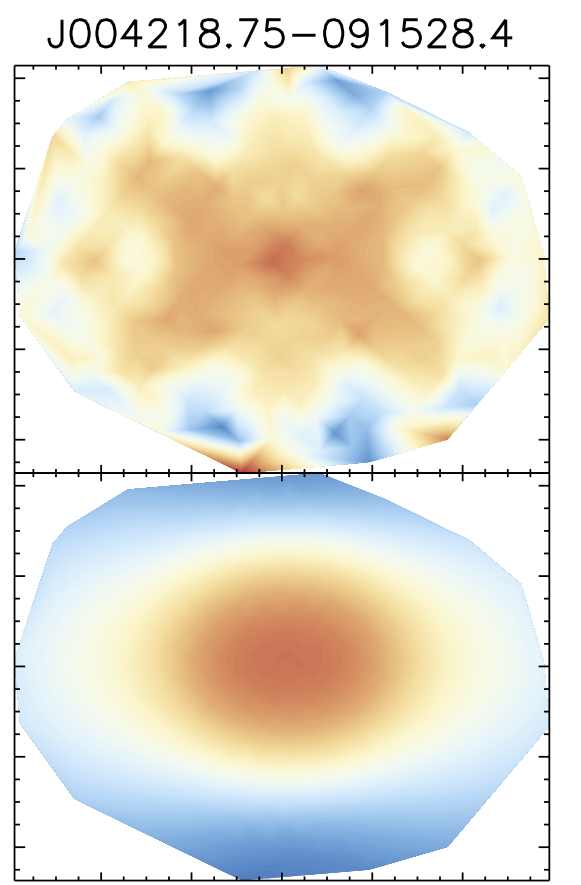}
\includegraphics[height=0.135\textheight]{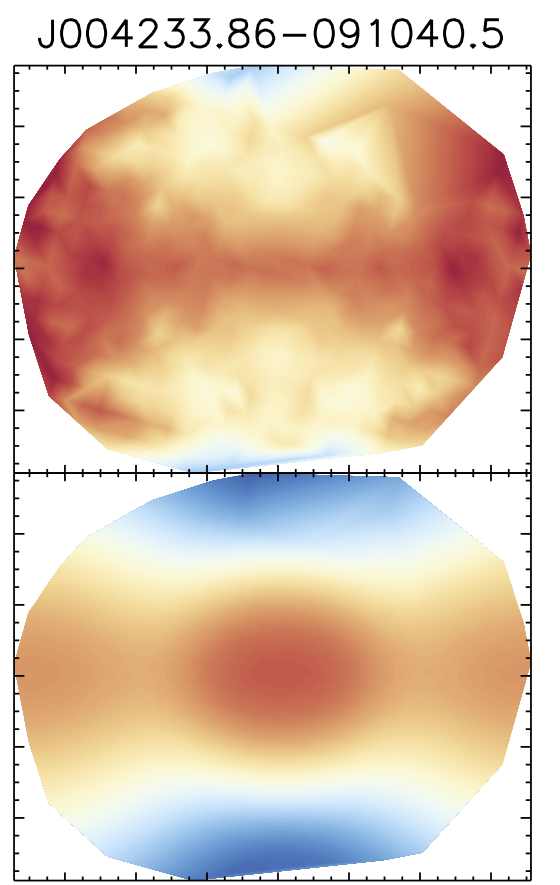}
\includegraphics[height=0.135\textheight]{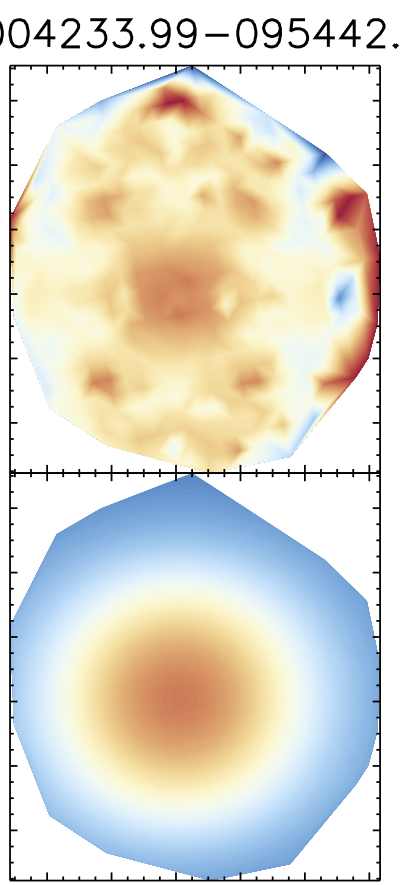}
\includegraphics[height=0.135\textheight]{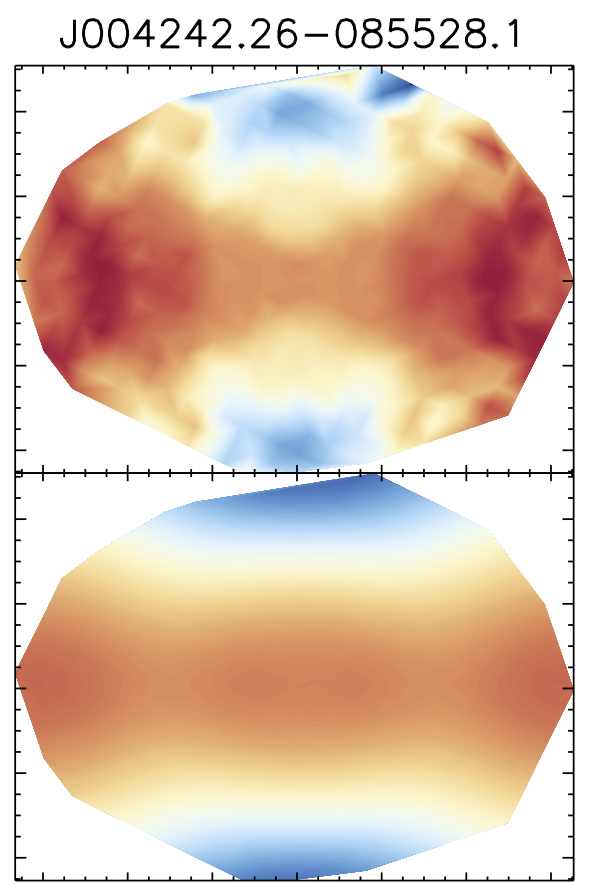}
\includegraphics[height=0.135\textheight]{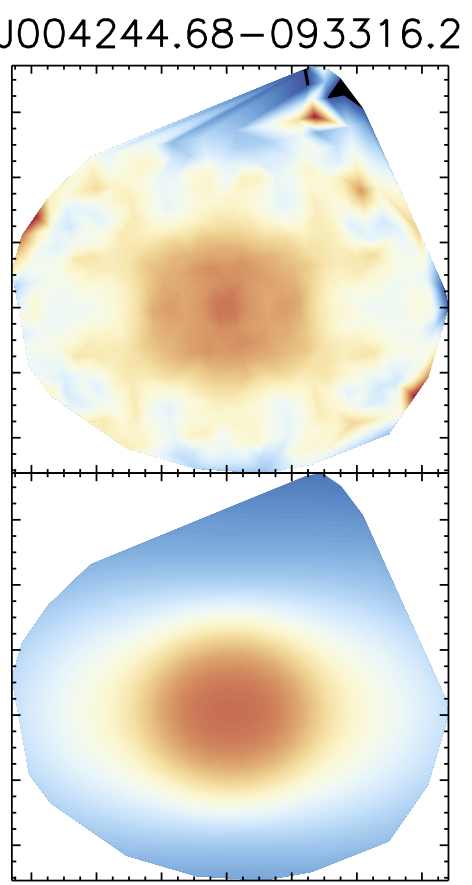}
\includegraphics[height=0.135\textheight]{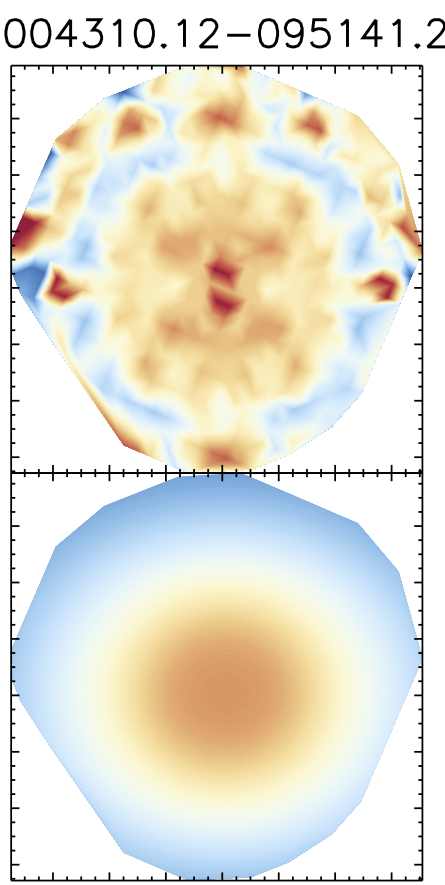}
\includegraphics[height=0.135\textheight]{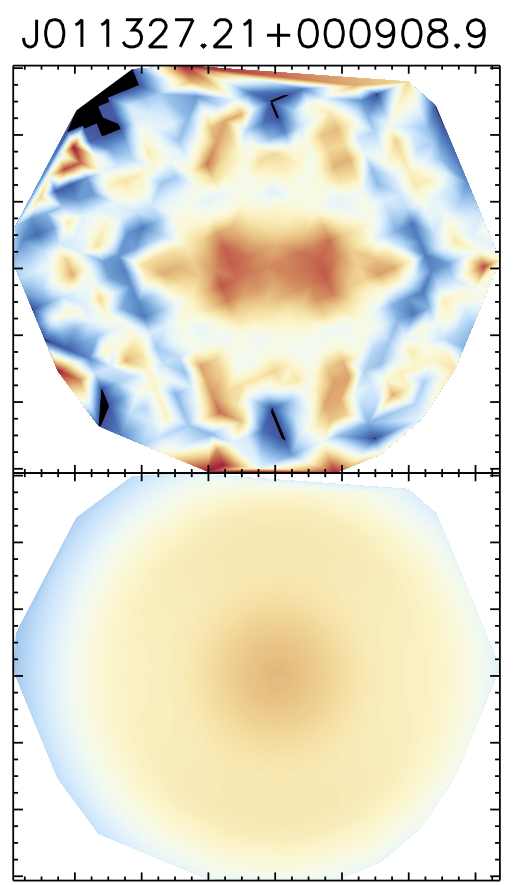}
\includegraphics[height=0.135\textheight]{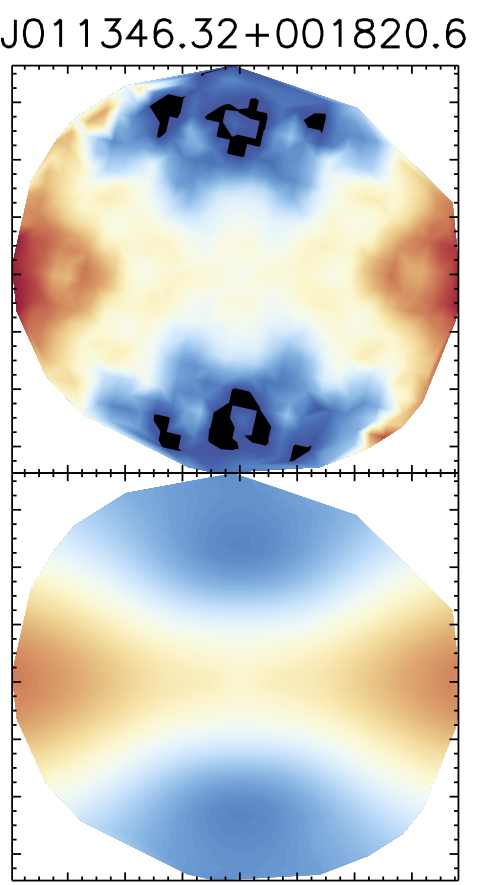}
\includegraphics[height=0.135\textheight]{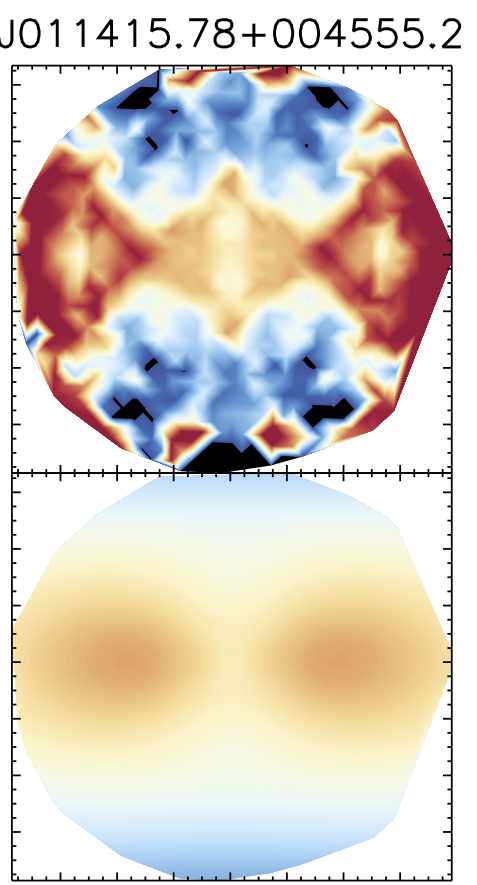}
\includegraphics[height=0.135\textheight]{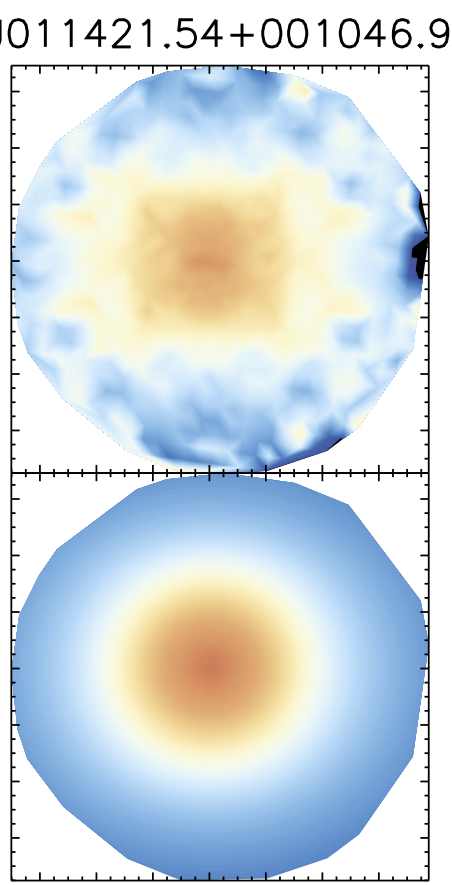}
\includegraphics[height=0.135\textheight]{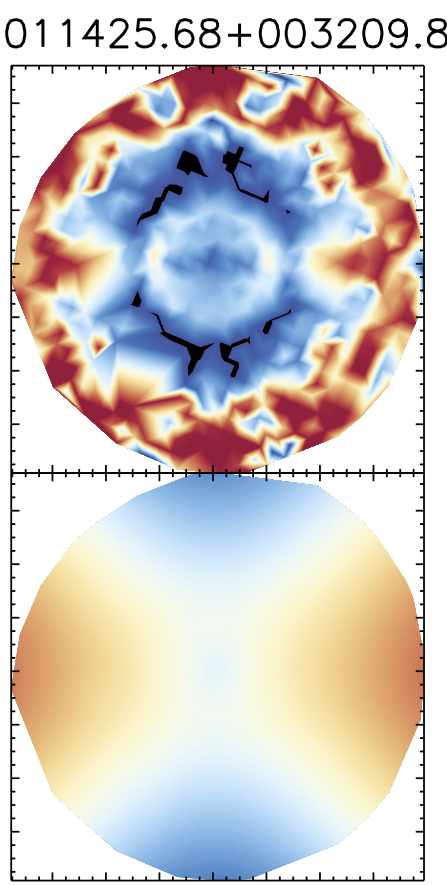}
\includegraphics[height=0.135\textheight]{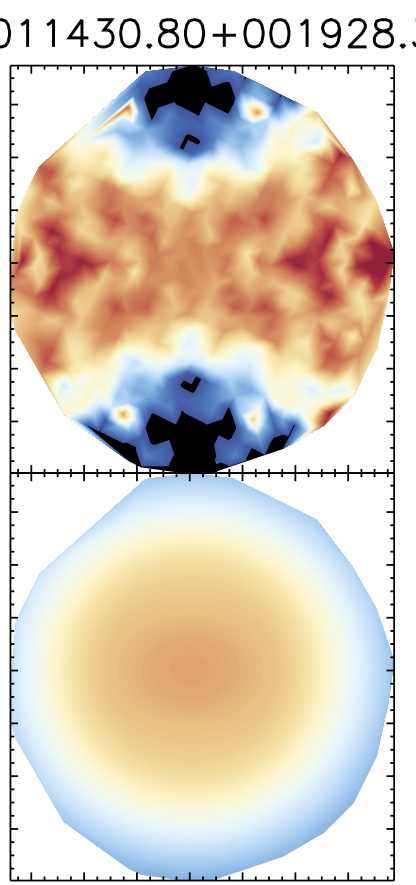}
\includegraphics[height=0.135\textheight]{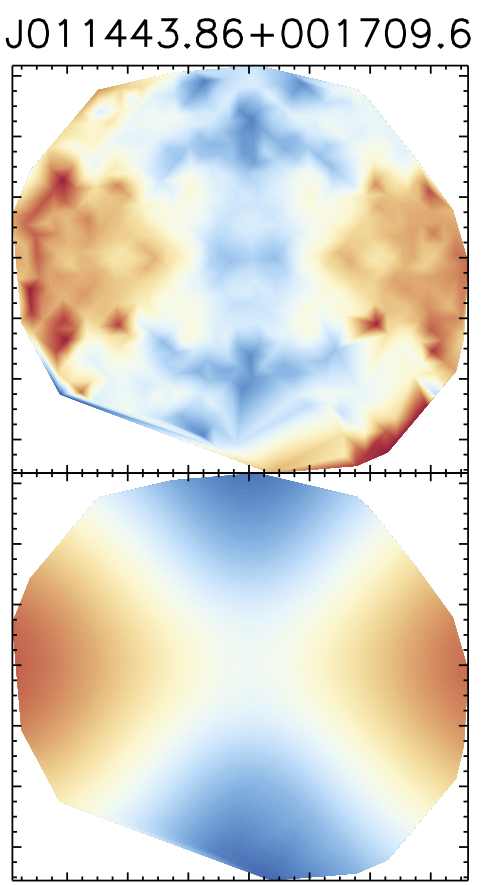}
\includegraphics[height=0.135\textheight]{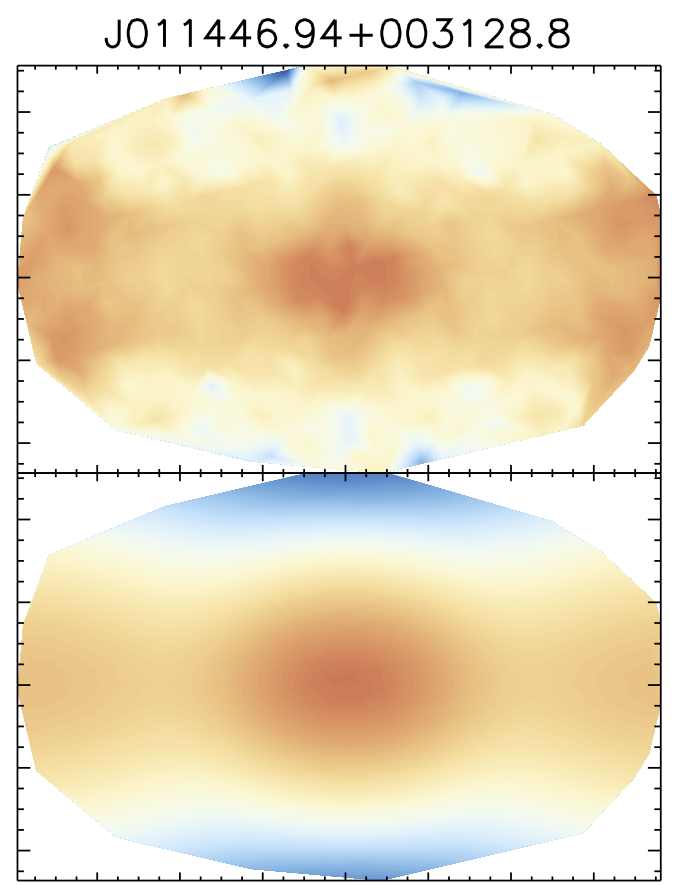}
\includegraphics[height=0.135\textheight]{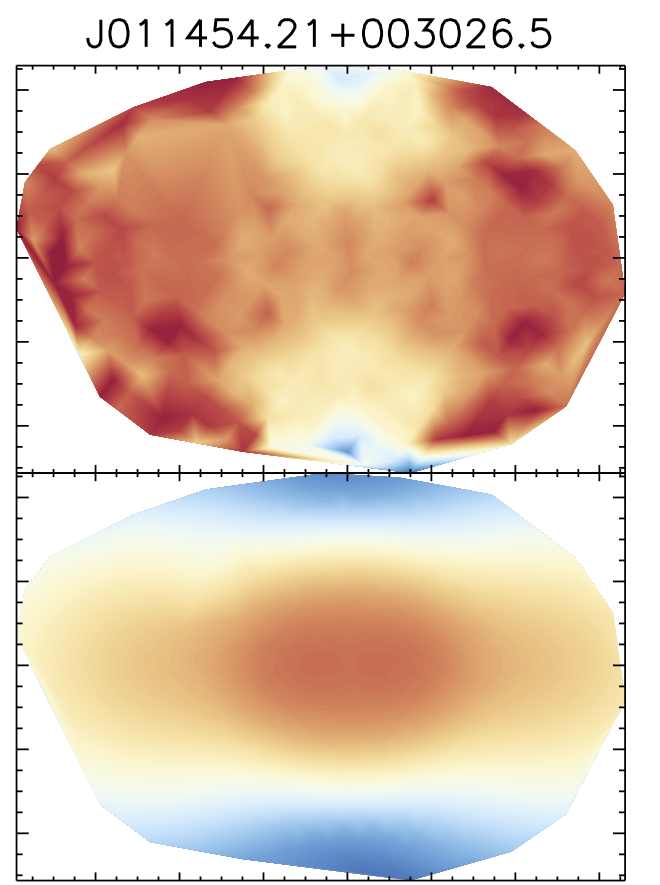}
\includegraphics[height=0.135\textheight]{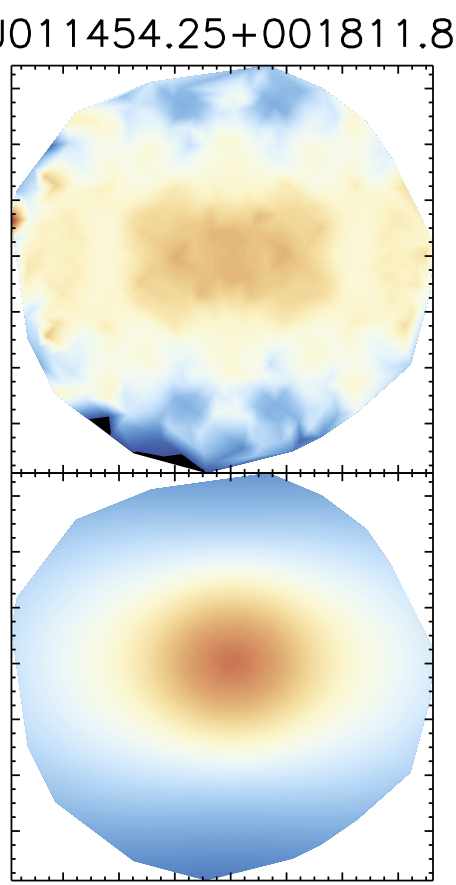}
\includegraphics[height=0.135\textheight]{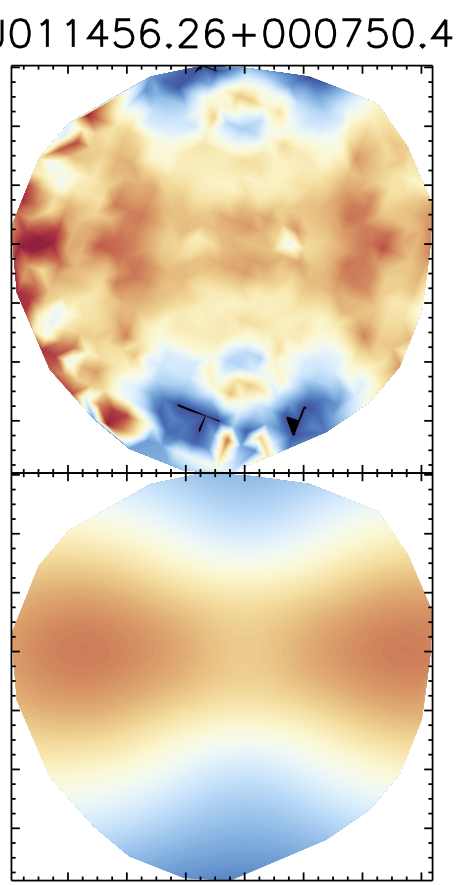}
\includegraphics[height=0.135\textheight]{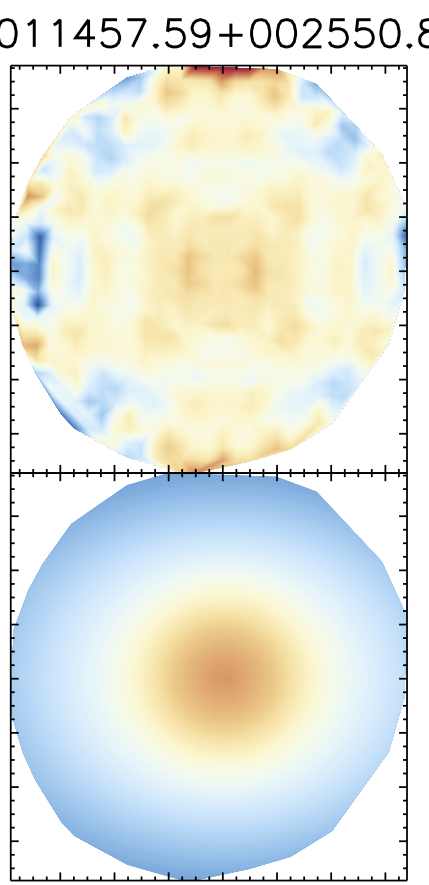}
\includegraphics[height=0.135\textheight]{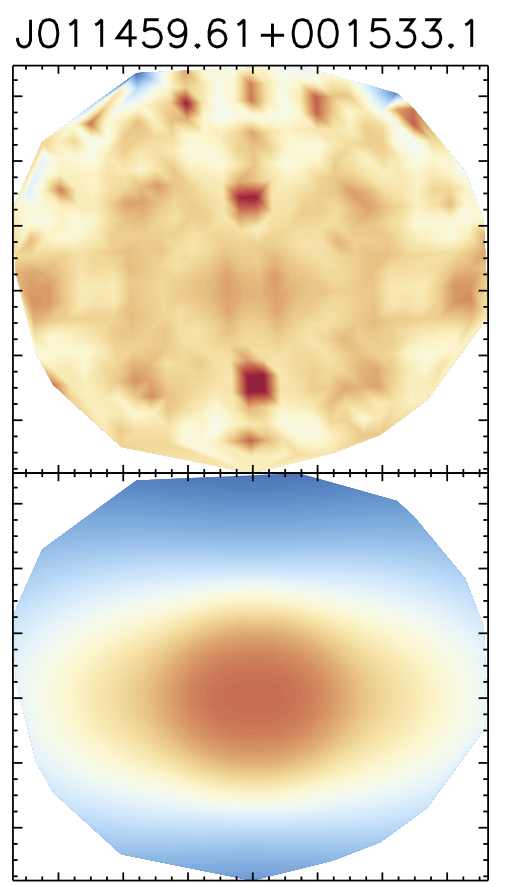}
\includegraphics[height=0.135\textheight]{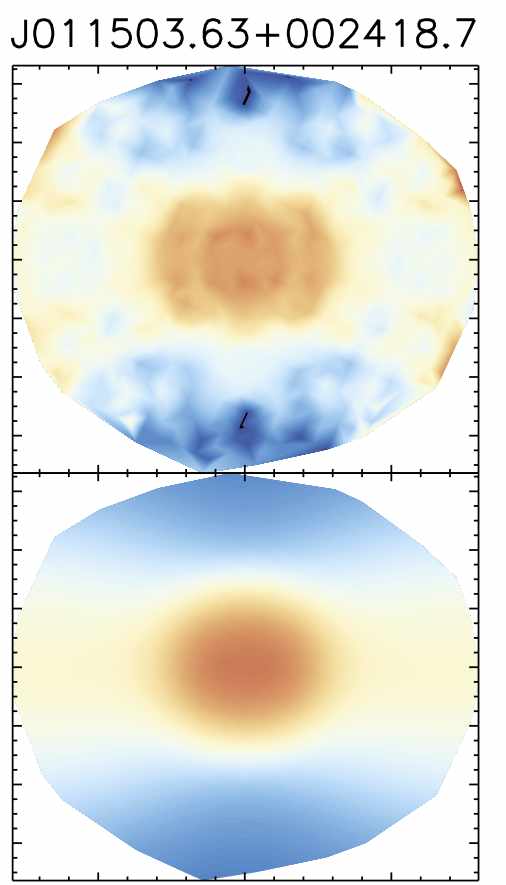}
\includegraphics[height=0.135\textheight]{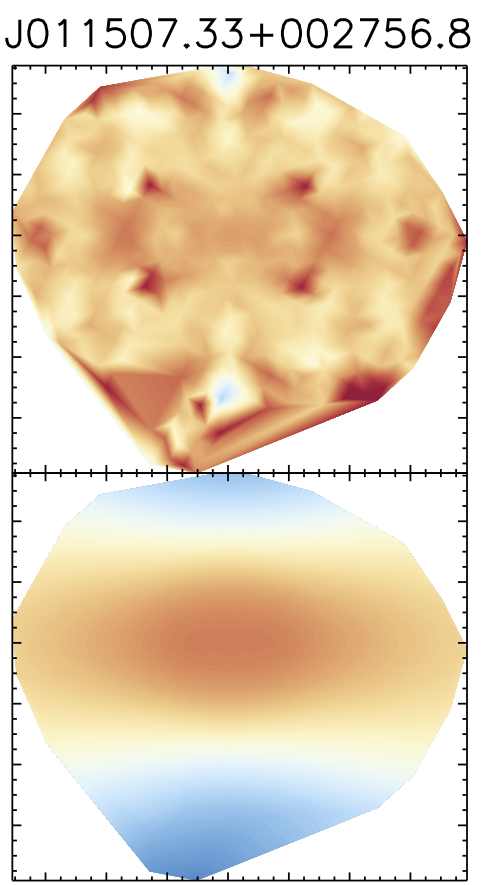}
\includegraphics[height=0.135\textheight]{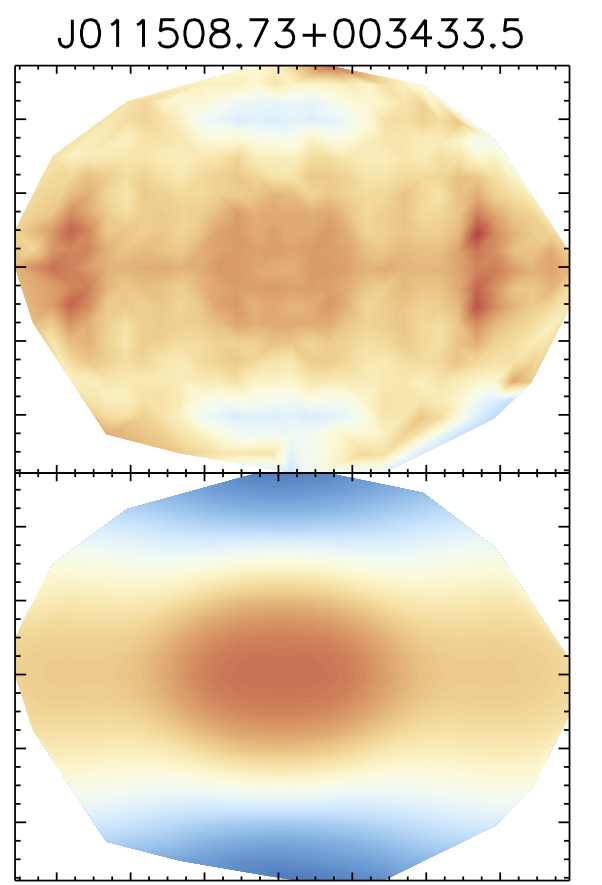}
\includegraphics[height=0.135\textheight]{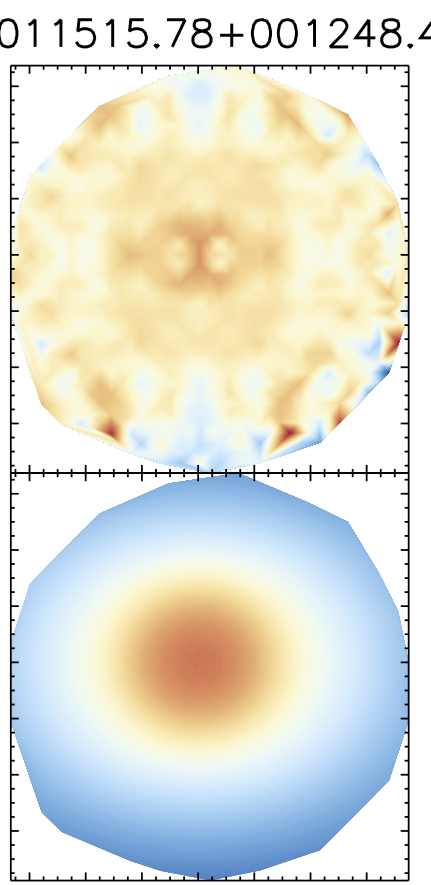}
\includegraphics[height=0.135\textheight]{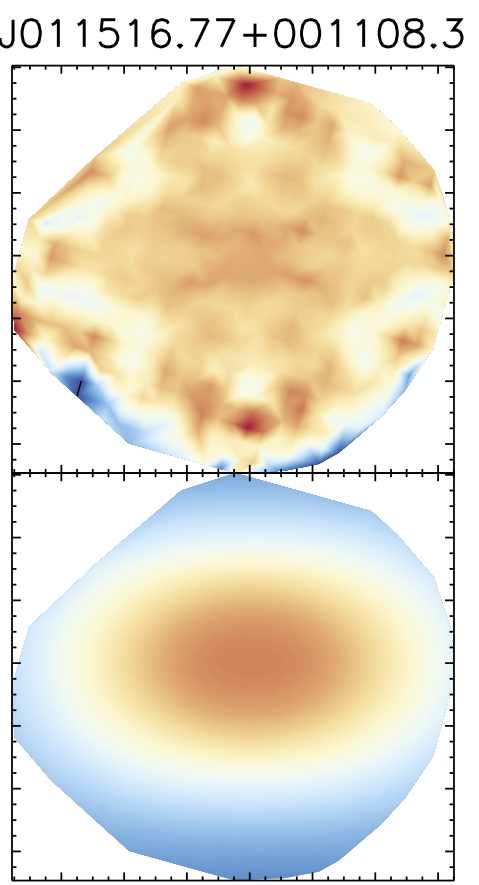}
\includegraphics[height=0.135\textheight]{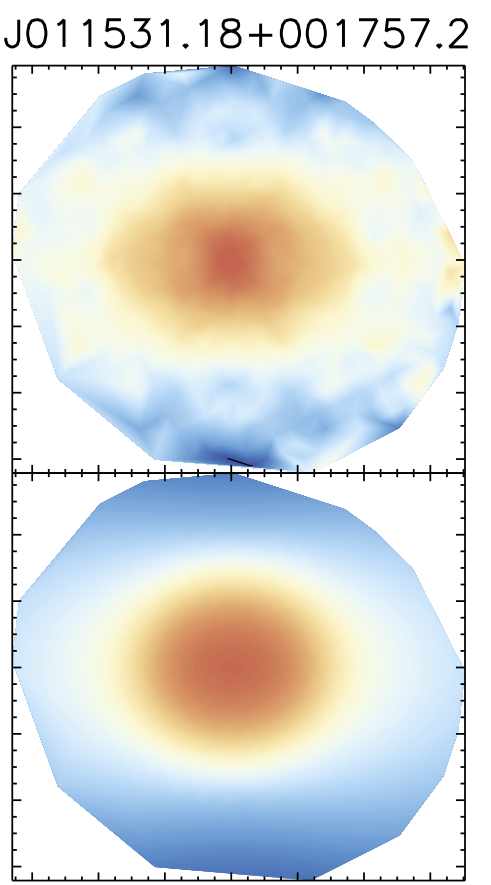}
\includegraphics[height=0.135\textheight]{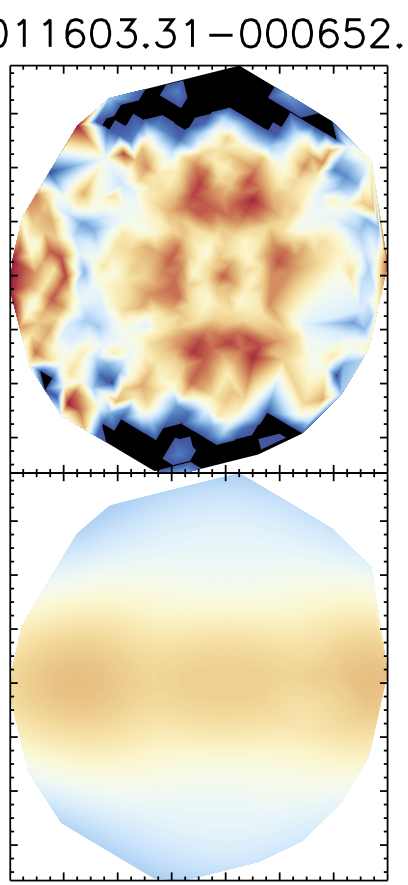}
\includegraphics[height=0.135\textheight]{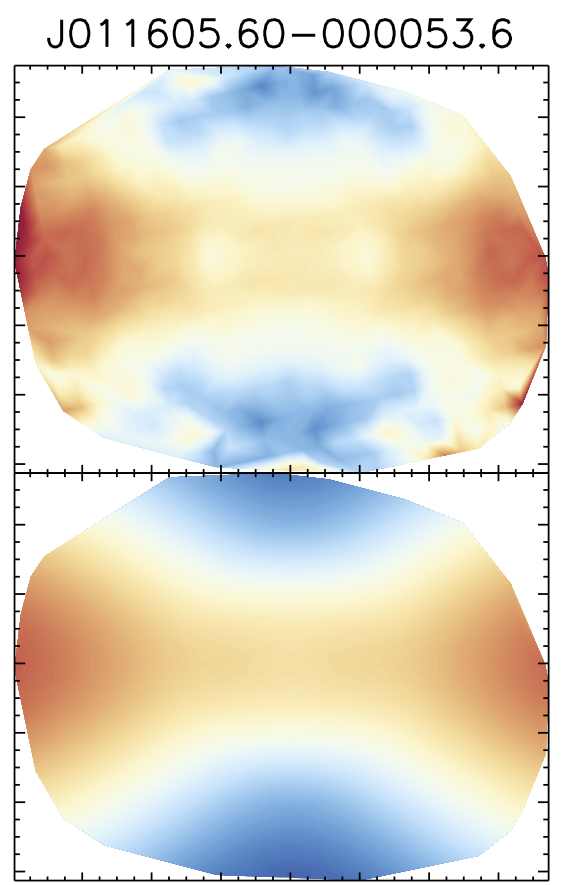}
\includegraphics[height=0.135\textheight]{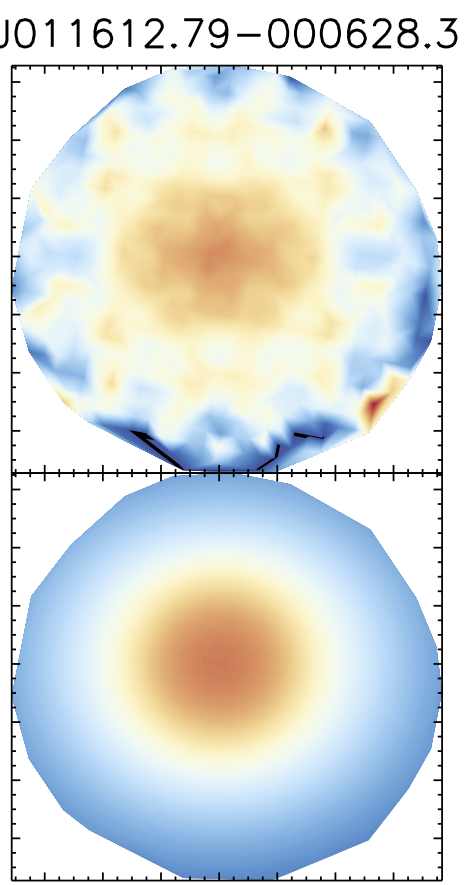}
\includegraphics[height=0.135\textheight]{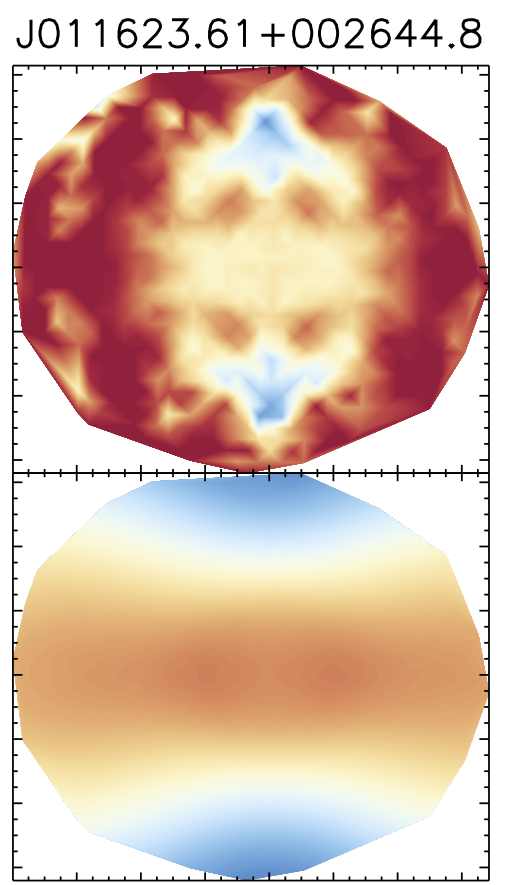}
\includegraphics[height=0.135\textheight]{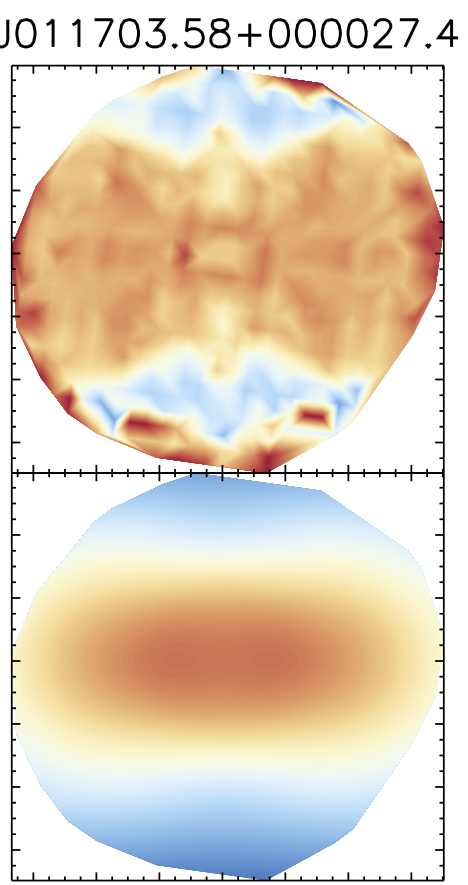}
\includegraphics[height=0.135\textheight]{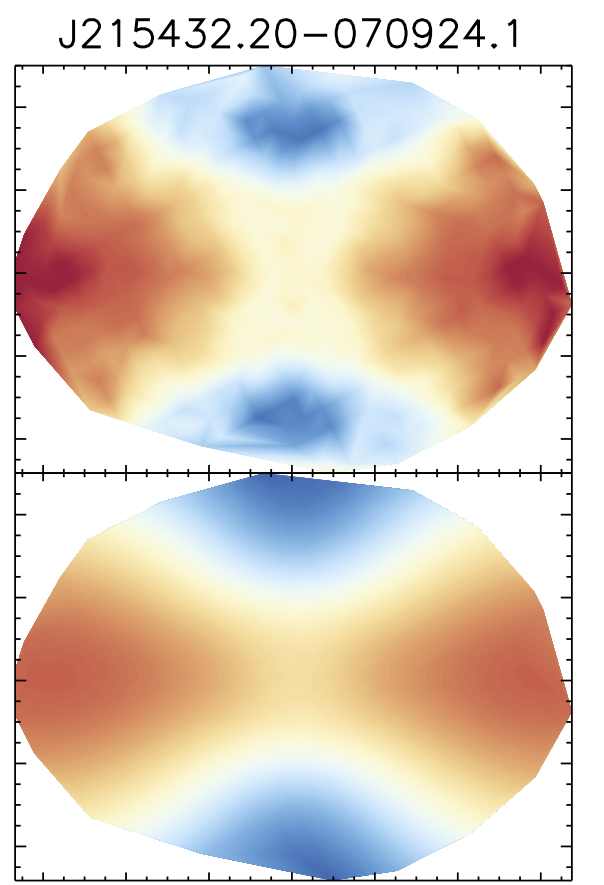}
\includegraphics[height=0.135\textheight]{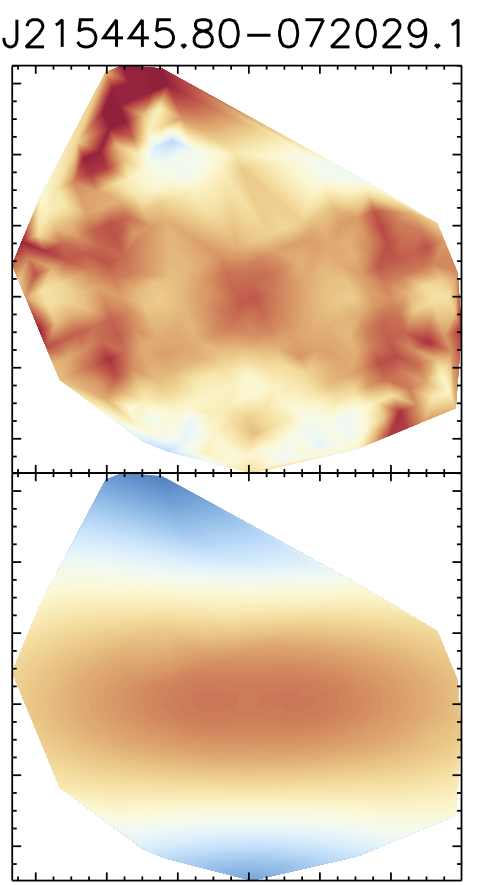}
\caption{Observed (top) and best-fitting JAM model (bottom) V$_{rms}$ maps for all 105 SAMI Pilot galaxies.}
\label{fig:all_jam_figures}
\end{figure*}
\begin{figure*}
\includegraphics[height=0.135\textheight]{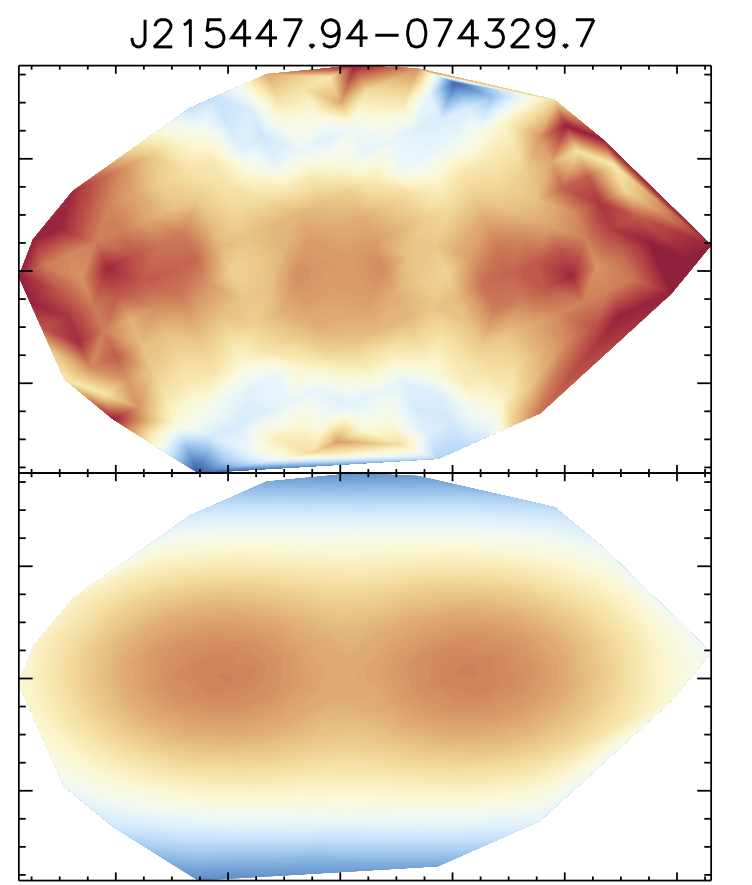}
\includegraphics[height=0.135\textheight]{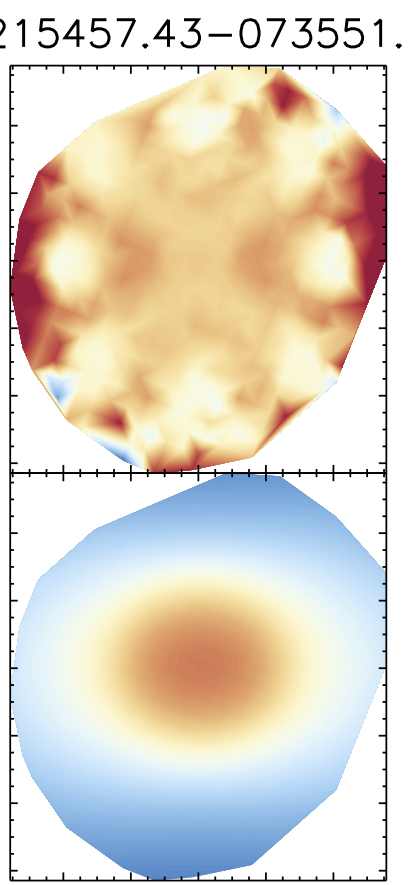}
\includegraphics[height=0.135\textheight]{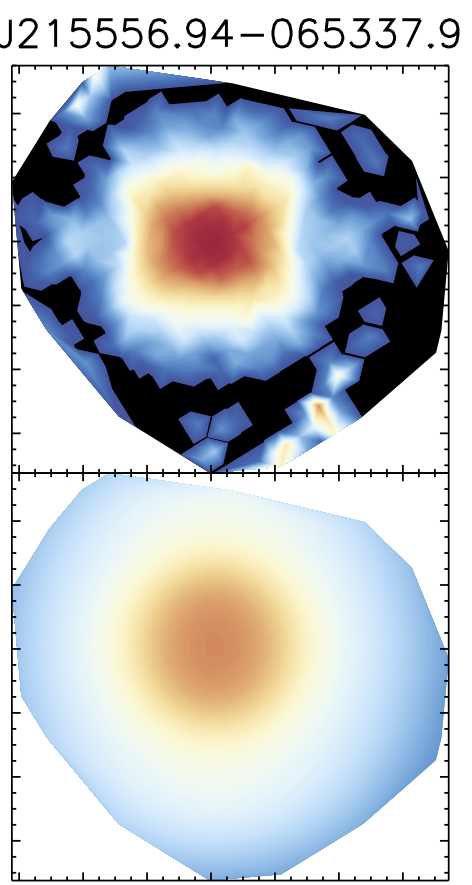}
\includegraphics[height=0.135\textheight]{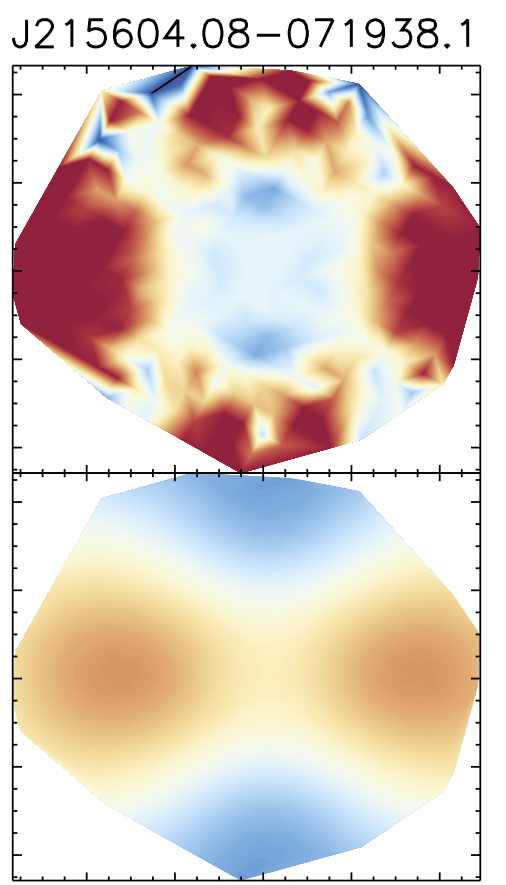}
\includegraphics[height=0.135\textheight]{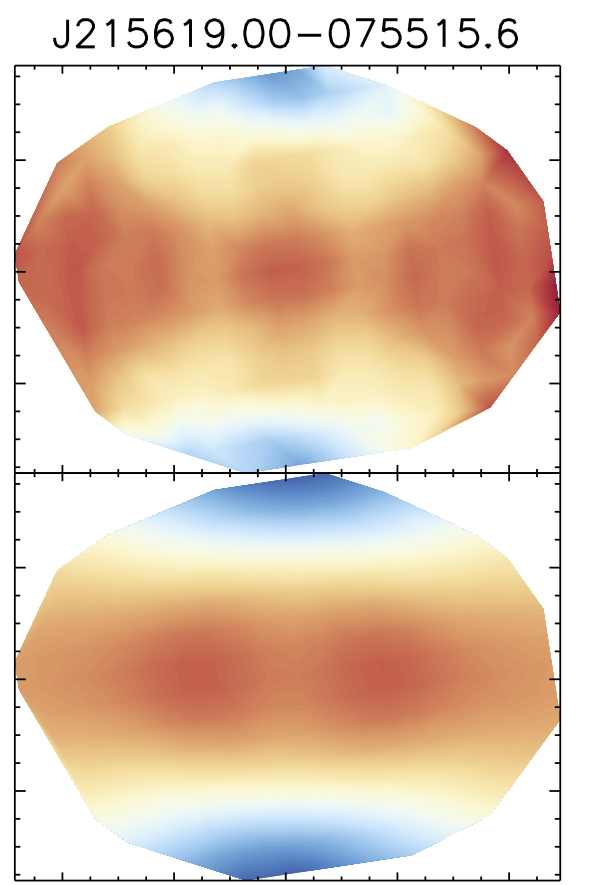}
\includegraphics[height=0.135\textheight]{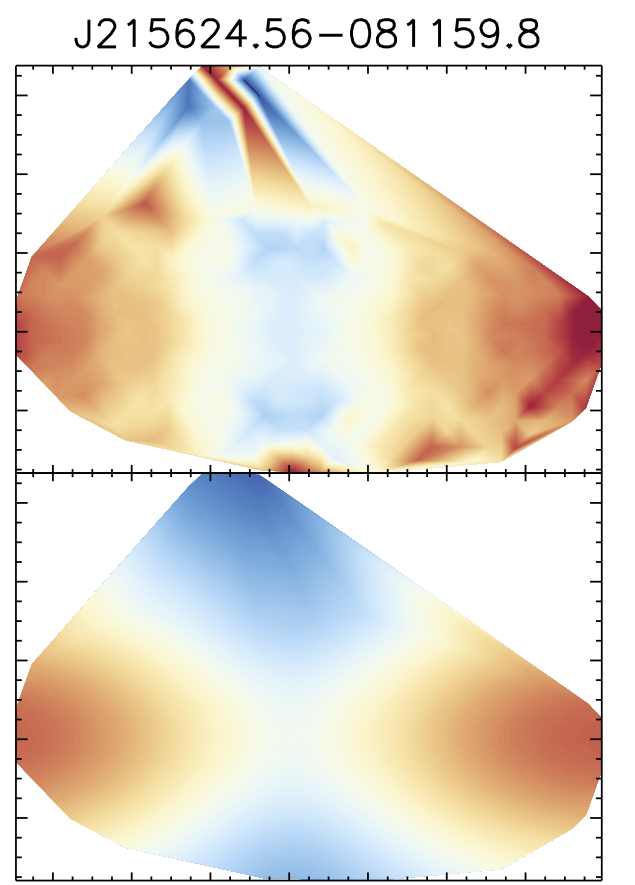}
\includegraphics[height=0.135\textheight]{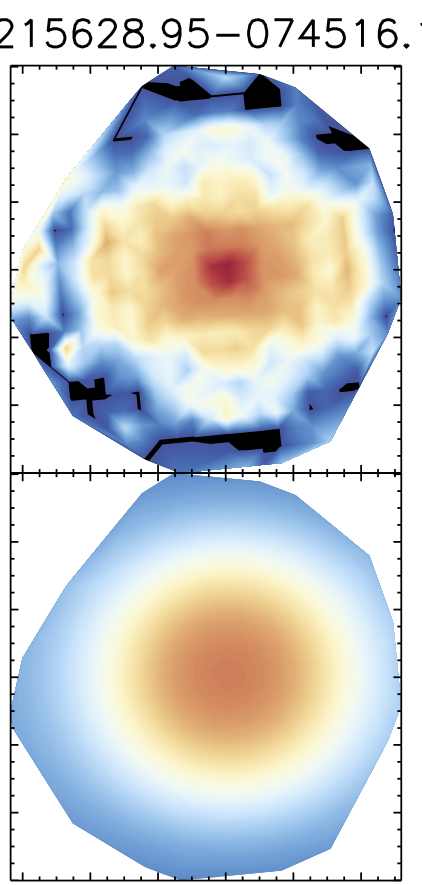}
\includegraphics[height=0.135\textheight]{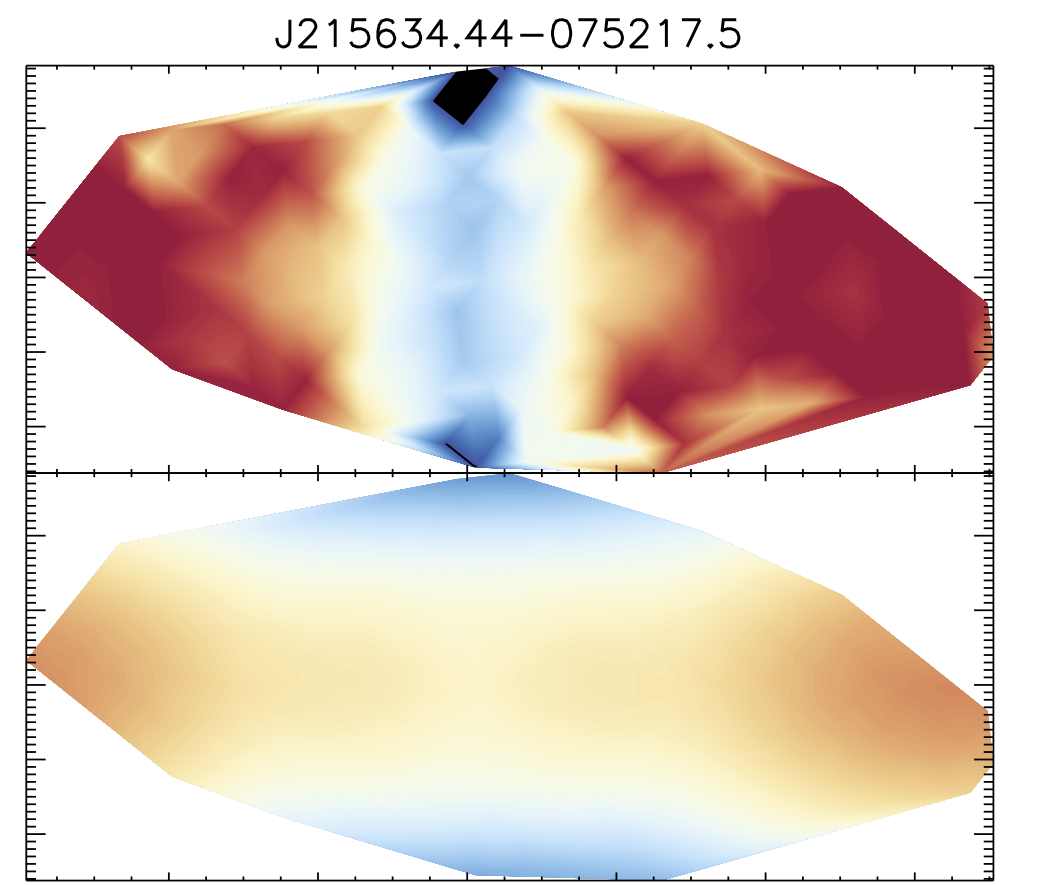}
\includegraphics[height=0.135\textheight]{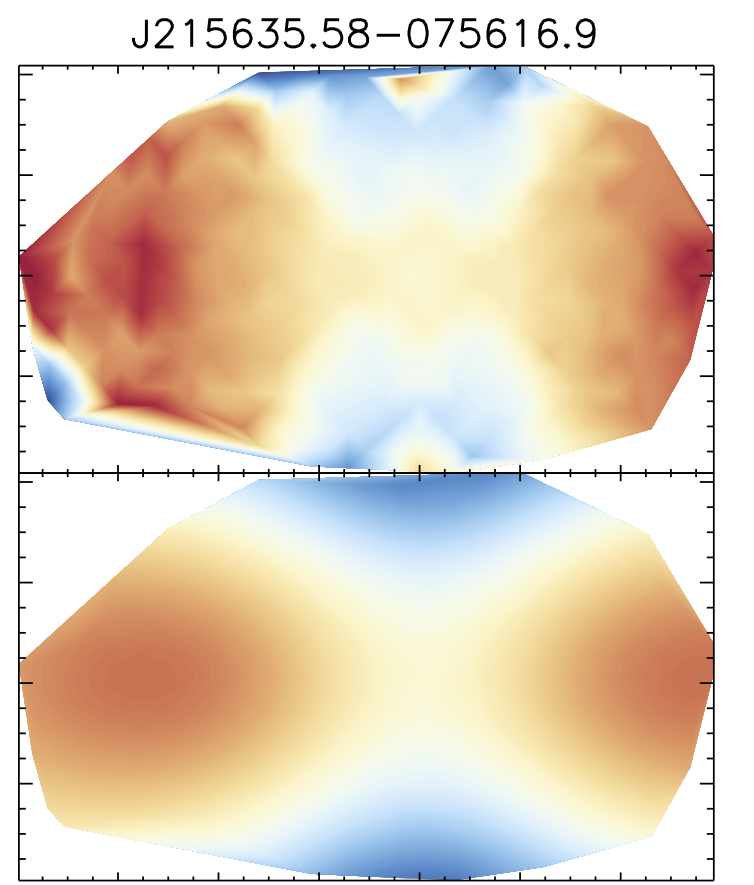}
\includegraphics[height=0.135\textheight]{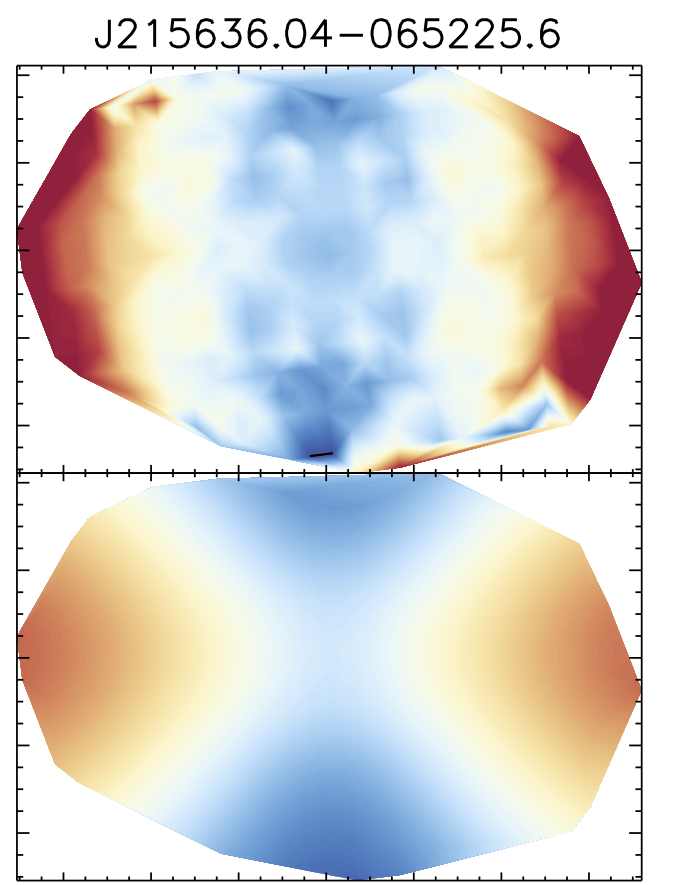}
\includegraphics[height=0.135\textheight]{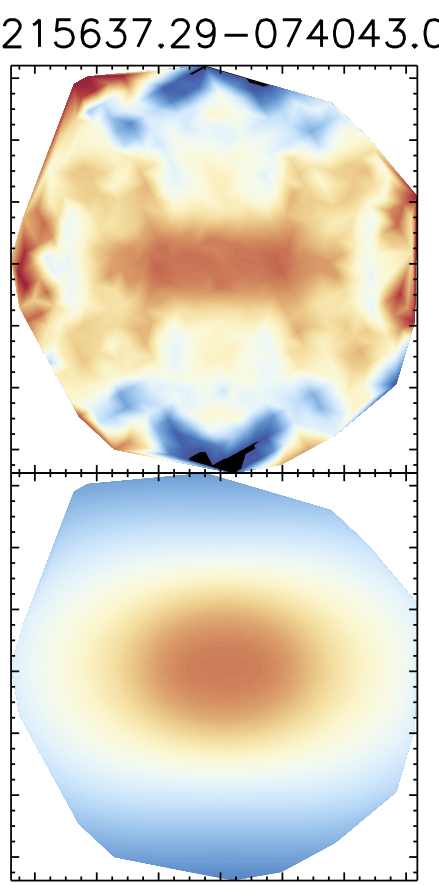}
\includegraphics[height=0.135\textheight]{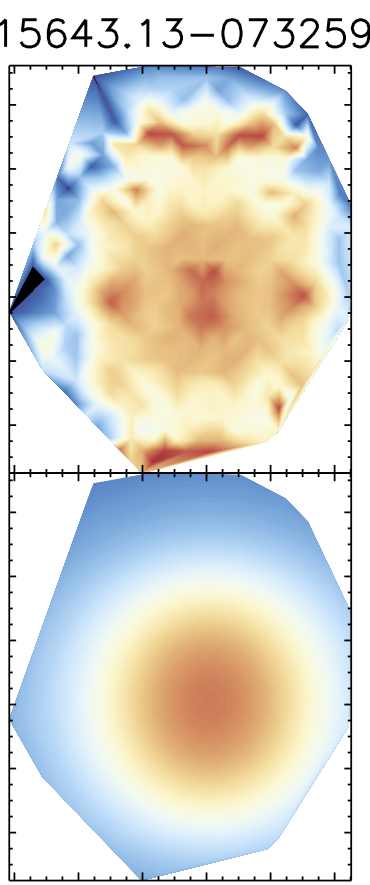}
\includegraphics[height=0.135\textheight]{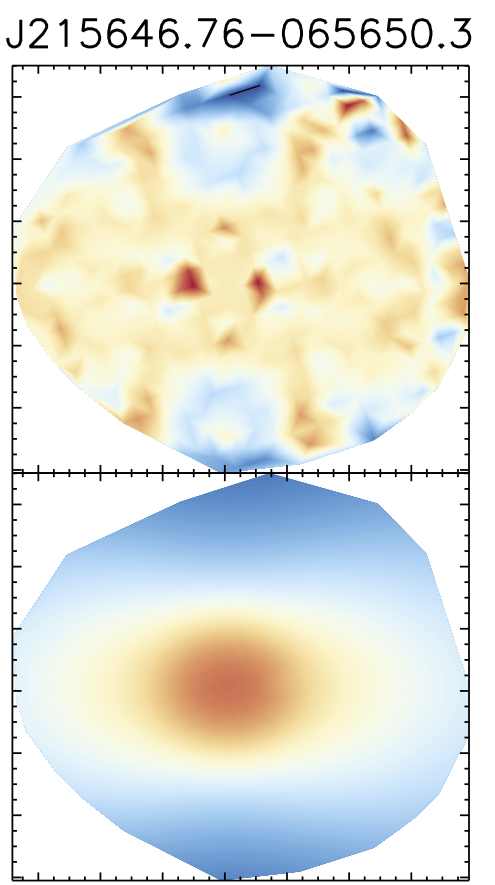}
\includegraphics[height=0.135\textheight]{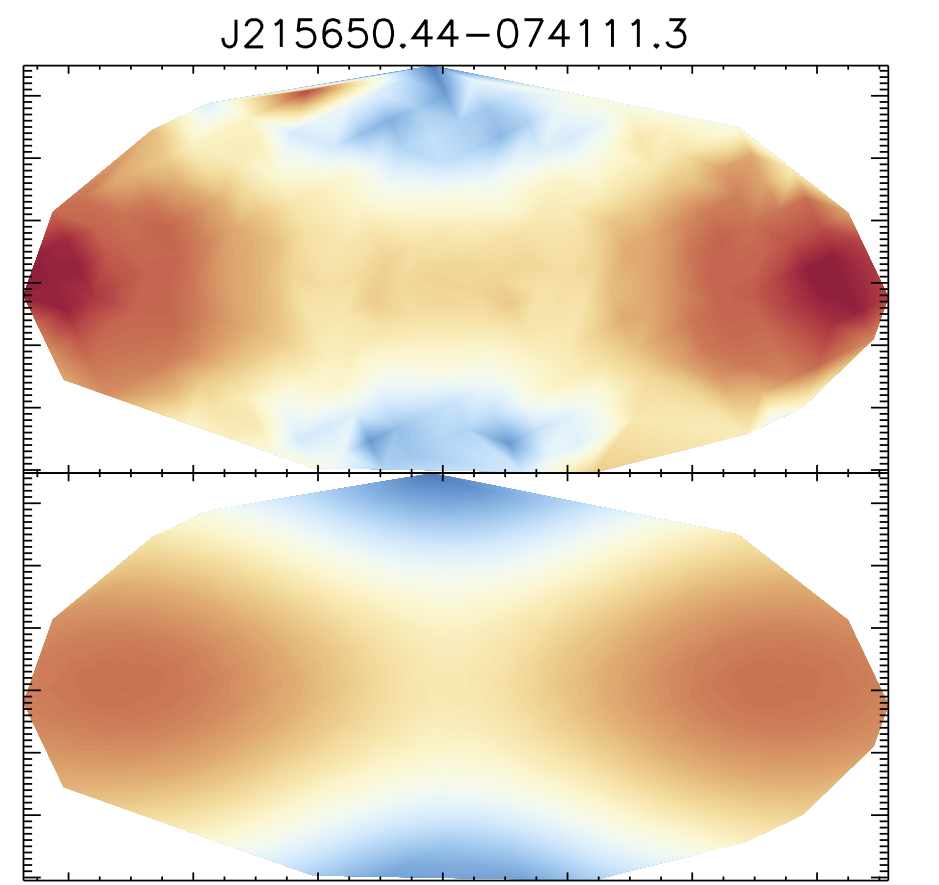}
\includegraphics[height=0.135\textheight]{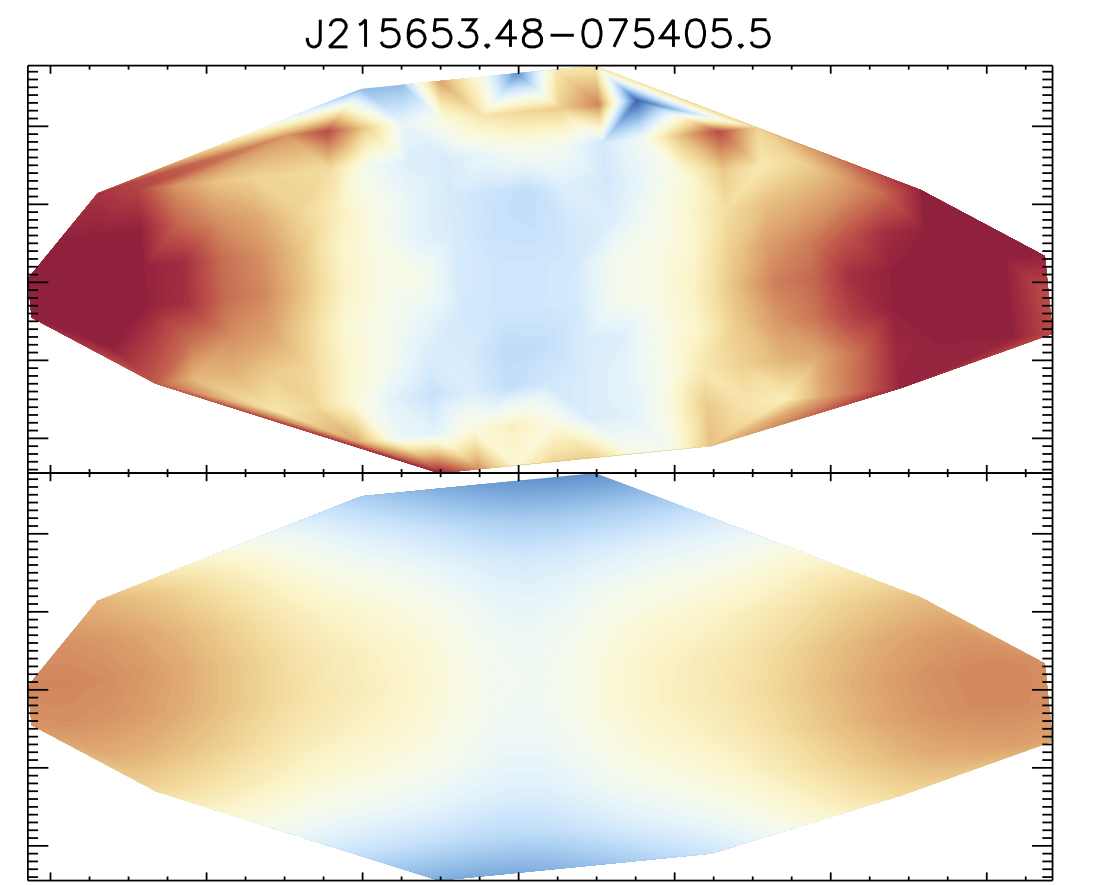}
\includegraphics[height=0.135\textheight]{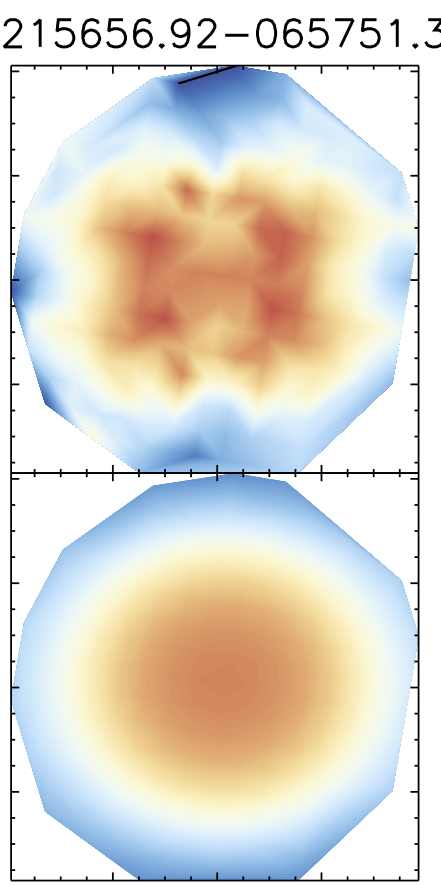}
\includegraphics[height=0.135\textheight]{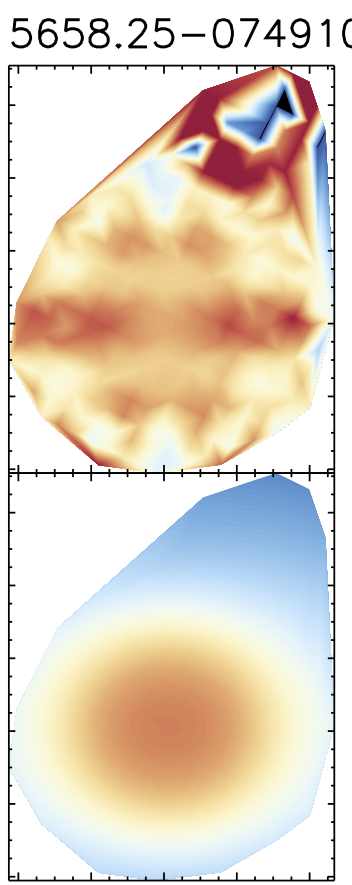}
\includegraphics[height=0.135\textheight]{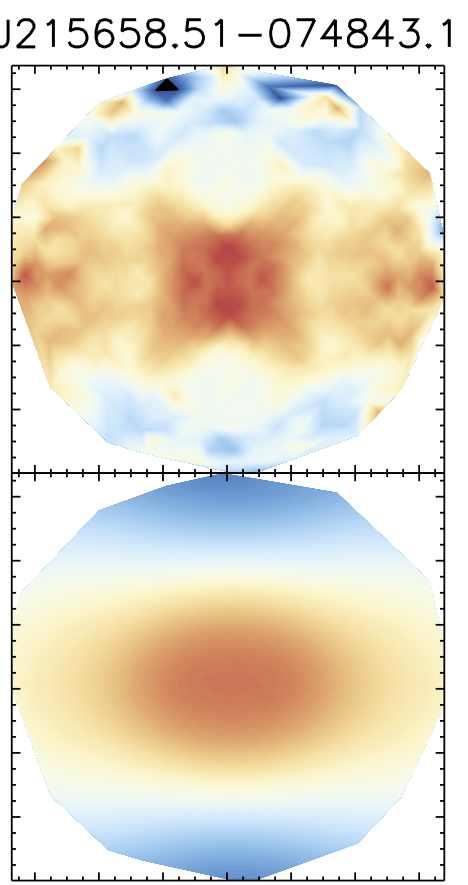}
\includegraphics[height=0.135\textheight]{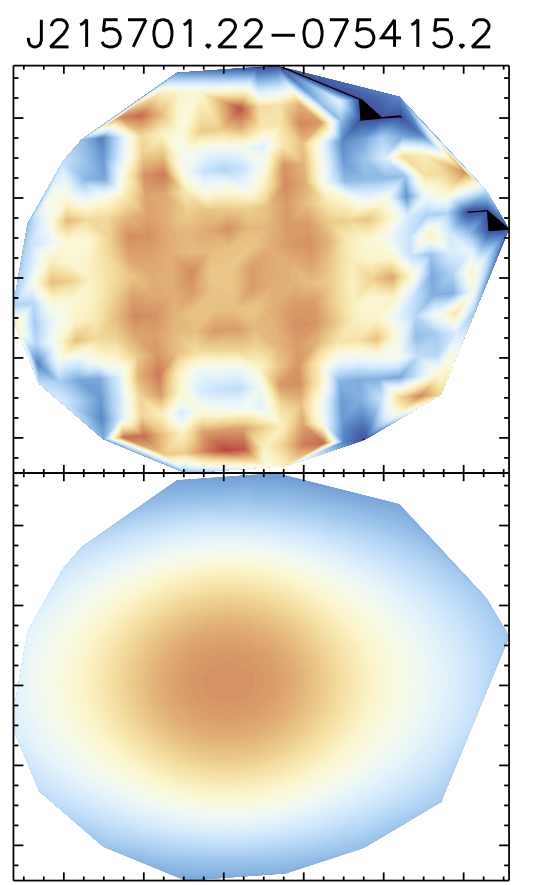}
\includegraphics[height=0.135\textheight]{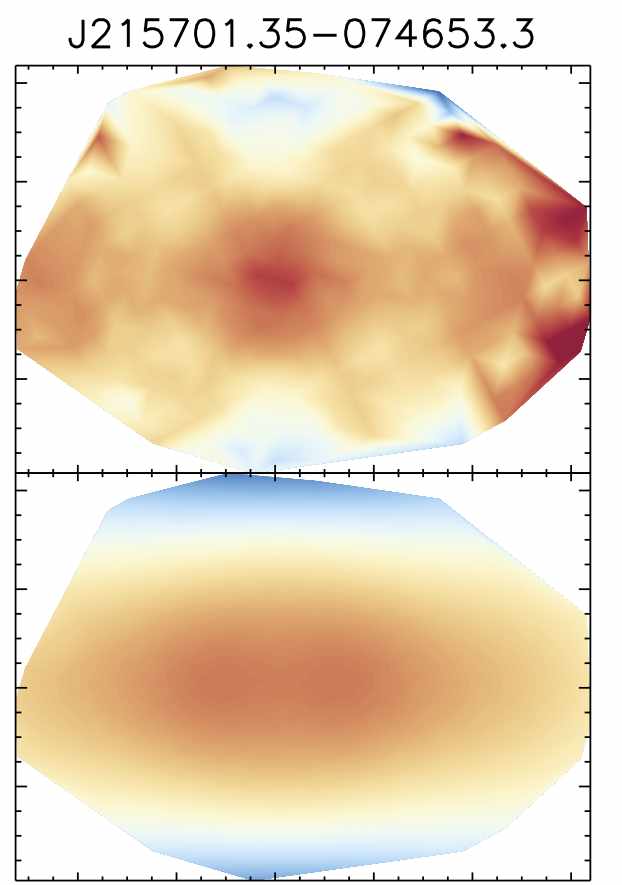}
\includegraphics[height=0.135\textheight]{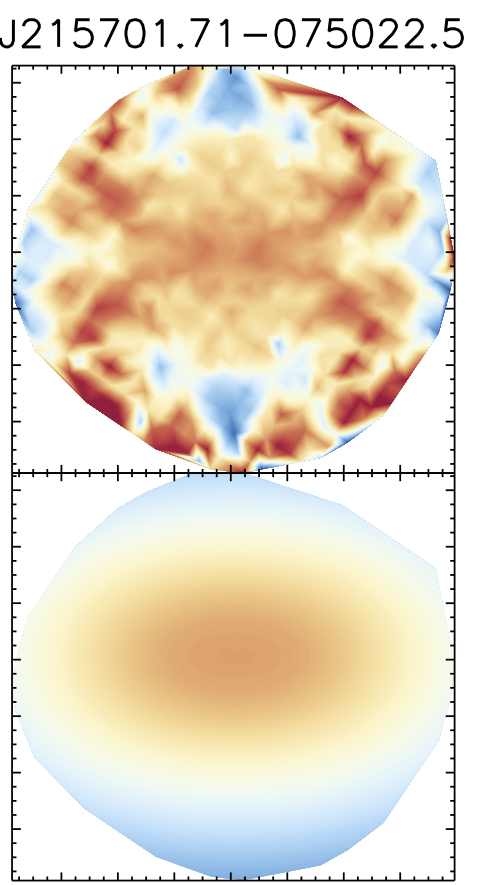}
\includegraphics[height=0.135\textheight]{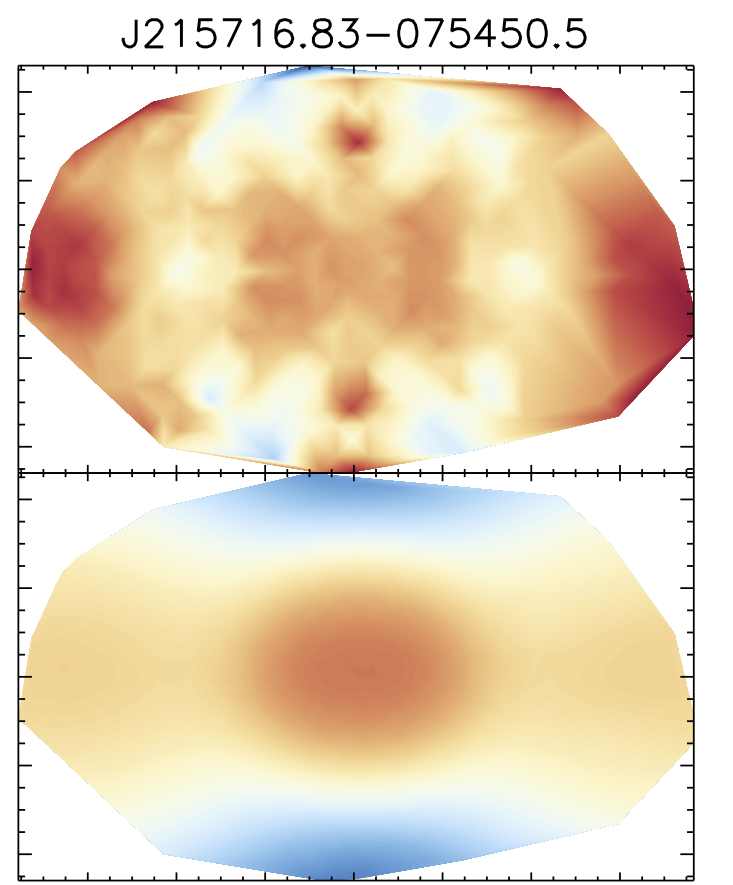}
\includegraphics[height=0.135\textheight]{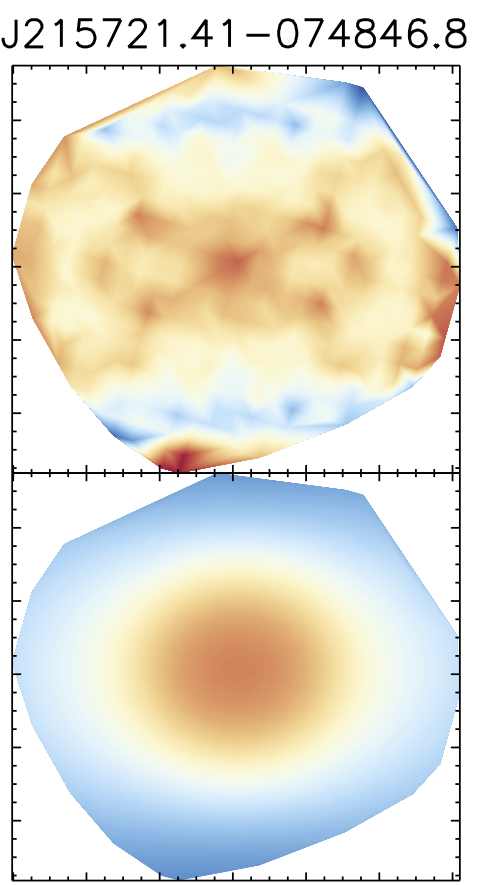}
\includegraphics[height=0.135\textheight]{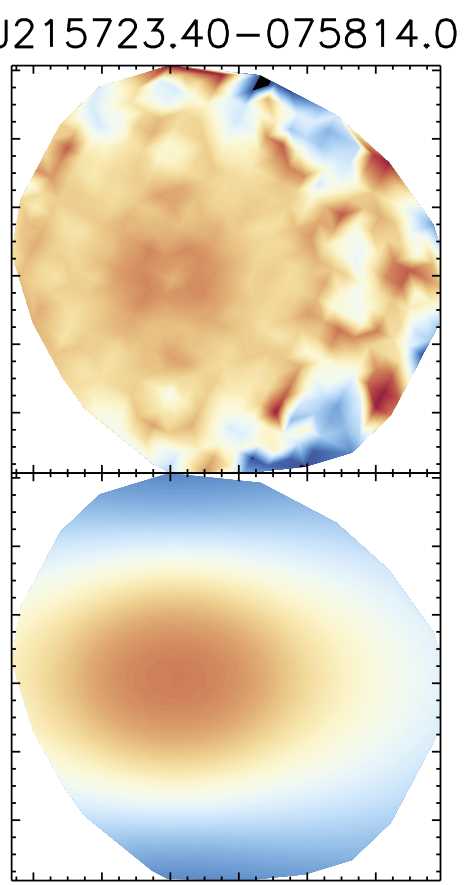}
\includegraphics[height=0.135\textheight]{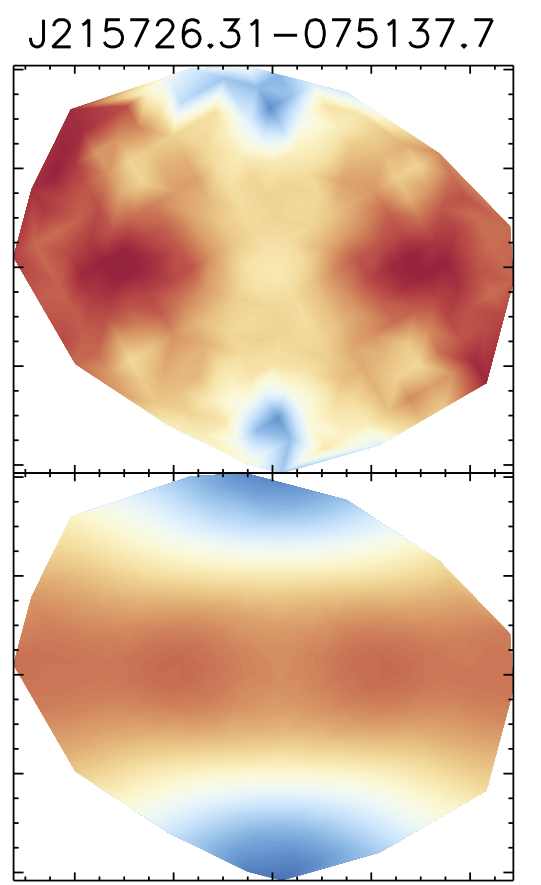}
\includegraphics[height=0.135\textheight]{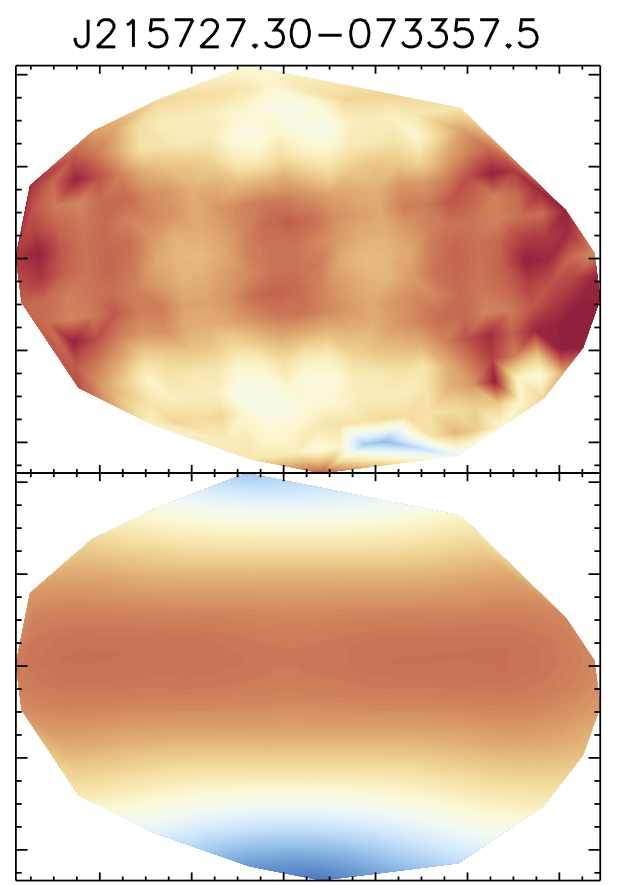}
\includegraphics[height=0.135\textheight]{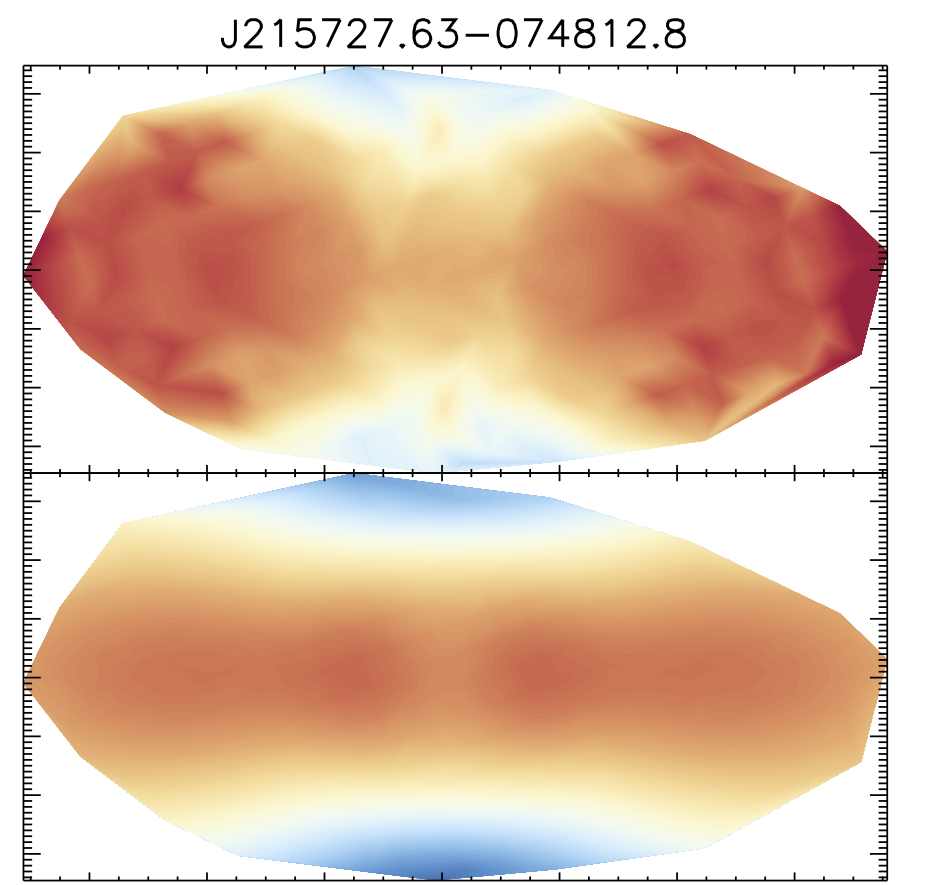}
\includegraphics[height=0.135\textheight]{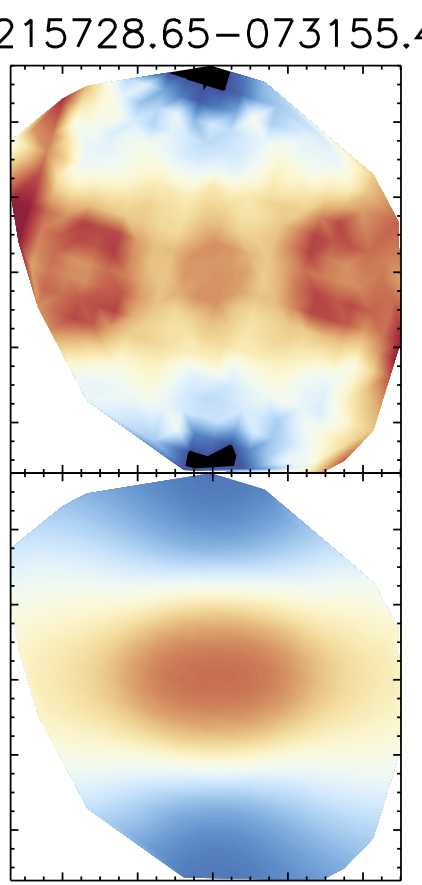}
\includegraphics[height=0.135\textheight]{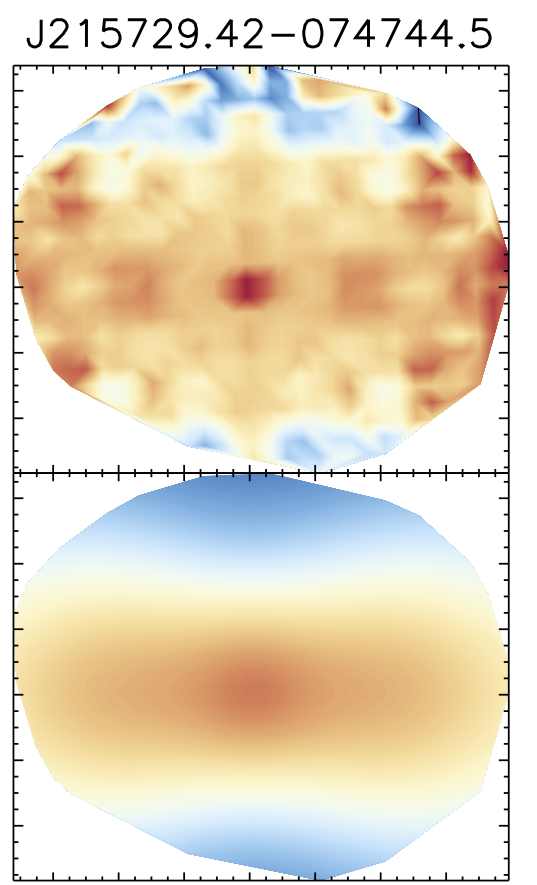}
\includegraphics[height=0.135\textheight]{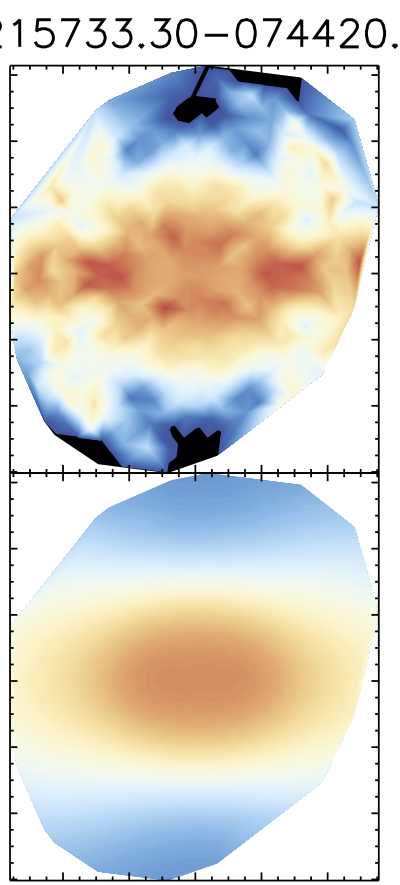}
\includegraphics[height=0.135\textheight]{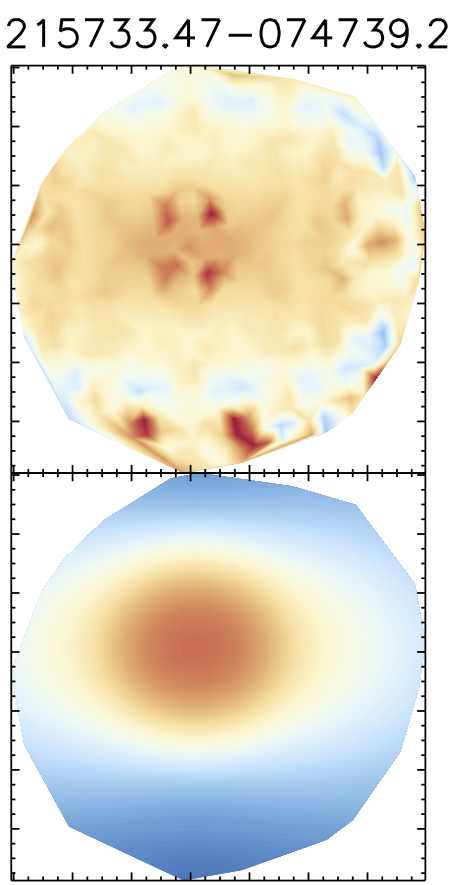}
\includegraphics[height=0.135\textheight]{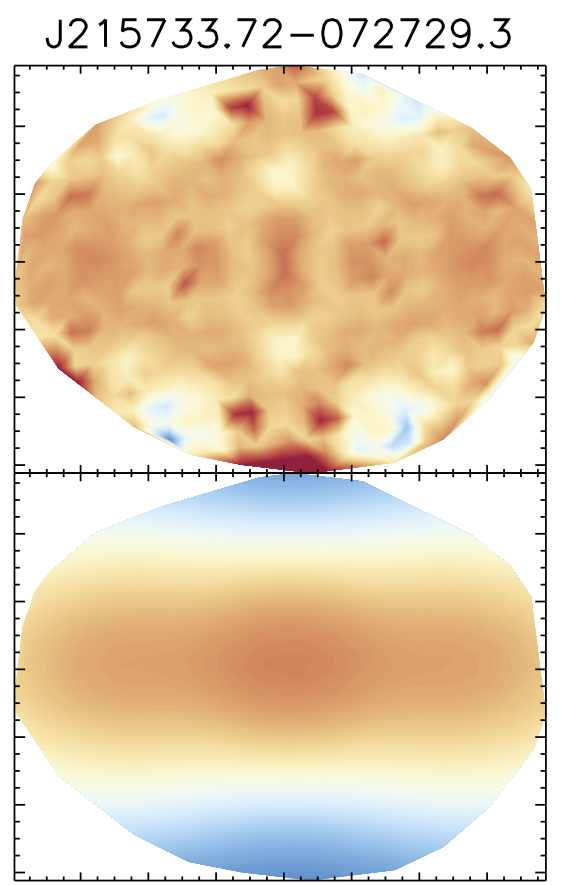}
\includegraphics[height=0.135\textheight]{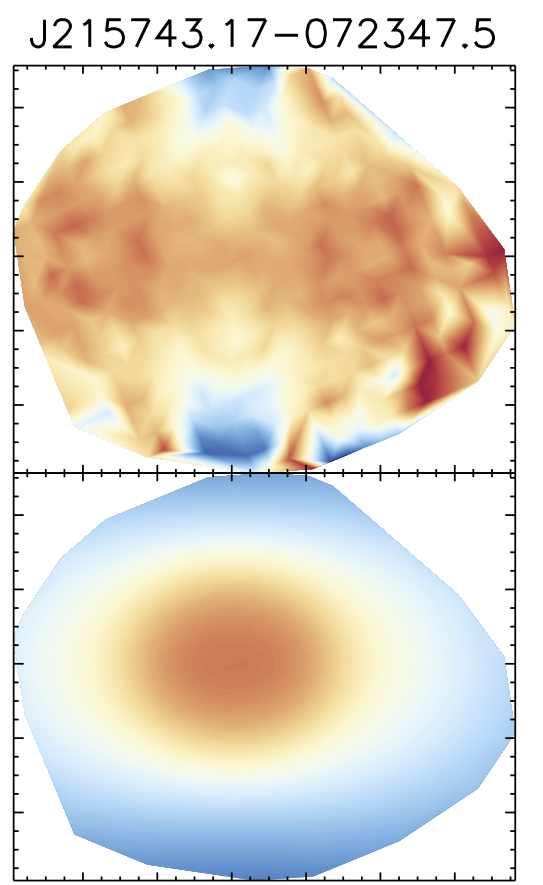}
\includegraphics[height=0.135\textheight]{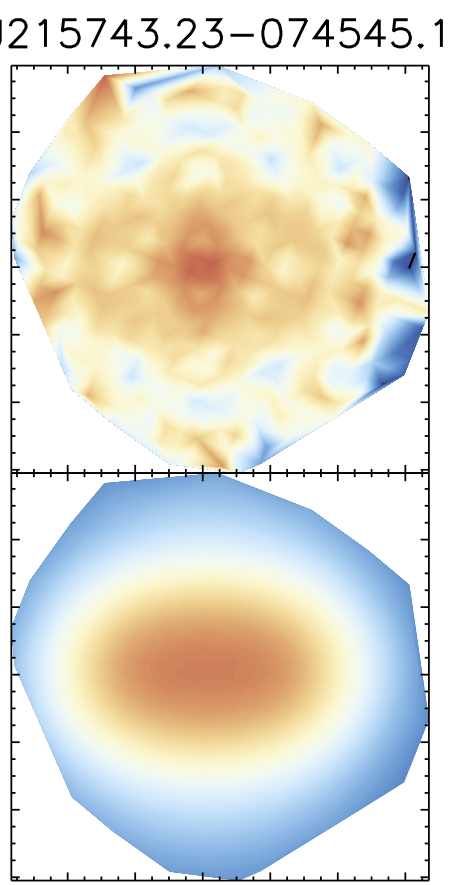}
\includegraphics[height=0.135\textheight]{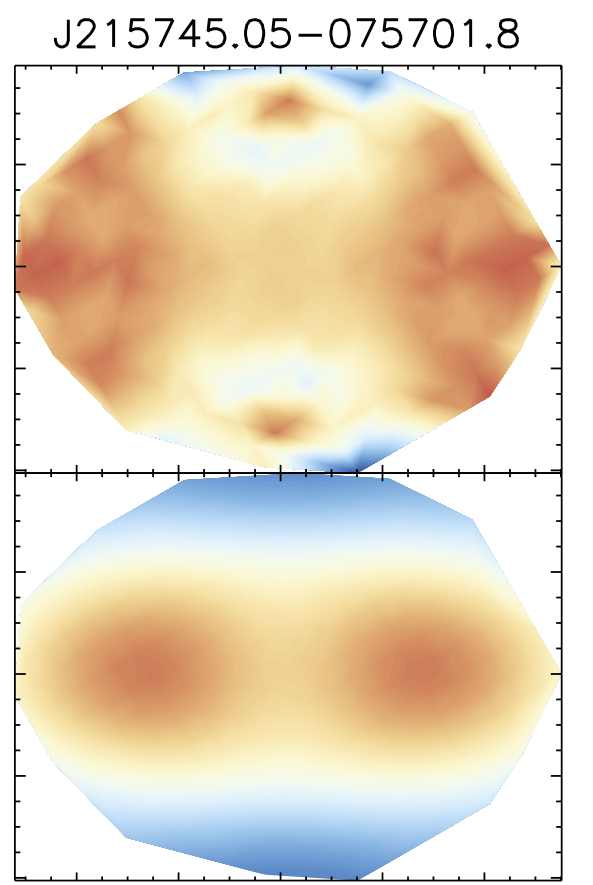}
\includegraphics[height=0.135\textheight]{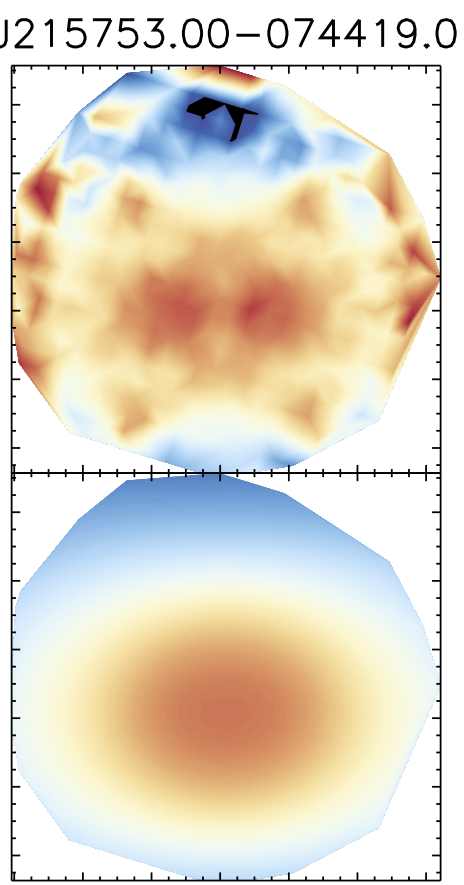}
\includegraphics[height=0.135\textheight]{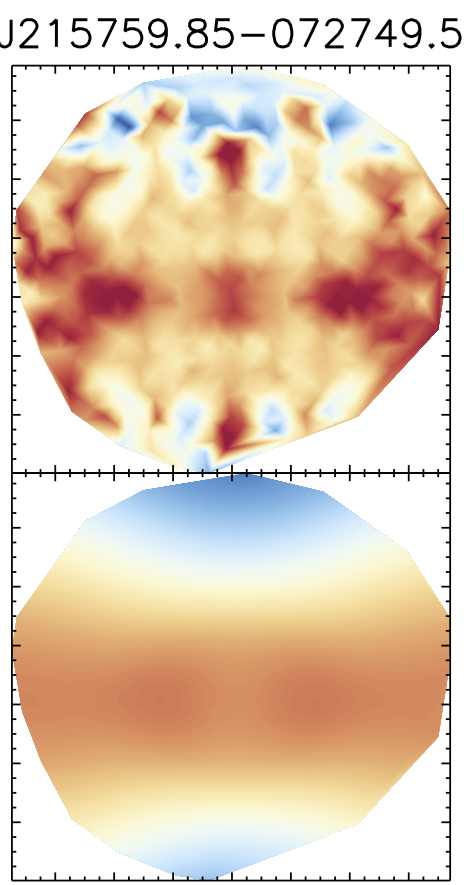}
\includegraphics[height=0.135\textheight]{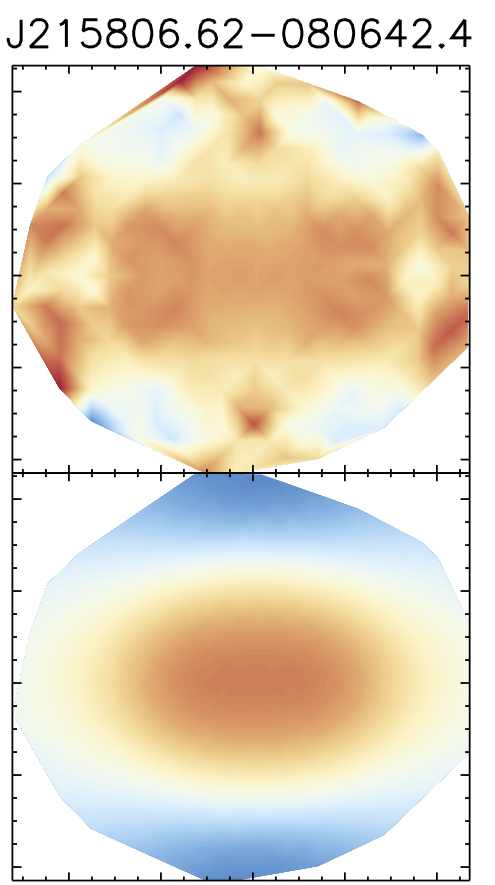}
\includegraphics[height=0.135\textheight]{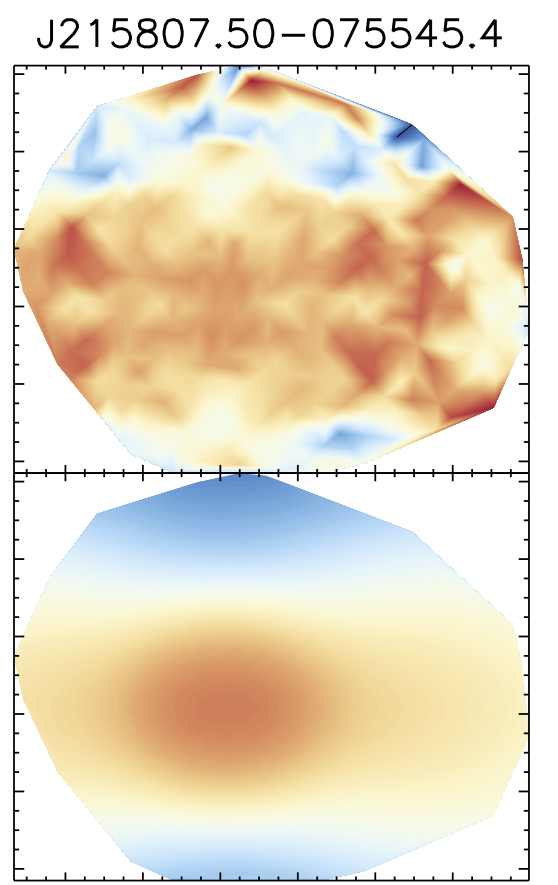}
\includegraphics[height=0.135\textheight]{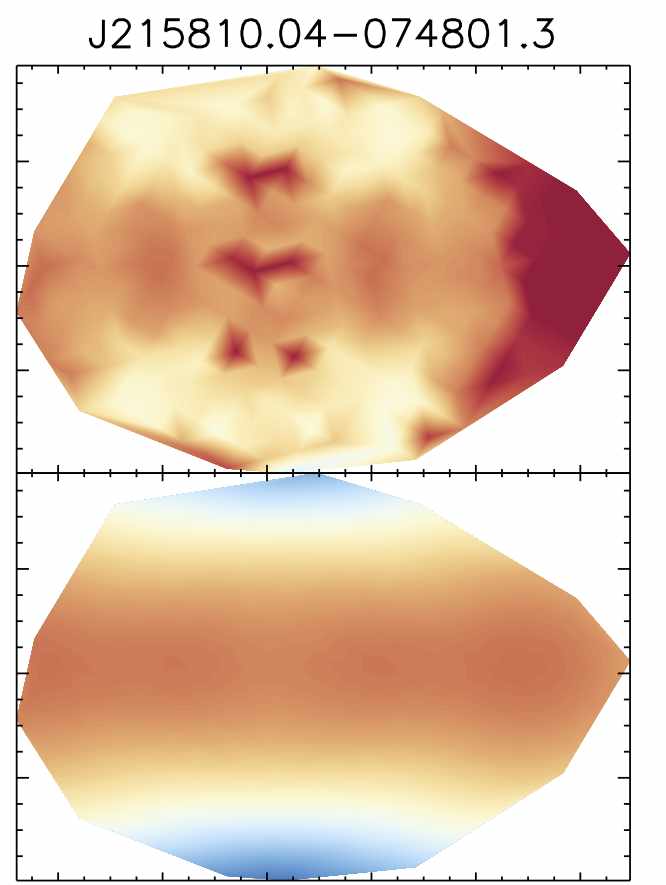}
\includegraphics[height=0.135\textheight]{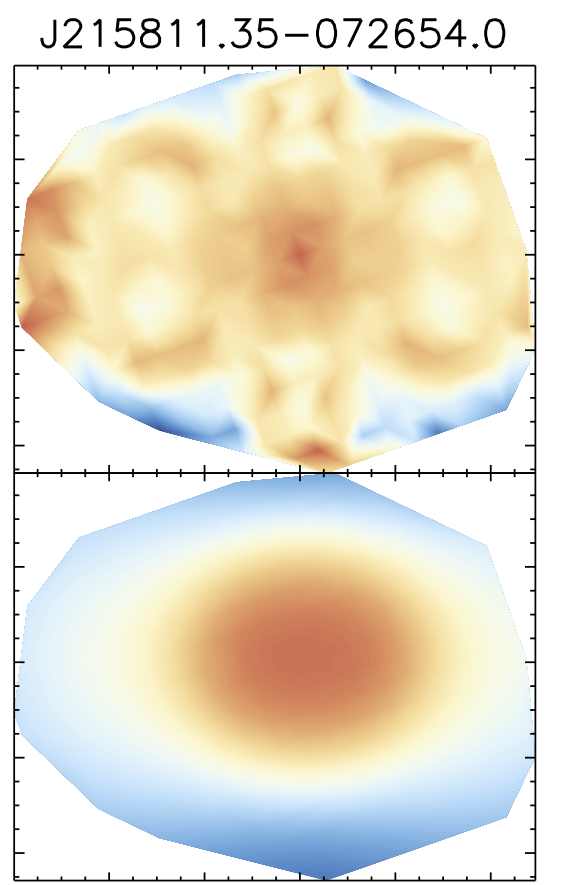}
\includegraphics[height=0.135\textheight]{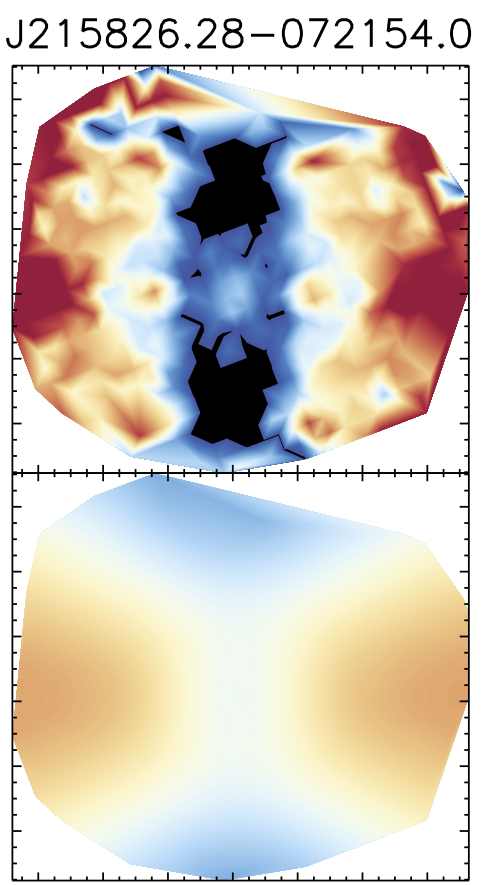}
\includegraphics[height=0.135\textheight]{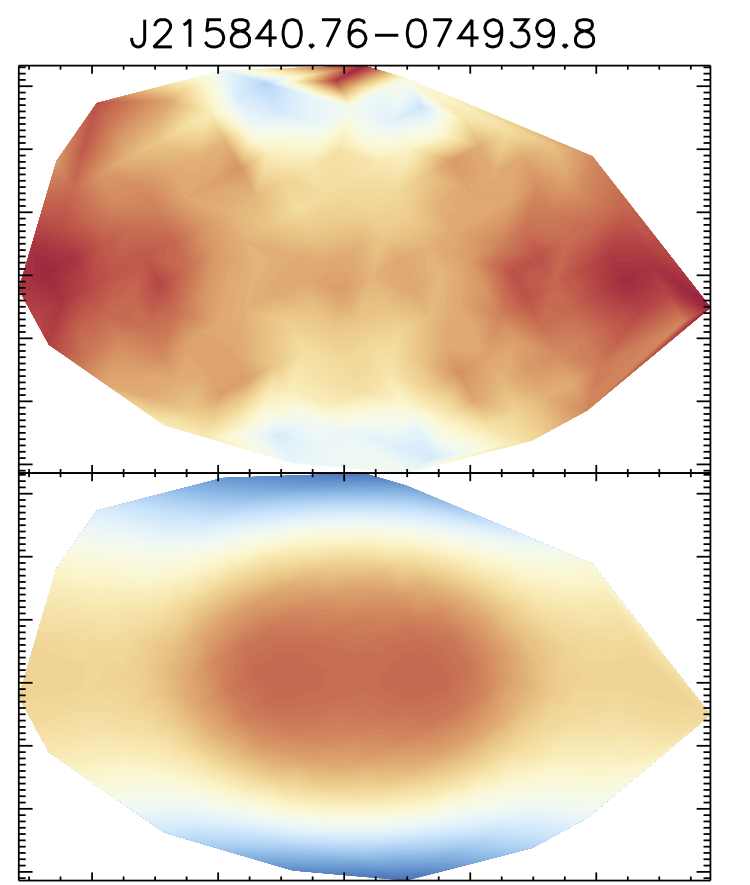}
\includegraphics[height=0.135\textheight]{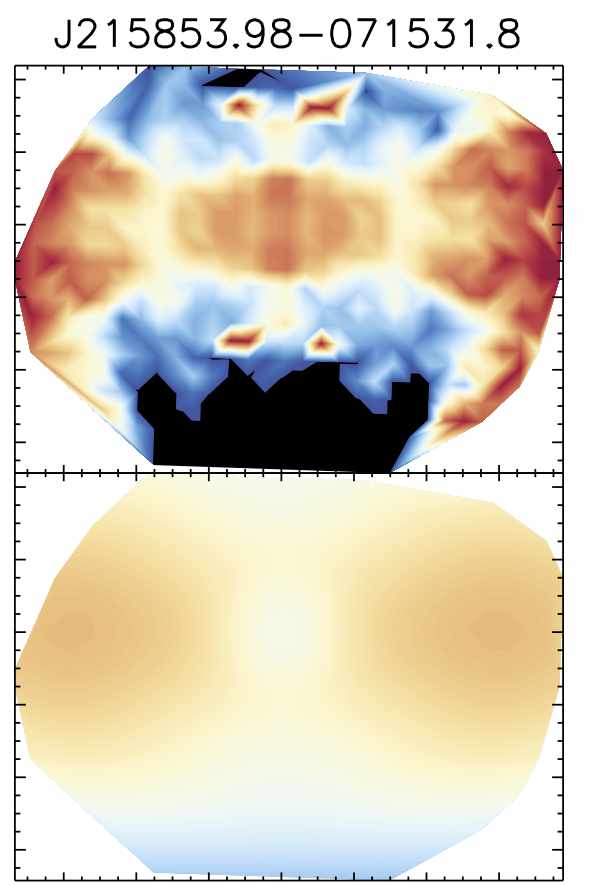}
\includegraphics[height=0.135\textheight]{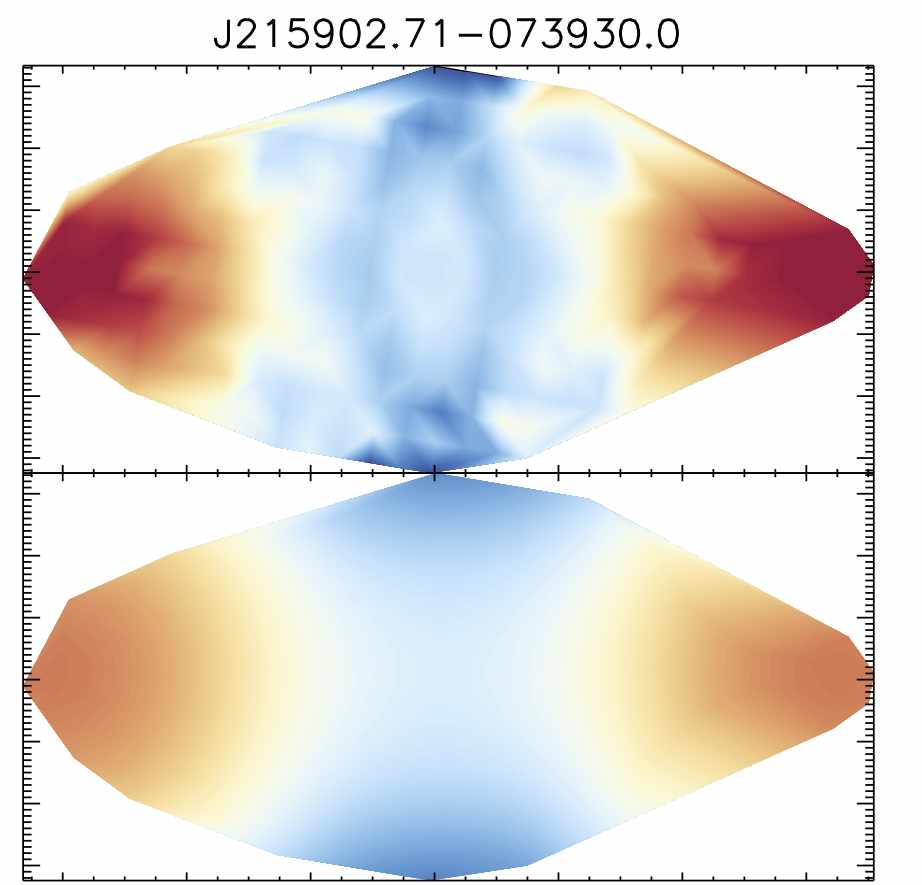}
\includegraphics[height=0.135\textheight]{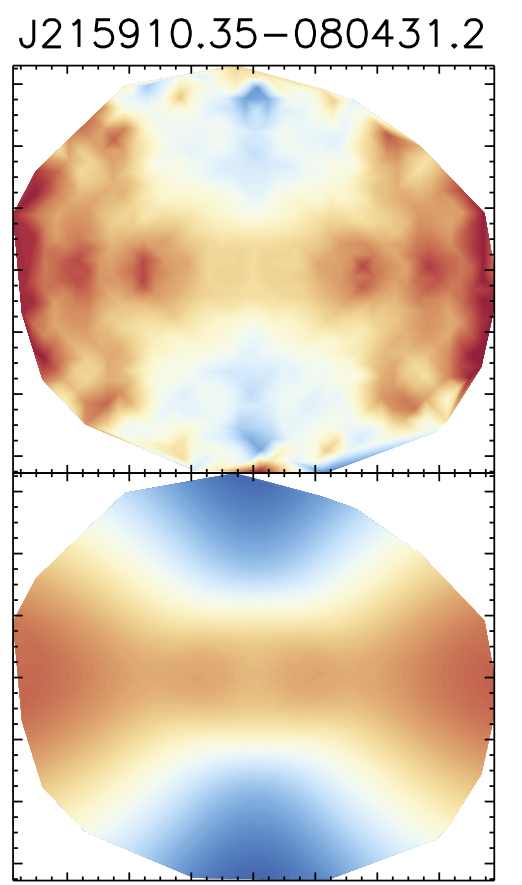}
\includegraphics[height=0.135\textheight]{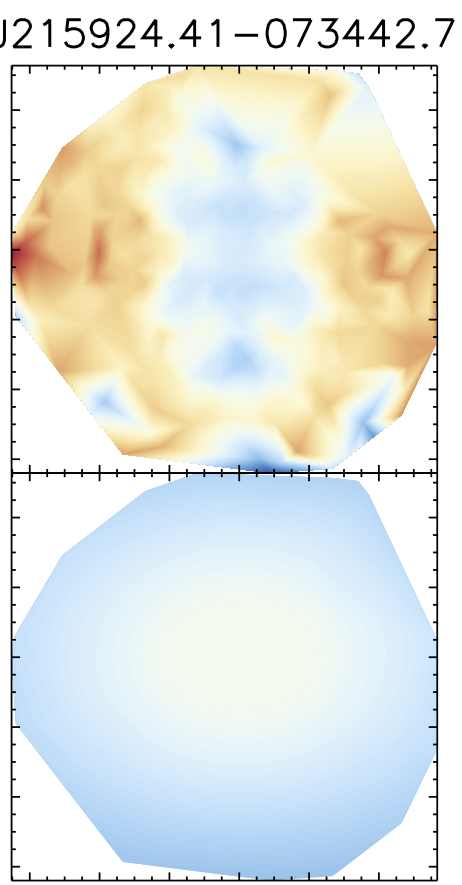}
\includegraphics[height=0.135\textheight]{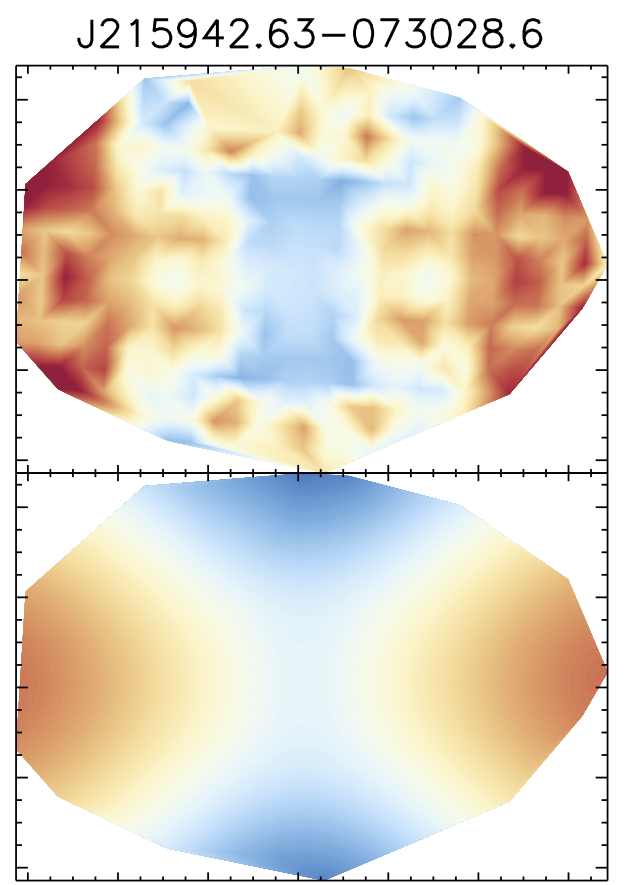}
\includegraphics[height=0.135\textheight]{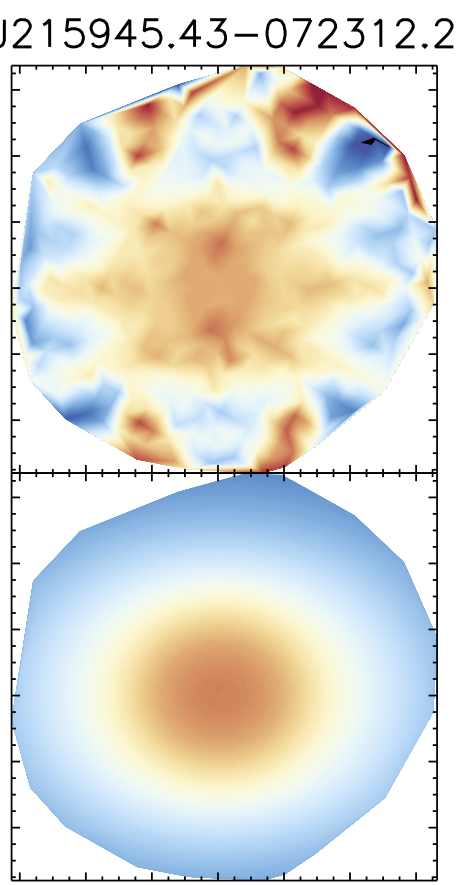}
\\
Figure \ref{fig:all_jam_figures} continued.
\end{figure*}
\label{lastpage}

\end{document}